\documentclass[a4paper, UKenglish]{article}

\usepackage[UKenglish]{babel}
\usepackage{amsthm, amsmath, amssymb, amsfonts}
\usepackage[noend]{algpseudocode}
\usepackage{algorithm}
\usepackage{algorithmicx}
\usepackage{etoolbox}
\usepackage{tikz}

\usetikzlibrary{tikzmark}
\usetikzlibrary{calc}
\usetikzlibrary{matrix,arrows.meta,decorations.pathreplacing}
\usepackage{enumerate}\usepackage{mathtools}
\usepackage{tabto}
\usepackage{graphicx}
\usepackage{calc}
\usepackage{scrextend}
\usepackage{xcolor}
\usepackage{float}
\usepackage{tabularx}
\usepackage{stackengine}
\usepackage{makeidx}
\usepackage{pgfplots}
\usepackage{cleveref}

\usetikzlibrary{shapes}
\usetikzlibrary{shapes.gates.logic.US}

\date{}

\title{Delay-Optimum Adder Circuits with Linear Size}

\author{Ulrich Brenner\footnote{brenner@or.uni-bonn.de, corresponding author},\, Benjamin David G\"org
\vspace*{3mm}\\ \vspace*{-3mm}
\normalsize Research Institute for Discrete Mathematics, University of Bonn, Lenn\'{e}str. 2, 53113 Bonn, Germany
}

\newtheorem{theorem}{Theorem}[section]
\newtheorem{lemma}[theorem]{Lemma}
\newtheorem{corollary}[theorem]{Corollary}
\newtheorem{proposition}[theorem]{Proposition}

\theoremstyle{definition}
\newtheorem{definition}[theorem]{Definition}
\newtheorem{example}[theorem]{Example}

\theoremstyle{remark}
\newtheorem*{remark}{Remark}

\newcommand{\alignedintertext}[1]{%
	\noalign{%
		\vskip\belowdisplayshortskip
		\vtop{\hsize=\linewidth#1\par
			\expandafter}%
		\expandafter\prevdepth\the\prevdepth
	}%
}

\errorcontextlines\maxdimen

\newcommand{\ALGtikzmarkcolor}{black}
\newcommand{\ALGtikzmarkextraindent}{4pt}
\newcommand{\ALGtikzmarkverticaloffsetstart}{-.5ex}
\newcommand{\ALGtikzmarkverticaloffsetend}{-.5ex}
\makeatletter
\newcounter{ALG@tikzmark@tempcnta}

\newcommand\ALG@tikzmark@start{%
	\global\let\ALG@tikzmark@last\ALG@tikzmark@starttext%
	\expandafter\edef\csname ALG@tikzmark@\theALG@nested\endcsname{\theALG@tikzmark@tempcnta}%
	\tikzmark{ALG@tikzmark@start@\csname ALG@tikzmark@\theALG@nested\endcsname}%
	\addtocounter{ALG@tikzmark@tempcnta}{1}%
}

\def\ALG@tikzmark@starttext{start}
\newcommand\ALG@tikzmark@end{%
	\ifx\ALG@tikzmark@last\ALG@tikzmark@starttext
	\else
	\tikzmark{ALG@tikzmark@end@\csname ALG@tikzmark@\theALG@nested\endcsname}%
	\tikz[overlay,remember picture] \draw[\ALGtikzmarkcolor] let \p{S}=($(pic cs:ALG@tikzmark@start@\csname ALG@tikzmark@\theALG@nested\endcsname)+(\ALGtikzmarkextraindent,\ALGtikzmarkverticaloffsetstart)$), \p{E}=($(pic cs:ALG@tikzmark@end@\csname ALG@tikzmark@\theALG@nested\endcsname)+(\ALGtikzmarkextraindent,\ALGtikzmarkverticaloffsetend)$) in (\x{S},\y{S})--(\x{S},\y{E});%
	\fi
	\gdef\ALG@tikzmark@last{end}%
}

\apptocmd{\ALG@beginblock}{\ALG@tikzmark@start}{}{\errmessage{failed to patch}}
\pretocmd{\ALG@endblock}{\ALG@tikzmark@end}{}{\errmessage{failed to patch}}
\makeatother

\DeclareMathOperator{\dplus}{+\kern -0.4em+}

\newlength{\mywidth}
\newcommand{\klgleich}[1][\rule{0pt}{0pt}]{\stackrel{\text{\parbox[t]{\mywidth}{\centering #1}}}{\leq}}
\newcommand{\gleich}[1][\rule{0pt}{0pt}]{\stackrel{\text{\parbox[t]{\mywidth}{\centering #1}}}{=}}

\newcommand{\inn}[1][\rule{0pt}{0pt}]{\stackrel{\text{\parbox[t]{\mywidth}{\centering #1}}}{\in}}

\newlength{\boxwidth}
\newcommand{\linebox}[1]{\begin{center}\framebox[\textwidth][c]{\parbox[t]{\boxwidth}{#1}}\end{center}}

\makeindex

\begin{document}

\maketitle







\begin{abstract}
We present efficient circuits for the addition of binary numbers.
We assume that we are given arrival times for all input bits and optimize
the delay of the circuits, i.e.\ the time when the last output bit is
computed. This contains the classical optimization of depth as a special 
case where all arrival times are $0$. In this model, we present, among other
results, the fastest adder circuits of sub-quadratic size and the fastest
adder circuits of linear size. In particular, for adding two $n$-numbers 
we get a circuits with linear size and delay
$\log_2W+3\log_2\log_2n+4\log_2\log_2\log_2n +const$ where $\log_2W$ is
a lower bound for the delay of any adder circuit (no matter what size it has).
\end{abstract}


\section{Introduction}

We consider the well-studied problem of constructing fast and small circuits 
for binary addition.
The standard method for addition binary numbers includes the computation of carry bits.
Given two numbers $ a = \sum_{i=0}^{n-1} a_i2^i$ and $ b = \sum_{i=0}^{n-1} a_i2^i$
in binary representation, the carry bits are defined recursively by 
\begin{align*}
		c_0=&0\\
		c_{i+1}=&(a_i\land b_i)\lor ((a_i\oplus b_i)\land c_i)
\end{align*}
for $ i\in\{0,\dots, n-1\} $, where $ a_i\oplus b_i $ is true if and only if 
exactly one of $ a_i $ and $ b_i $ is true. Hence, the carry bit $ c_{i+1} $ 
is true if and only if $ g_i:=a_i\land b_i $ is true (i.e., $ c_{i+1} $ is 
\textit{generated} at position $ i $), or $ p_i:=a_i\oplus b_i $ is true and 
$ c_i $ is true (i.e., $ c_{i+1} $ is \textit{propagated} from position $ i $).
By definition, each carry bit can be computed via the formula 
\begin{equation}\label{eq::carry_bits}
   c_{i+1}=g_i\lor (p_i\land(g_{i-1}\lor(p_{i-1}\land(\dots\lor(p_1\land g_0)\dots)))).
\end {equation}
Given all carry bits $ c_0,\dots, c_n $, the sum $ a + b $ is now easily computed via 
   \[(a+b)_i=\begin{cases}
      c_i\oplus p_i &\text{if } i\in\{0,\dots, n-1\},\\
      c_n &\text{if } i=n.
   \end{cases}\]
Hence, the binary addition is reduced to computing all the carry bits $ c_1,\dots,$ $ c_n $
which are realized by \textsc{And-Or} paths on the input pairs $ g_i,p_i,\dots, g_1,p_1,g_0 $
for all $ i\in\{1,\dots, n\} $. A circuit solving this problem is called adder circuit. 

We propose a new set of circuits computing the carry bits $c_0,\dots,c_{n+1}$,
thus also proposing a circuit for binary addition. In our model, we assume that 
not all input signals are necessarily available at the same time but that we are
given arrival times $a(a_0),\dots,a(a_{n-1})$ and $a(b_0),\dots,a(b_{n-1})$
which directly lead to arrival times $a(g_0),\dots,a(g_{n-1})$ and $a(p_0),\dots,a(p_{n-1})$.
We assume a delay of one for all \textsc{And}- and \textsc{Or}-gates, which is
reasonable since in practice the delay of both gates different only slightly on
real chips. Our goal is to construct circuits such that the last arrival time
on a carry bit is minimized. This model generalizes the optimization of circuit
depth (where all arrival times are assumed to be 0) and has been studied e.g.\
in \cite{Okl94}, \cite{Zim98}, \cite{Cho04}, \cite{HS17}, and \cite{BH19}.

Apart from the delay of a circuit, also its size (i.e.\ the number of gates)
is an important measurement of
quality. One could compute each carry bit separately but this would necessarily
lead to an at least quadratic size, which is unacceptable for large adder units.
The main task is to find a good tradeoff between the delay (depth) and the size 
of the circuit.
 	

\subsection{Our contributions}

We present the first circuits computing all carry bits with sub-quadratic size
and delay $\log_2 W + \mathcal{O}(\log_2 \log_2 n)$ where 
$ W=\sum_{i=0}^{n-1}\big(2^{a(p_i)}+2^{a(g_i)}\big) $
Note that $\log_2 W$ is a lower bound on the delay of 
{\it any} circuit depending on all input bits $(g_0,\dots,g_{n-1})$ and $(p_0,\dots,p_{n-1})$

\begin{table}[htb]
  \centering
\begin{tabular}[h]{|l|l|l|}\hline
          Delay & Size & Result \\ \hline\hline
			& & \\[-10pt]
 $ \log_2W+3\log_2\log_2n+5.007 $ & $ 2.422n\log_2^2n $ &  \Cref{mainthm1} \\ \hline
& & \\[-8pt]
 $ \log_2W+3\log_2\log_2n+3\log_2\log_2\log_2n+
    \begin{cases} 8.461, & \text{if } n\leq 2^{21}\\ 9.067 & \text{otherwise} \end{cases}$ 
&     \parbox{4.28cm}{$ 6.2n\log_2n $\tab \hspace{10pt} if $ n\leq 2^{21} $,\\
 $ 4.442n\log_2n $\tab \hspace{10pt} otherwise}
& \Cref{mainthm3}  \\[8pt] \hline
& & \\[-10pt]
$ \log_2W+3\log_2\log_2n+3\log_2\log_2\log_2n+11.085 $ & $ 24.436n\log_2\log_2n $ & \Cref{mainthm5} \\
\hline
& & \\[-8pt]
$ \log_2W+3\log_2\log_2n+4\log_2\log_2\log_2n+18.596 $ & $ 53.877n $ & \Cref{mainthm6} \\
\hline
& & \\[-10pt]
$ \log_2W + \sum_{i=2}^{j+1}3\log_2^{(i)}n + \mathcal{O}(1) $ & $ \mathcal{O}(n\log_2^{(j)}n) $ & \Cref{coralmostlinadders} \\[4pt]
\hline
& & \\[-10pt]
$ \log_2W + \sum_{i=2}^{j}3\log_2^{(i)}n + 4\log_2^{(j+1)}n + \mathcal{O}(1) $ & $ \mathcal{O}(n) $ & \Cref{corlinadders} \\[4pt]
\hline
\end{tabular}
\caption{Summary of the constructed adder circuits.}
\label{table_intro}
\end{table}
\Cref{table_intro} gives a detailed overview of the adder circuits presented in this paper.
These include
\begin{itemize}
   \item the fastest adder circuits with subquadratic size,
   \item the fastest adder circuits with size $\mathcal{O}(n \log_2^2 n$), $\mathcal{O}(n \log_2 n$) and
      $\mathcal{O}(n \log_2 \log_2 n$) each, and
   \item the fastest adder circuits with linear size.
\end{itemize}
Note that our delay bounds are in the range of the best known delay guarantees
for any circuit computing only the last carry bit, which is also 
$\log_2 W + \Theta(\log_2 \log_2 n)$ (see~\cite{BH19}).
The previously best known delay guarantee of an adder with a sub-quadratic size
was provided by Spirkl \cite{Sp14} who claimed a delay bound of order 
$ \lceil\log_2W\rceil + \mathcal{O}(\sqrt{\log_2n}) $ and a linear size.

\bigskip

The rest of the paper is organized as follows.
In Section~\ref{prelim}, the basic mathematical objects used here are introduced. 
Moreover, we present the main problems that we consider and give an overview of 
relevant previous results.
Section~\ref{analandprecirc} contains a detailed analysis of a group of circuits
proposed in~\cite{RSW}. In particular, we specify some constants in the bounds
on the delay and size. These constants have not been computed by the authors of
\cite{RSW} but they are important to us since we use these circuits in our
construction.
In Section~\ref{deloptadder}, we show (using the results by ~\cite{RSW} and 
\cite{BH19}) how to construct adder circuits with delay
at most $\log_2W+3\log_2\log_2n+5.007$ and size $\mathcal{O}(n \log_2^2 n)$.
In Section~\ref{sizereduction}, we describe an adder construction framework, 
presented in \cite{BS24}, which was used to linearize depth-optimizing
adders with sub-quadratic size while increasing the depth only by a small amount.
We will see that this framework is also suitable for the size reduction of
delay-optimizing adders. We apply the framework to our 
previously constructed adder to compute three more adder circuits, each having 
a larger delay bound and a smaller size bound than the last, asymptotically. 
These results include adder circuits of linear size.
Finally, Section~\ref{sec::conclusions} contains some concluding remarks.

\section{Preliminaries}
	\label{prelim}
In this section, we introduce the basic mathematical vocabulary, basic results,
and the main problems considered in this paper.

\subsection{Boolean Functions and Boolean Formulae}
	\label{Boofct}

 
A \textbf{Boolean variable}\index{Boolean variable} is a variable with values in $ \{0,1\} $. Given $ n\in\mathbb{N} $, a \textbf{Boolean function}\index{Boolean function} with $ n $ \textbf{Boolean input variables}\index{Boolean input variable} (short, \textbf{inputs}\index{inputs of a Boolean function}) is a function $ f:\{0,1\}^n\to \{0,1\} $. 
We call $ n $ the \textbf{arity}\index{Boolean function!arity of a} of $ f $ and 
$ f $ an \textbf{$ n $-ary} Boolean function. 
Often, 0 and 1 are not viewed as integers but rather 
as symbols representing truth values where 0 means \textbf{false}\index{false} 
and 1 stands for \textbf{true}\index{true}.
The inputs of a Boolean function $ f:\{0,1\}^n\to\{0,1\} $ are usually denoted by 
$ x_0,\dots, x_{n-1} $ or shorter as a vector $ x=(x_0,\dots,x_{n-1}) $.

\begin{definition}
		Let a Boolean function $ f:\{0,1\}^n\to\{0,1\} $, an input $ x_i $ with $ i\in \{0,\dots,n-1\} $ and $ \alpha\in\{0,1\} $ be given. The function $ f|_{x_i=\alpha}:\{0,1\}^{n-1}\to\{0,1\} $ defined by \[f|_{x_i=\alpha}((x_0,\dots,x_{i-1},x_{x+1},\dots,x_{n-1}))=f((x_0,\dots,x_{i-1},\alpha,x_{x+1},\dots,x_{n-1}))\] is called the \textbf{restriction}\index{Boolean function!restriction of a} of f to $ x_i=\alpha $.
We say that $f$ \textbf{depends essentially}\index{Boolean function!depends essentially} 
on $ x_i $ if $ f|_{x_i=0} $ and $ f|_{x_i=1} $ differ.
\end{definition}

There is another way to derive an equivalent definition of a Boolean function with the help of some elemental building blocks which will be defined next.
To this end, we consider the following three operations on $ \{0,1\} $.
The binary \textbf{\textsc{And}} operation $ \cdot\land\cdot:\{0,1\}\times\{0,1\}\to \{0,1\} $\index{and function@\textsc{And} function}  
is defined by 
$x\land y = 1$ if and only if $x=y=1$.
The binary \textbf{\textsc{Or}} operation $ \cdot\lor\cdot:\{0,1\}\times\{0,1\}\to \{0,1\} $\index{or function@\textsc{Or} function}
is defined by $x\lor y = 1$ if and only if $x=1$ or $y=1$.
The unary \textbf{\textsc{Not}} operation $ \bar{\cdot}:\{0,1\}\to \{0,1\} $\index{not function@\textsc{Not} function}, 
also called \textbf{Boolean negation}\index{Boolean negation}, 
is defined by $\bar{x} = 1$ if and only if $x=0$.

These building blocks can be seen as Boolean functions by themselves. 
	
\begin{definition}
Given $ n\in \mathbb{N} $ and Boolean variables $ x_0,\dots, x_{n-1} $, a \textbf{Boolean formula}\index{Boolean formula} on the input variables $ x_0,\dots, x_{n-1} $ is defined as follows:
\begin{enumerate}[(i)]
   \item The constants 0,1 and the variables $ x_0,\dots, x_{n-1} $ are Boolean formulae on $ x_0,\dots,x_{n-1} $.
   \item If $ \phi $ and $ \psi $ are Boolean formulae on $ x_0,\dots,x_{n-1} $, then $ (\phi\land\psi),(\phi\lor\psi) $ and $ \bar{\phi} $ are Boolean formulae on $ x_0,\dots,x_{n-1} $.
   \item Any Boolean formula $ \phi $ on $ x_0,\dots,x_{n-1} $ arises from finitely many applications of the rules (i) and (ii).
\end{enumerate}
		We also write $ \phi((x_0,\dots,x_{n-1})) $ or $ \phi(x) $ to denote a Boolean formula on input variables $ x=(x_0,\dots,x_{n-1}) $. We call $ n $ the \textbf{arity}\index{Boolean formula!arity of a} of $ \phi $.
\end{definition}

	The parentheses in part (ii) of the definition may be omitted if the formula is clear from the context.
	
\begin{definition}
Given Boolean input variables $ x=(x_0,\dots,x_{n-1}) $ and a Boolean formula $ \phi(x) $,
the Boolean function $ f_{\phi}:\{0,1\}^n\to\{0,1\} $ \textbf{realized by} $ \phi(x) $
is defined as follows: Given a point $ (\alpha_0,\dots,\alpha_{n-1})\in\{0,1\}^n $, 
the value $ f_\phi((\alpha_0,\dots,\alpha_{n-1})) $ is obtained by replacing $ x_i $
with $ \alpha_i $ for all $ i\in\{0,\dots,n-1\} $ in the formula $ \phi $ and then
recursively evaluating the intermediate formulas (starting at the inputs)
to compute the value of $ \phi $.
If a Boolean function $ f $ is realized by a Boolean formula $ \phi $, we say that
$ \phi $ is a \textbf{realization}\index{Boolean function!realization of a} of $ f $.
\end{definition}

A Boolean formula realizes a unique Boolean function, but a Boolean function might
have different realizations. 
%
%
%
%
Two Boolean formulae $ \phi $ and $ \psi $ are called \textbf{equivalent}\index{Boolean formula!equivalent} if $ \phi $ and $ \psi $ realize the same Boolean function. In this case, we write $ \phi=\psi $.
Since a Boolean formula $ \phi $ realizes a unique Boolean function $ f $, we may also view $ \phi $ as a Boolean function and write $ \phi=f $. 
The symbol $ = $ now is an equivalence relation on the union of the set of Boolean functions and the set of Boolean formulae with the same arity. 


	
Let Boolean functions $ f,g,h:\{0,1\}^n\to\{0,1\} $ be given. 
If $h(x)=f(x)\land g(x)$ for all $ x\in\{0,1\}^n $, then any realizations $ \phi_f $ and $ \phi_g $ of $ f $ and $ g $ yield a realization $ \phi_h=\phi_f\land\phi_g $ of $ h $
(analogously for $h(x)=f(x)\lor g(x)$).
If $ f(x)=\overline{g(x)} $ for all $ x\in\{0,1\}^n $, then any realization $ \phi_g $ of $ g $ yields a realization $ \phi_f=\overline{\phi_g} $ of $ f $.

	
\begin{definition}
		A Boolean function $ f:\{0,1\}^n\to\{0,1\} $ is called \textbf{monotone}\index{Boolean function!monotone} 
if for all $ \alpha,\beta\in\{0,1\}^n $ with $ \alpha\leq \beta $ 
(i.e.\ $\alpha_i\leq \beta_i $ for all $ i\in\{0,\dots,n-1\}$),
we have $ f(\alpha)\leq f(\beta) $. A Boolean formula $ \phi $ is called \textbf{monotone}\index{Boolean formula!monotone} if it does not contain any negations.
\end{definition}

The majority of Boolean functions and Boolean formulae that we consider are monotone.
We denote (without proofs) two easy results on Boolean functions and formulas in the following lemma.
	



\begin{lemma} \label{lemmanorm}
Any Boolean function can be realized by a Boolean formula.
Any monotone Boolean function can be realized by a monotone Boolean formula.
 \qed
\end{lemma}

The first statement of this lemma follows e.g.\ from Theorem~2.1.23 from 
Hermann \cite{Her20}. A proof of the second statement can be found in 
Crama and Hammer~\cite{CH11}.


Note that there are non-monotone Boolean formulae realizing a monotone Boolean function.


In the following, we summarize some notation and results 
on the important concept of Boolean duality.

	\begin{definition}
		\label{defdualform}
		Let $ \phi $ be a Boolean formula. The \textbf{dual Boolean formula}\index{dual Boolean formula} $ \phi^* $ of $ \phi $ is obtained from $ \phi $ by interchanging all $ \land $ and $ \lor $ operations and all 0 and 1 symbols.
	\end{definition}


	\begin{definition}
		\label{defdualfct}
		Let $ f:\{0,1\}^n\to\{0,1\} $ be a Boolean function. The \textbf{dual Boolean function}\index{dual Boolean function} $ f^*:\{0,1\}^n\to \{0,1\} $ of $ f $ is defined by \[f^*((x_0,\dots,x_{n-1}))=\overline{f((\overline{x_0},\dots,\overline{x_{n-1}}))}.\] To shorten notation, we write $ \bar{x}=(\overline{x_0},\dots,\overline{x_{n-1}}) $ and, thus, $ f^*(x)=\overline{f(\bar{x})} $.
	\end{definition}

These two definitions imply that for a Boolean formula 
$\phi$ we have $(\phi^*)^*=\phi$ and for a Boolean function $f$ we have
$(f^*)^*=f$.
The following well-known theorem shows that the dual of Boolean formulae 
and Boolean functions fit together. 
We omit the easy proof by induction here, 
it can be found e.g.\ in \cite{Her20} (Thm. 2.1.31).

\begin{theorem} \label{thmdual}
		For some $ n\in\mathbb{N} $, let $ \phi $ be an $ n $-ary Boolean formula and $ f_\phi $ its realized Boolean function. Then, $ \phi^* $ is a realization of $ f_\phi^* $, i.e., $ f_{\phi^*}=f_{\phi}^* $. \qed 
\end{theorem}

We can conclude form this theorem that if
$\phi $ and $ \psi$ are equivalent Boolean formulae then
the dual formulae $ \phi^* $ and $ \psi^* $ are also equivalent.
To see this note that if $\phi $ and $ \psi$ realize a function $f$
then by Theorem~\ref{thmdual} both both $ \phi^* $ and $ \psi^* $ 
realize $ f^* $. Hence, they are equivalent.



	


\subsection{Circuits}
	\label{circuits}

Circuits model the physical implementation of Boolean functions on a computer chip,
where devices computing elementary Boolean functions are combined in order to realize
more complex functions.

	
	\begin{definition}
		A \textbf{basis}\index{basis} is a set $ \Omega $ of Boolean formulae. Each element $ \phi\in\Omega $ is called \textbf{gate}\index{gate}.
	\end{definition}

	\begin{definition}
		A \textbf{circuit}\index{circuit} $ C=(\mathcal{V},\mathcal{E}) $ over the basis $ \Omega $ is an acyclic directed graph with labeled vertices $ \mathcal{V}=\mathcal{I}\dot\cup\mathcal{G} $ such that the following conditions hold:
		\begin{itemize}
			\item Each vertex $ v\in\mathcal{I} $ fulfills $ \delta^-(v)=\emptyset $ and $ \delta^+(v)\neq\emptyset $ and is labeled either with a distinct Boolean variable $ x_v $ or with a symbol 0 or 1. The vertices in $\mathcal{I}$ are called \textbf{inputs}\index{inputs of a circuit} of $ C $.
			\item Each vertex $ v\in\mathcal{G} $ fulfills $ k:=|\delta^-(v)|\geq 1 $ and is labeled with a $ k $-ary gate $ \phi\in\Omega $ together with a fixed ordering $ v_0,\dots, v_{k-1} $ of the predecessors of $ v $. The vertices in $\mathcal{G}$ are called \textbf{gates} of $ C $. The \textbf{gate type}\index{gate type} $ \phi\in\Omega $ of $ v $ is denoted by gt$ (v):=\phi $.
			\item There is a non-empty subset $ \mathcal{O}\subseteq\mathcal{V} $ whose elements are called \textbf{outputs}\index{output of a circuit} of $ C $. Every vertex $ v\in\mathcal{V} $ with $ \delta^+(v)=\emptyset $ is an output, but there may also be other outputs.
		\end{itemize}
	\end{definition}

	Given a circuit $ C $, we write $ \mathcal{V}(C),\mathcal{E}(C),\mathcal{I}(C),\mathcal{G}(C),\mathcal{O}(C) $ for its vertices, edges, inputs, gates, and outputs, respectively. If the circuit is clear from the context, we may omit the specifying appendage $ (C) $.
	

\begin{definition}
		Given a circuit $ C $ over a basis $ \Omega $ and a vertex $ v\in\mathcal{V} $, we call $ \phi_v $ the \textbf{Boolean formula corresponding to $ v $} and define it recursively as follows:
		\begin{itemize}
			\item If $ v\in\mathcal{I} $, then $ \phi_v=x_v $.
			\item If $ v\in\mathcal{G} $, let $ \phi=\text{gt}(v) $ and let $ v_0,\dots,v_{k-1} $ be the ordered predecessors of $ v $. Then $ \phi_v=\phi(\phi_{v_0},\dots, \phi_{v_{k-1}}) $.
		\end{itemize}
Note that $ \phi_v $ is a Boolean formula on the input variables $ (x_v)_{v\in\mathcal{I}} $.
If $ |\mathcal{O}(C)|=1 $, then out$ (C) $ denotes the unique output of 
$ C $ and $ \phi_C:=\phi_{\text{out}(C)} $ is called the 
\textbf{Boolean formula corresponding to the circuit $ C $}\index{circuit!Boolean formula corresponding to a}.
We then call $ f_C:=f_{\phi_C} $ the \textbf{Boolean function realized} 
(or, \textbf{computed}) \textbf{by $ C $}\index{circuit!Boolean function realized by a}.
\end{definition}
	
The bases considered in this paper consist only of the elementary Boolean formulae
\textsc{And}, \textsc{Or}, and \textsc{Not}.
	
	\begin{definition}
		The basis $ \Omega_\text{mon}:=\{\land, \lor\} $ is called the \textbf{standard monotone basis}\index{basis!standard monotone}. The basis $ \Omega_\text{nmon}:=\{\land,\lor,\bar{\cdot}\} $ is called the \textbf{standard non-monotone basis}\index{basis!standard non-monotone}. A circuit is called \textbf{monotone}\index{circuit!monotone} if each gate is labeled with a monotone Boolean formula. A circuit is called \textbf{binary}\index{circuit!binary} if each gate has an arity of at most 2.
	\end{definition}

Since every (monotone) Boolean function has a realizing 
(monotone) Boolean formula (Lemma~\ref{lemmanorm})
and every (monotone) Boolean formula can be represented by a (monotone)
circuit over $ \Omega_\text{nmon} $ ($ \Omega_\text{mon} $), 
we can conlude that
for every (monotone) Boolean function $ f:\{0,1\}^n\to \{0,1\} $, 
there is a (monotone) circuit over $ \Omega_\text{nmon} $ ($ \Omega_\text{mon} $)
realizing $ f $.
Each circuit with exactly one output corresponds to a unique Boolean function.
However, there might be several circuits corresponding to the same Boolean function.
Two circuits realizing the same Boolean function are called \textbf{equivalent}.
	

There is a canonical way to represent a Boolean formula by a unique circuit such that
each gate realizes exactly one of the symbols ``$ \cdot\land\cdot $'', ``$ \cdot\lor\cdot $''
and `` $ \bar{\cdot} $\,\,'' of the formula. Such a circuit is called a \textbf{formula circuit}.
It is easy to see that a circuit is a formula circuit if and only if each gate has
out-degree at most $1$.



		
We say that a circuit $C$ \textbf{depends essentially}\index{circuit!depends essentially} 
on an input $ x_i $, if $ f_{C_v} $ depends essentially on $ x_i $ for some output vertex 
$ v\in\mathcal{O}(C) $

We can also consider duality in the context of circuits.
Given a circuit $ C $, the dual circuit $ C^* $ arises from $ C $ 
by interchanging all \textsc{And}- and \textsc{Or}-gates and all 
0 and 1 symbols. This is compatible to duality in the context of
Boolean functions and formulae, i.e.
for any circuit $ C $, we have $ \phi_{C^*}=\phi_C^* $ 
and $ f_{C^*}=f_C^* $. Moreover, for any Boolean function $ \phi $, 
we have $ C_{\phi^*}=C_\phi^* $ (see e.g.\ Thm. 2.2.11 in \cite{Her20}
for the easy proof of this statement).





\subsection{Optimization Problems}
\label{optprob}

In this section, we will introduce a number of quality measures of 
circuits.

A natural measure for the 'speed' of a circuit in a computer chip is the maximum number of gates which an input has to travel through until an output is reached, i.e., the depth of the circuit.
	
\begin{definition}
Consider the circuit $ C $. The \textbf{depth}\index{depth of a vertex} of a vertex $ v\in\mathcal{V} $ is defined by 
\[
\text{d}(v):=\max\{|\mathcal{E}(P)| : P \text{ directed path in } C \text{ ending in } v\}.
\]
The \textbf{depth}\index{depth of a circuit} of $ C $ is the maximum depth of any vertex in $ C $ and is denoted by $ \text{d}(C) $.
\end{definition}

	However, it can not be assumed that all input signals of a circuit in a chip are accessible simultaneously. Some may have to be waited for, while others may already traverse through the circuit. This is, in particular, the case when circuits are embedded into other, more-complex circuits. 
For example, adder circuits, which we will define precisely in Section~\ref{addercirc}, are often used inside of multipliers which multiply binary numbers. The most common techniques involve the computation of partial products which are then summed up by adder circuits to produce the final product (see, e.g., \cite{Okl94} and \cite{Zim98}).
	
The problem of non-simultaneous accessibility of the inputs motivates the introduction of arrival times.
	
	\begin{definition}
		Consider a circuit with inputs $ x_0,\dots, x_{n-1} $. For each input $ x_i $ let a number $ a(x_i)\in\mathbb{R} $ be given for $ i\in\{0,\dots, n-1\} $. The numbers $ a(x_0),\dots, a(x_{n-1}) $ are the \textbf{arrival times}\index{arrival times} of $ x_0,\dots, x_{n-1} $. We define the \textbf{arrival time} of any gate $ v\in\mathcal{G} $ recursively by \[a(v):=\max_{w\in\delta^-(v)}\{a(w)\}+1.\]
		Then the \textbf{delay}\index{delay of a circuit} of $ C $ with respect to the arrival times $ a:\mathcal{I}\to \mathbb{R} $ is defined as the maximum arrival time of any output of $ C $, i.e. \[\text{del}(C;a):=\max_{v\in\mathcal{O}}a(v).\]When the arrival times are clear from the context, we also write $ \text{del}(C):=\text{del}(C;a) $.
	\end{definition}

Note that the depth is a special case of the delay where the arrival times of the inputs are all 0.
	
Since signals in a computer chip slow down when they are distributed too often, 
there is another measuring unit that has to be considered in order to predict the speed of a circuit. 

	\begin{definition}
		Given a circuit $ C $ and a vertex $ v\in\mathcal{V} $, the \textbf{fanout}\index{fanout of a vertex} of $ v $ is defined by fanout$ (v):=|\delta^+(v)| $. The \textbf{fanout}\index{fanout of a circuit} of $ C $ is the maximum fanout of any vertex in $ C $ and is denoted by fanout$ (C) $.
	\end{definition}

	Besides of the speed, the size of a circuit is also a very important factor in chip design. It controls not only the actual size of the circuit in the chip but also the power consumption and the production cost of the chip. 
	
\begin{definition}
Given a circuit $ C $, the \textbf{size}\index{size of a circuit} of $ C $ is the number of gates in $ C $ and is denoted by s$ (C) $.
\end{definition}

\begin{figure}[htb]
\centering
   \begin{tikzpicture}[scale=0.75]
      \node[outer sep=0pt] (i0) at (0, 0){$x_0$};
      \node[outer sep=0pt] (i1) at (1, 0){$x_1$};
      \node[outer sep=0pt] (i2) at (2, 0){$x_2$};
      \node[outer sep=0pt] (i3) at (3, 0){$x_3$};

      \node[scale=0.8, above=0.2cm] at (i0){\textcolor{blue}{5}};
      \node[scale=0.8, above=0.2cm] at (i1){\textcolor{blue}{2}};
      \node[scale=0.8, above=0.2cm] at (i2){\textcolor{blue}{1}};
      \node[scale=0.8, above=0.2cm] at (i3){\textcolor{blue}{2}};

      \node[fill=red, and gate US, draw, logic gate inputs=nn, rotate=270, thick, scale=0.6] at (0,-1) (a1){};
      \node[fill=red, and gate US, draw, logic gate inputs=nn, rotate=270, thick, scale=0.6] at (2,-1) (a2){};
      \node[fill=red, and gate US, draw, logic gate inputs=nn, rotate=270, thick, scale=0.6] at (0.8,-2) (a3){};
      \node[fill=green, or gate US, draw, logic gate inputs=nn, rotate=270, thick, scale=0.6] at (2.5,-2) (a4){};
      \node[fill=green, or gate US, draw, logic gate inputs=nn, rotate=270, thick, scale=0.6] at (1.5,-3) (o){};

      \node[scale=0.8, left=0.2cm] at (a1){\textcolor{blue}{6}};
      \node[scale=0.8, left=0.2cm] at (a2){\textcolor{blue}{3}};
      \node[scale=0.8, left=0.2cm] at (a3){\textcolor{blue}{7}};
      \node[scale=0.8, left=0.2cm] at (a4){\textcolor{blue}{4}};
      \node[scale=0.8, left=0.2cm] at (o){\textcolor{blue}{8}};

      \node at (1.5,-3.75) (l){};

      \node[below=0.25cm, align=flush center] at (l){\textbf{(a)} Circuit $ C_1 $.};

      \draw[thick] (i0)--(a1.input 2);
      \draw[thick] (i1)--(a1.input 1);
      \draw[thick] (i1)--(a2.input 2);
      \draw[thick] (i2)--(a3.input 1);
      \draw[thick] (i2)--(a2.input 1);
      \draw[thick] (i3)--(a4.input 1);
      \draw[thick] (a1.output)--(a3.input 2);
      \draw[thick] (a3.output)--(o.input 2);
      \draw[thick] (a2.output)--(a4.input 2);
      \draw[thick] (a4.output)--(o.input 1);
      \draw[->] (o)--(l);
\end{tikzpicture} \hspace{1cm}
\begin{tikzpicture}[scale=0.75]
      \node[outer sep=0pt] (i0) at (0, 0){$x_0$};
      \node[outer sep=0pt] (i1) at (1, 0){$x_1$};
      \node[outer sep=0pt] (i2) at (2, 0){$x_2$};
      \node[outer sep=0pt] (i3) at (3, 0){$x_3$};
      
      \node[scale=0.8, above=0.2cm] at (i0){\textcolor{blue}{5}};
      \node[scale=0.8, above=0.2cm] at (i1){\textcolor{blue}{2}};
      \node[scale=0.8, above=0.2cm] at (i2){\textcolor{blue}{1}};
      \node[scale=0.8, above=0.2cm] at (i3){\textcolor{blue}{2}};
      
      \node[fill=red, and gate US, draw, logic gate inputs=nn, rotate=270, thick, scale=0.6] at (1,-1) (a1){};
      \node[fill=red, and gate US, draw, logic gate inputs=nn, rotate=270, thick, scale=0.6] at (2,-1) (a2){};
      \node[fill=red, and gate US, draw, logic gate inputs=nn, rotate=270, thick, scale=0.6] at (0.8,-2) (a3){};
      \node[fill=green, or gate US, draw, logic gate inputs=nn, rotate=270, thick, scale=0.6] at (2.5,-2) (a4){};
      \node[fill=green, or gate US, draw, logic gate inputs=nn, rotate=270, thick, scale=0.6] at (1.5,-3) (o){};
      
      \node[scale=0.8, left=0.2cm] at (a1){\textcolor{blue}{3}};
      \node[scale=0.8, left=0.2cm] at (a2){\textcolor{blue}{3}};
      \node[scale=0.8, left=0.2cm] at (a3){\textcolor{blue}{6}};
      \node[scale=0.8, left=0.2cm] at (a4){\textcolor{blue}{4}};
      \node[scale=0.8, left=0.2cm] at (o){\textcolor{blue}{7}};
      
      \node at (1.5,-3.75) (l){};
      
      \node[below=0.25cm, align=flush center] at (l){\textbf{(b)} Circuit $ C_2 $.};
      
      \draw[thick] (i0)--(a3.input 2);
      \draw[thick] (i1)--(a1.input 2);
      \draw[thick] (i1)--(a2.input 2);
      \draw[thick] (i2)--(a1.input 1);
      \draw[thick] (i2)--(a2.input 1);
      \draw[thick] (i3)--(a4.input 1);
      \draw[thick] (a1.output)--(a3.input 2);
      \draw[thick] (a3.output)--(o.input 2);
      \draw[thick] (a2.output)--(a4.input 2);
      \draw[thick] (a4.output)--(o.input 1);
      \draw[->] (o)--(l);
\end{tikzpicture}\hspace{1cm}
\begin{tikzpicture}[scale=0.75]
      \node[outer sep=0pt] (i0) at (0, 0){$x_0$};
      \node[outer sep=0pt] (i1) at (1, 0){$x_1$};
      \node[outer sep=0pt] (i2) at (2, 0){$x_2$};
      \node[outer sep=0pt] (i3) at (3, 0){$x_3$};
      
      \node[scale=0.8, above=0.2cm] at (i0){\textcolor{blue}{5}};
      \node[scale=0.8, above=0.2cm] at (i1){\textcolor{blue}{2}};
      \node[scale=0.8, above=0.2cm] at (i2){\textcolor{blue}{1}};
      \node[scale=0.8, above=0.2cm] at (i3){\textcolor{blue}{2}};
      
      \node[fill=red, and gate US, draw, logic gate inputs=nn, rotate=270, thick, scale=0.6] at (1.5,-1) (a1){};
      \node[fill=red, and gate US, draw, logic gate inputs=nn, rotate=270, thick, scale=0.6] at (0.8,-2) (a2){};
      \node[fill=green, or gate US, draw, logic gate inputs=nn, rotate=270, thick, scale=0.6] at (2.5,-2) (a3){};
      \node[fill=green, or gate US, draw, logic gate inputs=nn, rotate=270, thick, scale=0.6] at (1.5,-3) (o){};
      
      \node[scale=0.8, left=0.2cm] at (a1){\textcolor{blue}{3}};
      \node[scale=0.8, left=0.2cm] at (a2){\textcolor{blue}{6}};
      \node[scale=0.8, left=0.2cm] at (a3){\textcolor{blue}{4}};
      \node[scale=0.8, left=0.2cm] at (o){\textcolor{blue}{7}};
      
      \node at (1.5,-3.75) (l){};
      
      \node[below=0.25cm, align=flush center] at (l){\textbf{(c)} Circuit $ C_3 $.};
      
      \draw[thick] (i0)--(a2.input 2);
      \draw[thick] (i1)--(a1.input 2);
      \draw[thick] (i2)--(a1.input 1);
      \draw[thick] (i3)--(a3.input 1);
      \draw[thick] (a1.output)--(a2.input 1);
      \draw[thick] (a2.output)--(o.input 2);
      \draw[thick] (a1.output)--(a3.input 2);
      \draw[thick] (a3.output)--(o.input 1);
      \draw[->] (o)--(l);
\end{tikzpicture}
\caption[Example of three equivalent circuits with arrival times.]
{Circuits realizing 
$((x_0\land x_1)\land x_2)\lor((x_1\land x_2)\lor x_3)$
from Example \ref{ex1} with arrival times in blue.}
\label{figex2}
\end{figure}
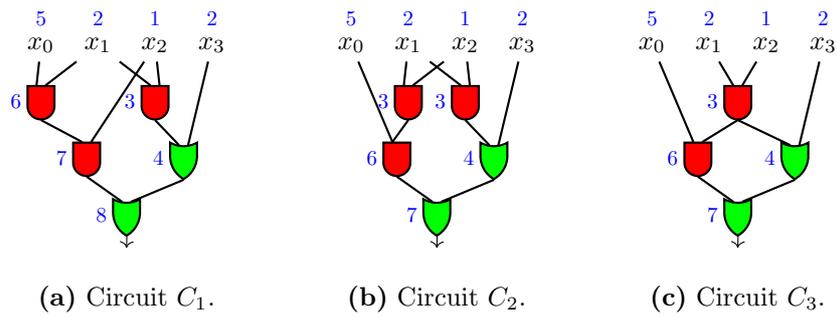

\begin{sloppypar}
\begin{example}
\label{ex1}
Figure~\ref{figex2} shows three circuits realizing the Boolean function given by
the Boolean formula
$ \phi((x_0,x_1,x_2,x_3))=((x_0\land x_1)\land x_2)\lor((x_1\land x_2)\lor x_3)$.
In pictures of circuits,
we always draw a circuit from top to bottom omitting the edge directions of the directed graph.
Inputs are drawn at the top. In this figure, the \textsc{And} gates are shown in red and the
\textsc{Or} gates in green. In further figures, the color of the
gates types might vary but the shapes are fixed.
We assume that we are given arrival times (shown in blue).
All three circuits have the same depth and the same fanout
(though $C_3$ is the only one with a fanout of $2$ for a gate).
With respect to the delay, we see that circuits
$ C_2 $ and $ C_3 $ are faster with a delay of 7, compared to $ C_1 $, which
has a delay of 8. In terms of size, $ C_3 $ outperforms the other
two circuits with a size of 4, compared to 5 for $ C_1 $ and $ C_2 $.
\end{example}
\end{sloppypar}

It is easy to see that 
all introduced quality measures are invariant under dualization, 
i.e., for a circuit $ C $ with arrival 
times $ a:\mathcal{I}\to \mathbb{R} $, 
we have $ \mathrm{d}(C^*)=\mathrm{d}(C)$, $ \mathrm{ del}(C^*)=\mathrm{del}(C)$, 
$ \mathrm{ fanout}(C^*)=\mathrm{fanout}(C) $ and $ \mathrm{s}(C^*)=\mathrm{s}(C) $.


Our main focus is minimizing the delay of a circuit, i.e., 
solving the following problem.\vspace{5pt}\\
	\fbox{\parbox{\textwidth-6.8pt}{\textsc{\Large Circuit Delay Optimization Problem\vspace{5pt}}\\ 
	\begin{tabularx}{\textwidth-6.8pt}{l X}
		Instance: & A Boolean function $ f:\{0,1\}^n\to \{0,1\} $ on inputs $ x=(x_0,\dots, x_{n-1}) $, arrival times $ a(x_0),\dots, a(x_{n-1})\in\mathbb{R} $, and a basis $ \Omega $.\\
		Task: & Compute a circuit over $ \Omega $ realizing $ f(x) $ with minimum possible delay.
	\end{tabularx}
	}}\index{circuit delay optimization problem@\textsc{Circuit Delay Optimization Problem}}

\begin{proposition}
   The \textsc{Circuit Delay Optimization Problem} is NP-hard.
\end{proposition}

	\begin{proof}
We reduce the NP-hard \textsc{Satisfiability Problem} 
to the \textsc{Circuit Delay Optimization Problem}. Let a Boolean formula $ \phi(x) $ on the input variables $ x=(x_0,\dots, x_{n-1}) $ be given and set $ a(x_0)=\dots=a(x_{n-1})=0 $. Then $ \phi $ is not satsifiable if and only if $ f_{\phi}(\alpha)=0 $ for all $ \alpha\in\{0,1\}^n $. Hence, $ \phi $ is satisfiable if and only if a delay minimum circuit $ C $ realizing $ f_\phi $ fulfills $ C\neq 0 $, i.e., the circuit whose only output vertex is the constant 0.
	\end{proof}

Note that we have proven NP-hardness of the problem even for the special case of depth optimization. 

In order to check the quality of a circuit, we need lower bounds for 
the delay of an optimal solution for the problem. 
The next theorem provides such a lower bound.
It is a specialization of a result by Golumbic \cite{Gol76} who gave a lower bound for the delay of any circuit whose basis contains only Boolean formulae constant arity.
	
\begin{theorem}[Cf. Golumbic \cite{Gol76}] \label{thmlowerbound}
Consider a monotone Boolean function $ f:\{0,1\}^n\to \{0,1\} $ on inputs $ x_0,\dots, x_{n-1} $ with arrival times $ a(x_0),\dots, a(x_{n-1})\in \mathbb{N} $. Assume that $ f $ depends essentially on all its inputs. Then for any circuit $ C $ over $ \Omega_\text{mon} $ computing $ f $, we have 
\[
\mathrm{del}(C)\geq \bigg\lceil\log_2\bigg(\sum_{i=0}^{n-1}2^{a(x_i)}\bigg)\bigg\rceil.\tag*{\qed}
\]
\end{theorem}

For a proof of this theorem, we refer to Golumbic~\cite{Gol76}.

	Note that if we do not assume natural numbers as arrival times, we get the slightly weaker result $ \text{del}(C)\geq \log_2\big(\sum_{i=0}^{n-1}2^{a(x_i)}\big) $. 
For the special case of depth optimization, Theorem~\ref{thmlowerbound} implies $\text{d}(C)\geq \lceil\log_2n\rceil$.
The theorem motivates the following definition that will be used often for 
delay bounds of a circuit.
\begin{definition} \label{defweight}
The \textbf{weight}\index{weight of inputs} of inputs $ x=(x_0,\dots, x_{n-1}) $ with arrival times $ a(x_0),\dots, a(x_{n-1})\in\mathbb{R} $ is defined by 
\[W(x;a):=\sum_{i=0}^{n-1}2^{a(x_i)}.\] When the inputs and their arrival times are clear from the context, we may write $ W:=W(x;a) $. Given a subset $ P\subseteq\{x_0,\dots, x_{n-1}\} $ of the inputs, we write $ W_P:=W_P(x,a):=\sum_{x_i\in P}2^{a(x_i)} $.
\end{definition}

\subsection{Adder Circuits}
	\label{addercirc}
	
In this section, we finally introduce binary addition formally and reduce it 
to solving special kinds of circuits. 
We want to construct circuits for computing the \textbf{summation function}
$s_n:\{0,1\}^{2n}\to \{0,1\}^{n+1}$ for adding two $n$-bit binary numbers
(with $n\in\mathbb{N}$). 
This means for $a=(a_0,\dots, a_{n-1})$ and 
$ b=(b_0,\dots, b_{n-1}) $ we want to compute an $(n+1)$-bit binary number
$ s_n((a,b))$ encoding $ \sum_{i=0}^{n-1} (a_i + b_i)2^i$.
The sum of two given binary numbers can be easily computed via carry bits.
	
\begin{definition}
		\label{defcb}
		Let $ n\in\mathbb{N} $ and let $ a=(a_0,\dots, a_{n-1}) $ and
                $ b=(b_0,\dots,$ $ b_{n-1}) $ be two binary numbers with most
                significant bit $ n-1 $. For $ i\in\{0,\dots,n-1\} $, we define the $ i $-th \textbf{generate signal}\index{generate signal} for $ a $ and $ b $ as $ g_i:=a_i\land b_i $ and the $ i $-th \textbf{propagate signal}\index{propagate signal} for $ a $ and $ b $ as $ p_i:=a_i\oplus b_i $, 
where $ \cdot\oplus\cdot:~\{0,1\}^2\to\{0,1\} $ is the 
\textbf{\textsc{Xor} function}\index{xor function@\textsc{Xor} function},
i.e.\ $x_0\oplus x_1$ is $1$ if and only if exactly one of $x_0$ and $x_1$ is $1$.
The \textbf{carry bits}\index{carry bit} $ c_0,\dots, c_n $ are defined recursively 
\begin{align*}
   c_0&=0,\\
   c_{i+1}&=g_i\lor(p_i\land c_i)
\end{align*}
for $ i\in\{0,\dots,n-1\} $.
\end{definition}

Hence, we have a $c_{i+1} = 1$ for some $ i\in\{0,\dots, n-1\} $ if and only if one of the following statements hold:
	\begin{itemize}
		\item $ g_i $ is true, i.e., both $ a_i $ and $ b_i $ are 1 ($ c_{i+1} $ is \textit{generated} at position $ i $),
		\item $ p_i\land c_i $ is true, i.e., exactly one of $ a_i $ and $ b_i $ is 1 and we have a carry bit at position $ i $ ($ c_{i+1} $ is \textit{propagated} from position $ i-1 $).
	\end{itemize}
	
	Given the carry bits $ c_0,\dots, c_{n} $ corresponding to some binary numbers $ a=(a_0,\dots, a_{n-1}) $ and $ b=(b_0,\dots, b_{n-1}) $, the summation function on $ a $ and $ b $ can now be computed easily via 
	\[(s_n((a,b)))_i=
	\begin{cases}
		c_i\oplus (a_i\oplus b_i) = c_i\oplus p_i &\text{if }i\in\{0,\dots, n-1\},\\
		c_n &\text{if }i=n.
	\end{cases}\]
	
	The computation of the generate and propagate signals of two given binary numbers as well as the computation of the summation function for given carry bits and propagate signals can by done by circuits of linear size and constant depth. Hence, computing the summation function can be reduced to solving the following problem.\vspace{5pt}\\ 
	\fbox{\parbox{\linewidth-6.8pt}{\textsc{\Large Adder Optimization Problem\vspace{5pt}}\\
		\begin{tabularx}{\textwidth-6.8pt}{l X}
			Instance: & $ n\in\mathbb{N} $.\\
			Task: & Construct a circuit over $ \Omega_\text{mon}=\{\land,\lor\} $ on $ n $ input pairs $ p_0,g_0,\dots, p_{n-1},g_{n-1} $ computing all the carry bits $ c_1,\dots, c_n $.
		\end{tabularx}
	}}\index{adder optimization problem@\textsc{Adder Optimization Problem}}
	
	\begin{definition}
		A circuit solving the \textsc{Adder Optimization Problem} for a given $ n\in\mathbb{N} $ is called an \textbf{adder circuit}\index{adder circuit} (or, \textbf{adder}\index{adder}). A \textbf{family of adder circuits}\index{family of adder circuits} is a family of circuits $ (A_n)_{n\in\mathbb{N}} $ where each $ A_n $ is an adder circuit on $ n $ input pairs. Given an adder circuit $ A_n $ in $ n $ input pairs and $ i\in\{1,\dots, n\} $, we denote the output of $ A_n $ computing the carry bit $ c_i $ by $ \text{out}_i(A_n) $.
	\end{definition}

	Note that Definition~\ref{defcb} implies that $ c_1=g_0 $. Thus, an adder on input pairs $ p_0,g_0,\dots, p_{n-1},g_{n-1} $ does not depend essentially on $ p_0 $. However, to simplify notation, we keep $ p_0 $ as part of the input. Whenever arrival times $ a(p_0),a(g_0),\dots, a(p_{n-1}),a(g_{n-1})\in\mathbb{N} $ are given, we may assume $ a(p_0)=0 $.
	
	\begin{remark}
		By Definition~\ref{defcb}, the carry bits can be computed via monotone Boolean formulae and implemented by circuits over $ \Omega_{\text{mon}}=\{\land,\lor\} $. It is an open question whether the Boolean negation can help to find better adders regarding the quality measures compared to using only monotone circuits.
	\end{remark}

Each carry bit $ c_{i+1} $ can be computed by a path-like formula on the inputs 
$ g_i,p_i,\dots, g_1,p_1,g_0 $
with alternating \textsc{Or} and \textsc{And} gates (see (\ref{eq::carry_bits}))
The problem of finding good circuits for these kinds of formulae plays a crucial role in 
finding good adder circuits. 
Our main goal is to find an adder which is almost as fast as a single \textsc{And-Or} path circuit 
but still has (nearly) linear size. We formally define and analyze
\textsc{And-Or} paths in the next section.
	
	We conclude this section by giving an example of two very basic approaches for constructing an adder circuit.
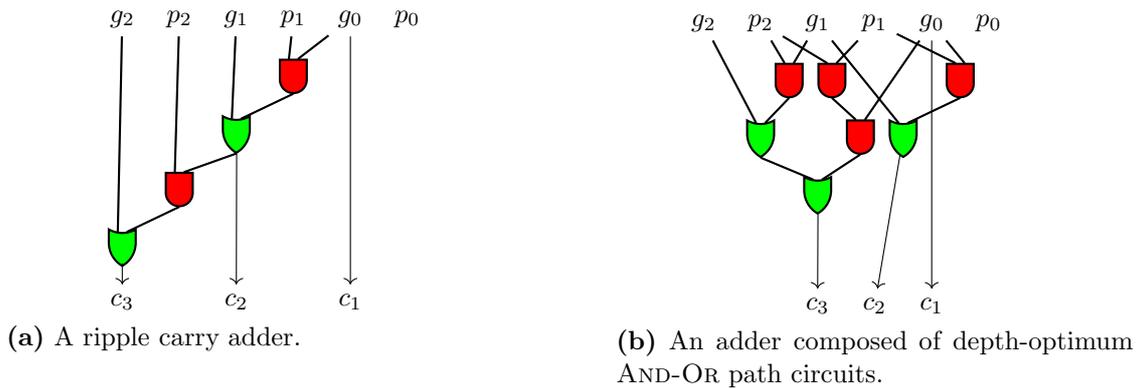
\begin{figure}[htb]
\centering
   \begin{tikzpicture}[scale=0.75]
      \node at (0,0) (a) {$ g_2 $};
      \node at (1,0) (b) {$ p_2 $};
      \node at (2,0) (c) {$ g_1 $};
      \node at (3,0) (d) {$ p_1 $};
      \node at (4,0) (e) {$ g_0 $};
      \node at (5,0) (f) {$ p_0 $};

      \node[fill=red, and gate US, draw, logic gate inputs=nn, rotate=270, thick, scale=0.6] at (3,-1) (g) {};
      \node[fill=green, or gate US, draw, logic gate inputs=nn, rotate=270, thick, scale=0.6] at (2,-2) (h) {};
      \node[fill=red, and gate US, draw, logic gate inputs=nn, rotate=270, thick, scale=0.6] at (1,-3) (i) {};
      \node[fill=green, or gate US, draw, logic gate inputs=nn, rotate=270, thick, scale=0.6] at (0,-4) (j) {};

      \node at (0,-5) (c3) {$ c_3 $};
      \node at (2,-5) (c2) {$ c_2 $};
      \node at (4,-5) (c1) {$ c_1 $};

      \node[below=0.2cm, align=flush center] at (2.5,-5){\parbox{6.8cm}{\textbf{(a)} A ripple carry adder.\vspace{13pt}}};

      \draw[thick]  (a)--(j.input 2);
      \draw[thick]  (b)--(i.input 2);
      \draw[thick]  (c)--(h.input 2);
      \draw[thick]  (d)--(g.input 2);
      \draw[thick]  (e)--(g.input 1);
      \draw[thick]  (g.output)--(h.input 1);
      \draw[thick]  (h.output)--(i.input 1);
      \draw[thick]  (i.output)--(j.input 1);
      \draw [->] (j)--(c3);
      \draw [->] (h)--(c2);
      \draw [->] (e)--(c1);
\end{tikzpicture} \hspace{0.75cm}
\begin{tikzpicture}[scale=0.75]
      \node at (0,0) (g2) {$ g_2 $};
      \node at (1,0) (p2) {$ p_2 $};
      \node at (2,0) (g1) {$ g_1 $};
      \node at (3,0) (p1) {$ p_1 $};
      \node at (4,0) (g0) {$ g_0 $};
      \node at (5,0) (p0) {$ p_0 $};

      \node[fill=red, and gate US, draw, logic gate inputs=nn, rotate=270, thick, scale=0.6] at (1.5,-1) (a1) {};
      \node[fill=red, and gate US, draw, logic gate inputs=nn, rotate=270, thick, scale=0.6] at (2.25,-1) (a2) {};
      \node[fill=red, and gate US, draw, logic gate inputs=nn, rotate=270, thick, scale=0.6] at (4.5,-1) (a3) {};
      \node[fill=red, and gate US, draw, logic gate inputs=nn, rotate=270, thick, scale=0.6] at (2.75,-2) (a4) {};
      \node[fill=green, or gate US, draw, logic gate inputs=nn, rotate=270, thick, scale=0.6] at (1,-2) (o1) {};
      \node[fill=green, or gate US, draw, logic gate inputs=nn, rotate=270, thick, scale=0.6] at (3.5,-2) (o2) {};
      \node[fill=green, or gate US, draw, logic gate inputs=nn, rotate=270, thick, scale=0.6] at (2,-3) (o3) {};

      \node at (2,-5) (c3) {$ c_3 $};
      \node at (3,-5) (c2) {$ c_2 $};
      \node at (4,-5) (c1) {$ c_1 $};

      \node[below=0.2cm, align=flush center] at (c2){\parbox{6.8cm}{\textbf{(b)} An adder composed of depth-optimum \textsc{And-Or} path circuits.}};
			
      \draw[thick]  (g2)--(o1.input 2);
      \draw[thick]  (p2)--(a1.input 2);
      \draw[thick]  (p2)--(a2.input 2);
      \draw[thick]  (g1)--(a1.input 1);
      \draw[thick]  (g1)--(o2.input 2);
      \draw[thick]  (p1)--(a2.input 1);
      \draw[thick]  (p1)--(a3.input 2);
      \draw[thick]  (g0)--(a4.input 1);
      \draw[thick]  (g0)--(a3.input 1);
      \draw[thick]  (a1.output)--(o1.input 1);
      \draw[thick]  (a2.output)--(a4.input 2);
      \draw[thick]  (a3.output)--(o2.input 1);
      \draw[thick]  (o1.output)--(o3.input 2);
      \draw[thick]  (a4.output)--(o3.input 1);
      \draw [->] (o3)--(c3);
      \draw [->] (o2)--(c2);
      \draw [->] (g0)--(c1);
\end{tikzpicture}
\caption[Two trivial adder circuits on 3 input pairs]{Two adder circuits on $ n=3 $ input pairs, see Example~\ref{extrem_adders}.} \label{figac}
\end{figure}

\begin{example}
\label{extrem_adders}
Figure~\ref{figac} shows two different adder circuits for the inputs
$p_0,g_0,p_1,g_1,p_2,g_2$. In~(a), we see the so-called 
\textbf{ripple carry adder}\index{ripple carry adder} which has for $n$ input pairs
a depth and a size of $2n-2$. While the size is best possible, the 
linear depth is undesirable in most cases.
Instead, one could use a depth-optimizing (or delay-optimizing) \textsc{And-Or} path
circuit $ AOP_i $ realizing Equation~\eqref{eq::carry_bits} for each carry $ c_i $ separately,
and then combine the circuits $ AOP_1,\dots, AOP_n $ to receive an adder circuit. 
An example for such an adder circuit (optimizing depth) is given in Figure~\ref{figac}~(b).
The depth (delay) of such an adder on $ n $ inputs would be equal to the depth (delay)
of $ AOP_n $. However, its size would be $ \sum_{i=1}^ns(AOP_i) $ and, thus, at least
quadratic in $ n $ since the size of a single \textsc{And-Or} path circuit is at
least linear in $ n $.
\end{example}

\subsection{\textsc{And-Or} Path Circuits}
\label{andorpathcirc}

\begin{definition}
		\label{defaop}
For Boolean input variables $ x=(x_0,\dots, x_{m-1}) $,
we define the Boolean function $ g(t) $ by \[g(x)=\begin{cases}
			x_0 & m=1,\\
			x_0\land x_1 & m=2,\\
			x_0\land x_1\lor(g((x_2,\dots,x_{m-1}))) & m\geq 3.
		\end{cases}\]
		Then $ g(x) $ and its dual Boolean function $ g^*(x) $ are called \textbf{\textsc{And-Or} paths}\index{and-or path@\textsc{And-Or} path} on $ m $ inputs. We call $ m $ the \textbf{length}\index{and-or path@\textsc{And-Or} path!length of an} of the \textsc{And-Or} paths $ g(x) $ and $ g^*(x) $.
\end{definition}
	
	By Theorem~\ref{thmdual}, we can describe $ g^*(x) $ by 
	\begin{equation}
		\label{eqoap}
		g^*(x)=
		\begin{cases}
			x_0 & m=1,\\
			x_0\lor x_1 & m=2,\\
			x_0\lor x_1\land(g^*((x_2,\dots,x_{m-1}))) & m\geq 3
		\end{cases}
	\end{equation} 
	for $ m\in\mathbb{N}_{>0} $.
Moreover, we have 
$g(x) = x_0\land g^*((x_1,\dots, x_{m-1}))$ 
and
$g^*(x) = x_0\lor g((x_1,\dots, x_{m-1}))$ 
for $ m\in\mathbb{N}_{>1} $.
	
\begin{definition}
		A circuit realizing an \textsc{And-Or} path function is called \textbf{\textsc{And-Or} path circuit}\index{and-or path circuit@\textsc{And-Or} path circuit}. A \textbf{family of \textsc{And-Or} path circuits}\index{family of \textsc{And-Or} path circuits} is a family of circuits $ (AOP_m)_{m\in\mathbb{N}} $ where each $ AOP_m $ is an \textsc{And-Or} path circuit on $ m $ inputs. Given an \textsc{And-Or} path $ h(x) $ (either $ g(x) $, or $ g^*(x) $), the Boolean formula for $ h(x) $ emerging from Definition~\ref{defaop} or Equation \eqref{eqoap} is called a \textbf{standard \textsc{And-Or} path realization}\index{and-or path@\textsc{And-Or} path!standard realization of an}. The corresponding circuit is called a \textbf{standard \textsc{And-Or} path circuit}\index{and-or path circuit@\textsc{And-Or} path circuit!standard} for $ h(x) $. 
\end{definition}

\begin{figure}[htb]
\centering
\begin{tikzpicture}[scale=0.41]
 \node[scale=0.7] at (0,0) (x0) {$ x_0 $};
 \node[scale=0.7] at (1,0) (x1) {$ x_1 $};
 \node[scale=0.7] at (2,0) (x2) {$ x_2 $};
 \node[scale=0.7] at (3,0) (x3) {$ x_3 $};
 \node[scale=0.7] at (4,0) (x4) {$ x_4 $};
 \node[scale=0.7] at (5,0) (x5) {$ x_5 $};
 \node[scale=0.7] at (6,0) (x6) {$ x_6 $};
 \node[scale=0.7] at (7,0) (x7) {$ x_7 $};
 \node[scale=0.7] at (8,0) (x8) {$ x_8 $};
 \node[scale=0.7] at (9,0) (x9) {$ x_9 $};
 \node[scale=0.7] at (10,0) (x10) {$ x_{10} $};
 \node[scale=0.7] at (11,0) (x11) {$ x_{11} $};
			
 \node[fill=green, or gate US, draw, logic gate inputs=nn, rotate=270, thick, scale=0.4] at (0.5,-11) (o1) {};
 \node[fill=red, and gate US, draw, logic gate inputs=nn, rotate=270, thick, scale=0.4] at (1.5,-10) (a1) {};
 \node[fill=green, or gate US, draw, logic gate inputs=nn, rotate=270, thick, scale=0.4] at (2.5,-9) (o2) {};
 \node[fill=red, and gate US, draw, logic gate inputs=nn, rotate=270, thick, scale=0.4] at (3.5,-8) (a2) {};
 \node[fill=green, or gate US, draw, logic gate inputs=nn, rotate=270, thick, scale=0.4] at (4.5,-7) (o3) {};
 \node[fill=red, and gate US, draw, logic gate inputs=nn, rotate=270, thick, scale=0.4] at (5.5,-6) (a3) {};
 \node[fill=green, or gate US, draw, logic gate inputs=nn, rotate=270, thick, scale=0.4] at (6.5,-5) (o4) {};
 \node[fill=red, and gate US, draw, logic gate inputs=nn, rotate=270, thick, scale=0.4] at (7.5,-4) (a4) {};
 \node[fill=green, or gate US, draw, logic gate inputs=nn, rotate=270, thick, scale=0.4] at (8.5,-3) (o5) {};
 \node[fill=red, and gate US, draw, logic gate inputs=nn, rotate=270, thick, scale=0.4] at (9.5,-2) (a5) {};
 \node[fill=green, or gate US, draw, logic gate inputs=nn, rotate=270, thick, scale=0.4] at (10.5,-1) (o6) {};

 \node at (0.5,-12) (l) {};
			
\node[below=4.9cm, align=flush center] at (x5){\parbox{5.5cm}{\textbf{(a)} Standard \textsc{And-Or} path circuit for $ g^*((x_0,\dots, x_{11})) $.\vspace{0.0pt}}};

 \draw[thick]  (x0)--(o1.input 2);
 \draw[thick]  (x1)--(a1.input 2);
 \draw[thick]  (x2)--(o2.input 2);
 \draw[thick]  (x3)--(a2.input 2);
 \draw[thick]  (x4)--(o3.input 2);
 \draw[thick]  (x5)--(a3.input 2);
 \draw[thick]  (x6)--(o4.input 2);
 \draw[thick]  (x7)--(a4.input 2);
 \draw[thick]  (x8)--(o5.input 2);
 \draw[thick]  (x9)--(a5.input 2);
 \draw[thick]  (x10)--(o6.input 2);
 \draw[thick]  (x11)--(o6.input 1);

 \draw[thick]  (o6.output)--(a5.input 1);
 \draw[thick]  (a5.output)--(o5.input 1);
 \draw[thick]  (o5.output)--(a4.input 1);
 \draw[thick]  (a4.output)--(o4.input 1);
 \draw[thick]  (o4.output)--(a3.input 1);
 \draw[thick]  (a3.output)--(o3.input 1);
 \draw[thick]  (o3.output)--(a2.input 1);
 \draw[thick]  (a2.output)--(o2.input 1);
 \draw[thick]  (o2.output)--(a1.input 1);
 \draw[thick]  (a1.output)--(o1.input 1);

 \draw[->]  (o1)--(l);
\end{tikzpicture}
\hspace{10mm}
\begin{tikzpicture}[scale=0.41]
 \node[scale=0.7] at (0,0) (x0) {$ x_0 $};
 \node[scale=0.7] at (1,0) (x1) {$ x_1 $};
 \node[scale=0.7] at (2,0) (x2) {$ x_2 $};
 \node[scale=0.7] at (3,0) (x3) {$ x_3 $};
 \node[scale=0.7] at (4,0) (x4) {$ x_4 $};
 \node[scale=0.7] at (5,0) (x5) {$ x_5 $};
 \node[scale=0.7] at (6,0) (x6) {$ x_6 $};
 \node[scale=0.7] at (7,0) (x7) {$ x_7 $};
 \node[scale=0.7] at (8,0) (x8) {$ x_8 $};
 \node[scale=0.7] at (9,0) (x9) {$ x_9 $};
 \node[scale=0.7] at (10,0) (x10) {$ x_{10} $};
 \node[scale=0.7] at (11,0) (x11) {$ x_{11} $};
			
 \node[fill=green, or gate US, draw, logic gate inputs=nn, rotate=270, thick, scale=0.4] at (3,-6) (o1) {};
 \node[fill=red, and gate US, draw, logic gate inputs=nn, rotate=270, thick, scale=0.4] at (3.5,-5) (a1) {};
 \node[fill=green, or gate US, draw, logic gate inputs=nn, rotate=270, thick, scale=0.4] at (4,-4) (o2) {};
 \node[fill=red, and gate US, draw, logic gate inputs=nn, rotate=270, thick, scale=0.4] at (4.5,-3) (a2) {};
 \node[fill=green, or gate US, draw, logic gate inputs=nn, rotate=270, thick, scale=0.4] at (5,-2) (o3) {};
 \node[fill=red, and gate US, draw, logic gate inputs=nn, rotate=270, thick, scale=0.4] at (5.5,-1) (a3) {};
			
 \node[fill=green, or gate US, draw, logic gate inputs=nn, rotate=270, thick, scale=0.4] at (9.5,-3) (o4) {};
 \node[fill=red, and gate US, draw, logic gate inputs=nn, rotate=270, thick, scale=0.4] at (10,-2) (a4) {};
 \node[fill=green, or gate US, draw, logic gate inputs=nn, rotate=270, thick, scale=0.4] at (10.5,-1) (o5) {};
			
 \node[fill=yellow, and gate US, draw, logic gate inputs=nn, rotate=270, thick, scale=0.4] at (5.5,-3) (a5) {};
 \node[fill=yellow, and gate US, draw, logic gate inputs=nn, rotate=270, thick, scale=0.4] at (6,-2) (a6) {};
 \node[fill=yellow, and gate US, draw, logic gate inputs=nn, rotate=270, thick, scale=0.4] at (6.5,-1) (a7) {};
			
 \node[fill=blue, and gate US, draw, logic gate inputs=nn, rotate=270, thick, scale=0.4] at (7.5,-5) (a8) {};
 \node[fill=blue, or gate US, draw, logic gate inputs=nn, rotate=270, thick, scale=0.4] at (5,-7) (o6) {};
			
 \node at (5,-8) (l) {};
			
\node[below=4.9cm, align=flush center] at (x6){\parbox{6.8cm}{\textbf{(b)} Circuit realizing  $ g((x_0,\dots,$ $ x_{11})) $ via the alternating split (Theorem~\ref{thmaltsplit}) with $ k=7 $.\vspace{0.0pt}}};
			
 \draw[thick]  (x0)--(o1.input 2);
 \draw[thick]  (x1)--(a1.input 2);
 \draw[thick]  (x2)--(o2.input 2);
 \draw[thick]  (x3)--(a2.input 2);
 \draw[thick]  (x4)--(o3.input 2);
 \draw[thick]  (x5)--(a3.input 2);
 \draw[thick]  (x6)--(a3.input 1);

 \draw[thick]  (x8)--(o4.input 2);
 \draw[thick]  (x9)--(a4.input 2);
 \draw[thick]  (x10)--(o5.input 2);
 \draw[thick]  (x11)--(o5.input 1);

 \draw[thick]  (x1)--(a5.input 2);
 \draw[thick]  (x3)--(a6.input 2);
 \draw[thick]  (x5)--(a7.input 2);
 \draw[thick]  (x7)--(a7.input 1);

 \draw[thick]  (o5.output)--(a4.input 1);
 \draw[thick]  (a4.output)--(o4.input 1);
 \draw[thick]  (o4.output)--(a8.input 1);
 \draw[thick]  (a3.output)--(o3.input 1);
 \draw[thick]  (o3.output)--(a2.input 1);
 \draw[thick]  (a2.output)--(o2.input 1);
 \draw[thick]  (o2.output)--(a1.input 1);
 \draw[thick]  (a1.output)--(o1.input 1);
 \draw[thick]  (o1.output)--(o6.input 2);

 \draw[thick]  (a7.output)--(a6.input 1);
 \draw[thick]  (a6.output)--(a5.input 1);
 \draw[thick]  (a5.output)--(a8.input 2);
 \draw[thick]  (a8.output)--(o6.input 1);

 \draw[->]  (o6)--(l);
\end{tikzpicture}
\caption[Visualization of the alternating split.]{Visualization of the alternating split (Theorem~\ref{thmaltsplit}).}
\label{figaltsplit}
\end{figure}
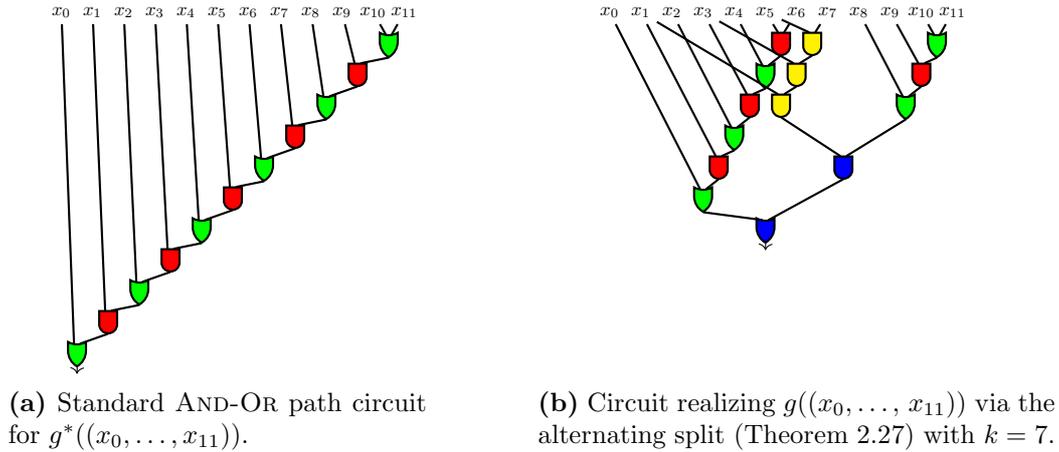

In Figure~\ref{figaltsplit}~(a), we see a standard \textsc{And-Or} path circuit.

By dualization circuits, we see that any circuit over $ \Omega_{\text{mon}} $ realizing 
$ g(x) $ for given inputs $ x $ including arrival times can be transformed into 
a circuit over $ \Omega_{\text{mon}} $ realizing $ g^*(x) $ with the same delay and 
size by interchanging all \textsc{And} and \textsc{Or} gates and vice versa.
Thus, the following naturally arising problem is well defined.

\fbox{\parbox{\linewidth-6.8pt}{\textsc{\Large And-Or Path Circuit Optimization Problem\vspace{5pt}}\\
			\begin{tabularx}{\textwidth-6.8pt}{l X}
				Instance: & $ m\in\mathbb{N} $, Boolean input variables $ x=(x_0,\dots, x_{m-1}) $, arrival times $ a(x_0),\dots, a(x_{m-1})\in\mathbb{N} $.\\
				Task: & Compute a circuit over $ \Omega_\text{mon} $ realizing $ g(x) $ or $ g^*(x) $ with minimum possible delay.
			\end{tabularx}
	}}\index{and-or path circuit optimization problem@\textsc{And-Or Path Circuit Optimization Problem}}

	\begin{remark}
		The \textsc{And-Or Path Circuit Optimization Problem} is equivalent to finding an adder circuit with minimum delay as every adder circuit yields an \textsc{And-Or} path circuit with the same delay and vice versa (see Equation \eqref{eq::carry_bits} and 
Figure~\ref{figac}~(b)). 
	\end{remark}
	
In Figure~\ref{figac}~(b), we have already seen (as a sub-circuit)
a non-trivial \textsc{And-Or} path circuit 
on 5 inputs $g_0, p_1, g_1, p_2, g_2$ computing the output $c_3$.
%
%
%
%
It is indeed an \textsc{And-Or} path circuit which is verified via
\begin{align*}
(g_2\lor(p_2\land g_1))\lor((p_2\land p_1)\land g_0)
        &=g_2\lor\big((p_2\land g_1)\lor(p_2\land (p_1\land g_0))\big)
	=g_2\lor\big((p_2\land (g_1\lor (p_1\land g_0))\big).
\end{align*}
Moreover, it is depth-optimum since any circuit for $ g^*(x) $ on 5 inputs has a depth of at least $ \lceil\log_25\rceil=3 $ by Theorem~\ref{thmlowerbound} 
(since, obviously, an \textsc{And-Or} path depends essentially on all its inputs).


We can adapt Definition~\ref{defcb} to \textsc{And-Or} paths.

\begin{definition}
		Let $ h(x) $ be an \textsc{And-Or} path ($ g(x) $ or $ g^*(x) $) on $ m\in\mathbb{N}_{>1} $ inputs $ x=(x_0,\dots, x_{m-1}) $. We call an input $ x_i $ a \textbf{generate signal}\index{generate signal} (respectively, \textbf{propagate signal}\index{propagate signal}) if the unique successor of $ x_i $ in the standard \textsc{And-Or} path circuit for $ h(x) $ is an \textsc{Or} gate (respectively, \textsc{And} gate).
\end{definition}
	
Note that the inputs of an \textsc{And-Or} path, except for the last input, are 
alternating generate and propagate signals. The last two inputs are both generate 
or both propagate signals.

\noindent{\it Notation:}
We define the $n$-ary \textsc{And} function as $\bigwedge(x)=\bigwedge_{i=0}^{n-1}x_i=x_0\land x_1\land\dots\land x_{n-2}\land x_{n-1}$. 
Similarly, we define the $ n $-ary \textsc{Or} function as
$\bigvee(x)=\bigvee_{i=0}^{n-1}x_i=x_0\lor x_1\lor\dots\lor x_{n-2}\lor x_{n-1}$.

Now we can state a result that turns out to be very useful for \textsc{And-Or} path optimization.
Any \textsc{And-Or} path can be realized by computing two shorter \textsc{And-Or} paths on disjoint subsets of the inputs. We call this kind of realization an \textbf{alternating split}\index{alternating split}. This gives rise to a recursive strategy used in several different variants in the past (see, e.g., Grinchuk \cite{Gri09}) to optimize \textsc{And-Or} path circuits as well as adder circuits. 
The alternating split is also the basis of every adder construction described in this paper.
We omit the proof of the following theorem which can be found in \cite{Gri09}, Lemma~3, 
or in \cite{BH19}, Lemma~2.5.
	
\begin{theorem}[Alternating split. Cf. Grinchuk \cite{Gri09}] \label{thmaltsplit}
Let $ m\in\mathbb{N}_{>1} $, input variables $ x=(x_0,\dots, x_{m-1}) $ and 
an odd integer $ k $ with $ 1\leq k<m $ be given. Then, we have 
\begin{align*}
   g^*(x)&=g^*((x_0,\dots,x_{k-1}))\lor
          \Big(\bigwedge((x_1,x_3,\dots, x_{k}))\land g^*((x_{k+1},\dots, x_{m-1}))\Big) \intertext{and}
   g(x)&=g((x_0,\dots,x_{k-1}))\land
          \Big(\bigvee((x_1,x_3,\dots, x_{k}))\lor g((x_{k+1},\dots, x_{m-1}))\Big).\tag*{\qed}
\end{align*}
\end{theorem}

Though we do not give a full proof here,
we sketch the idea how to prove it. Consider as an example Figure~\ref{figaltsplit}
which shows a circuit realizing $g^*((x_0,\dots, x_{11}))$ via the alternating split
described in Theorem~\ref{thmaltsplit} for $ k=7 $. 
The red and green gates form the sub-circuits realizing the two functions 
$ g^*((x_0,\dots, x_6)) $ and $ g^*((x_8,\dots, x_{11})) $ and the sub-circuit consisting 
of the yellow gates realizes the function $ \bigwedge((x_1,x_3,x_5,x_7)) $. 
The two blue gates are the extra gates needed to combine these sub-circuits.
To see that both circuits are equivalent check that the output of the circuit in (b)
is $1$ if and only if
\begin{itemize}
   \item $g^*((x_0,\dots, x_{6}))$ computes $1$ or
   \item $g^*((x_8,\dots, x_{11}))$ computes $1$ and the propagating 
         inputs $x_1$, $x_3$, $x_5$, and $x_7$ are all $1$.
\end{itemize}
However, it is not to difficult to see that
in this case, and only in this case, the output of $g^*((x_0,\dots, x_{11}))$ is $1$.
This is due to the fact that the output of $g^*((x_0,\dots, x_{11}))$ is one if and
only if there is a generate input $x_i$ with value $1$ for which all propagate inputs
$x_j$ with $j < i$ have value $1$.

Note that the sub-circuit in Figure~\ref{figac}~(b) is also an example for 
the alternating split applied to $ g^*(g_2,p_2,g_1,p_1,g_0) $ with $ k=3 $.

We now state well-known lower bounds on the delay 
of \textsc{And-Or} path circuits and adder circuits. 
The first result is implied by Theorem~\ref{thmlowerbound} 
and the fact that \textsc{And-Or} paths depend essentially on all their inputs.

\begin{proposition}
\label{proplbaop}
		The delay of any \textsc{And-Or} path circuit over $ \Omega_{\text{mon}} $ on inputs $ x=(x_0,\dots, x_{m-1}) $ with arrival times $ a(x_0),\dots, a(x_{m-1})\in\mathbb{N} $ is at least $ \lceil\log_2W\rceil $, where $ W:=\sum_{i=0}^{m-1}2^{a(x_i)} $. \qed
\end{proposition}

This is the best known lower bound for the delay of \textsc{And-Or} path circuits and, hence, also of adder circuits.

\begin{proposition} \label{proplb}
The delay of any adder circuit over $ \Omega_{\text{mon}} $ on the input pairs $ p_0,g_0,\dots, p_{n-1},g_{n-1} $ with arrival times $ a(p_i),a(g_i)\in \mathbb{N} $ for $ i\in\{0,\dots,n-1\} $ is at least $ \lceil\log_2(W-1)\rceil $, where $ W:=\sum_{i=0}^{n-1}(2^{p_i}+2^{g_i}) $.
\end{proposition}
	
\begin{proof}
The delay of an adder circuit is at least the delay of an \textsc{And-Or} path circuit on $ (g_{n-1},p_{n-1},\dots, g_1,p_1,g_0) $ 
(see Eq.~\eqref{eq::carry_bits}), which has a delay of at least 
$ \lceil\log_2(\sum_{i=0}^{n-1}(2^{p_i}+2^{g_i})-2^0)\rceil=\lceil\log_2(W-1)\rceil $
by Proposition~\ref{proplbaop}.
\end{proof}

\subsection{Previous Work}
	\label{prevwork}

In this section, we present some previous work on depth- and delay-optimizing adder circuits, 
\textsc{And-Or} path circuits and \textsc{And}-prefix circuits. 
All results presented in this section are adapted to our notation.

\subsubsection{Depth-Optimizing Algorithms} \label{depthoptalg}

Going back to the 1960s, there is a vast literature on depth-optimization 
of adder circuits. Here, we restrict ourselves on some more recent approaches
that will be useful for our delay-optimizing adder circuits. For an overview of the
historic development of adder circuits, we refer to \cite{Sp14} and \cite{BS24}.

	Hermann \cite{Her20} constructed an adder circuit based on the recursive strategy implied by the alternating split, presented in Theorem~\ref{thmaltsplit}, as follows.
	Given some $ n\in\mathbb{N} $ and input pairs $ p_0,g_0,\dots,p_{n-1},g_{n-1} $, the inputs are split in two (almost) equal consecutive parts $ P_r:=(p_0,g_0,\dots, p_{s-1},g_{s-1}) $ and $ P_l:=(p_s,g_s,\dots, p_{n-1},g_{n-1}) $ for $ s:=\lceil\frac{n}{2}\rceil $. Recursively, an adder $ A_r $ (respectively, $ A_l $) is then computed on the two halves $ P_r $ (respectively, $ P_l $). Then, all the carry bits $ c_i $ for $ i\leq s $ are computed directly by $ A_r $, i.e., $ c_i=\text{out}_i(A_r) $. The other carry bits are now computed as follows. Equation~\eqref{eq::carry_bits} implies that any carry bit $ c_{i+1} $ is given by $ c_{i+1}=g^*((g_i,p_i,\dots, g_1,$ $p_1,g_0)) $. Theorem~\ref{thmaltsplit} with $ k=2(i-s)+1 $ implies that
	\begin{align}
		c_{i+1}=\,&g^*((g_i,p_i,\dots,g_{s+1},p_{s+1},g_{s}))\,\lor\, 
		\Big(\bigwedge((p_{s},\dots, p_{i}))\land g^*((g_{s-1},p_{s-1},\dots, g_{1},p_1,g_0))\Big) \label{eqqadderconstr}
	\end{align} 
	for all $ s\leq i\leq n $. The term $ g^*((g_i,p_i,\dots,g_{s+1},p_{s+1},g_{s})) $ in \eqref{eqqadderconstr} is given by the adder $ A_l $ for each $ s\leq i\leq n $, i.e., $ g^*((g_i,p_i,\dots,g_{s+1},p_{s+1},g_{s}))=\text{out}_{i-s+1}(A_l) $. Hence, the addition is reduced to computing the single \textsc{And-Or} path $ g^*((g_{s-1},p_{s-1},\dots, g_{1},p_1,g_0)) $ as well as the $ (i-s+1) $-ary \textsc{And} function $ \bigwedge((p_{s},\dots, p_{i})) $ for all $ s\leq i\leq n $.
	 
	We formalize the latter problem in a more general way.\vspace{5pt}\\
	\fbox{\parbox{\linewidth-6.8pt}{\textsc{\Large Parallel And-Prefix Problem\vspace{5pt}}\\
			\begin{tabularx}{\textwidth-6.8pt}{l X}
				Instance: & $ n\in\mathbb{N} $.\\
				Task: & Construct a circuit over $ \Omega=\{\land\} $ on inputs $ x_0,\dots, x_{n-1} $ computing the \textsc{And}-prefixes $ x_i\land x_{i-1}\land\dots\land x_0 $ for all $ i=0,\dots, n-1 $.
			\end{tabularx}
	}}\index{parallel and-prefix problem@\textsc{Parallel And-Prefix Problem}}

	\begin{definition}
		A circuit solving the \textsc{Parallel And-Prefix Problem} is called \textbf{\textsc{And}-prefix circuit}\index{and-prefix circuit@\textsc{And}-prefix circuit}. A \textbf{family of \textsc{And}-prefix circuits}\index{family of \textsc{And}-prefix circuits} is a family of circuits $ (S_n)_{n\in\mathbb{N}} $ where each $ S_n $ is an \textsc{And}-prefix circuit on $ n $ inputs. Given an \textsc{And}-prefix circuit $ S_n $ on $ n $ inputs and $ i\in\{1,\dots, n\} $, we denote the output of $ x_{i-1}\land\dots\land x_0 $ by $ \text{out}_i(S_n) $.
	\end{definition}

	There have been several works dedicated to the \textsc{Parallel And-Prefix Problem}. The first construction combining an optimal depth and a linear size was made by Ladner and Fischer \cite{LF80}.
	
\begin{theorem}[Ladner and Fischer \cite{LF80}]
		\label{thmlf}
For any $ n\in\mathbb{N}_{>0} $ and any constant $ 0\leq f\leq \lceil\log_2n\rceil $, one can construct in polynomial time
an \textsc{And}-prefix circuit $ S_n^f $ on $ n $ inputs with depth 
		\begin{align*}
			\mathrm{d}(S_n^f)\leq&\lceil\log_2n\rceil + f \intertext{and size}
			\mathrm{s}(S_n^f)\leq &2(1+2^{-f})n.
		\end{align*}
\end{theorem}

Note that the depth of this construction is optimal for $ f=0 $ due to Theorem~\ref{thmlowerbound}. 

The circuit from Theorem~\ref{thmlf} was be used to compute the terms 
$\bigwedge((p_{s},\dots, p_{i})) $ in \eqref{eqqadderconstr} for all 
$s\leq i\leq n$, and a circuit for the \textsc{And-Or} path 
$g^*((g_{s-1},p_{s-1},\dots, g_{1},p_1,g_0))$ in \eqref{eqqadderconstr} 
was provided in \cite{BS24}.


\begin{theorem}[Brenner and Silvanus \cite{BS24}] \label{thmdaop}
For any $ m\in\mathbb{N}_{\geq 2} $, one can construct in running time
$ \mathcal{O}(m\log_2m) $ an \textsc{And-Or} path circuit $ AOP_m $ on 
$ m $ inputs with depth
   \begin{align*}
			\mathrm{d}(AOP_m)\leq& \log_2m+\log_2\log_2m+0.65 \intertext{and size}
			\mathrm{s}(AOP_m)\leq& 3.67m-2.
   \end{align*}
\end{theorem}

These \textsc{And-Or} path circuits are the best known regarding the depth. 
The following result of Commentz-Walter \cite{CW79} even shows that the
circuits from Theorem~\ref{thmdaop} are optimal in depth 
up to a an additive constant.
	
\begin{theorem}[Cf. Commentz-Walter \cite{CW79}] \label{thmlbaop}
There is some $ \alpha\in\mathbb{R} $ and some $ N_\alpha\in\mathbb{N} $ such that for all $ n\geq N_\alpha $, any circuit $ C $ over $ \Omega_{\text{mon}} $ for $ g((x_0,\dots, x_{2n-1})) $ fulfills \[\mathrm{d}(C)\geq \log_2n+\log_2\log_2n+\alpha.\]
\end{theorem}

	\begin{remark}
		Hitschke \cite{Hit18} showed that the constant $ \alpha $ in the previous theorem can be chosen arbitrarily close to $ -4 $, asymptotically. For example, a lower bound of $ \log_2n+\log_2\log_2n-4.01 $ was shown
for all $ n\geq $\vspace{2pt} $N_{-4.01}=~2^{2^{18}} $.
	\end{remark}
	
Using the \textsc{And-Or} path circuit from Theorem~\ref{thmdaop} and the \textsc{And}-prefix circuit by Ladner and Fischer (Theorem~\ref{thmlf}) as sub-circuits in Formula \eqref{eqqadderconstr}, 
Brenner and Silvanus \cite{BS24} achieved an adder with a sub-quadratic size whose 
depth is only by 2 larger than the depth of the \textsc{And-Or} path circuit from Theorem~\ref{thmdaop}.
		

\begin{theorem}[Brenner and Silvanus \cite{BS24}] \label{daddercirc}
For any $ n\in \mathbb{N}_{\geq 3} $, one can construct 
in running time $ \mathcal{O}(n\log_2^2n) $
an adder circuit $ A_{n} $ on $ n $ input pairs with depth 
		\begin{align*}
			\mathrm{d}(A_{n})&\leq \log_2n + \log_2\log_2n + 2.65 \intertext{and size} \mathrm{s}(A_n)&\leq 6.2n\log_2n.
		\end{align*} 
\end{theorem}

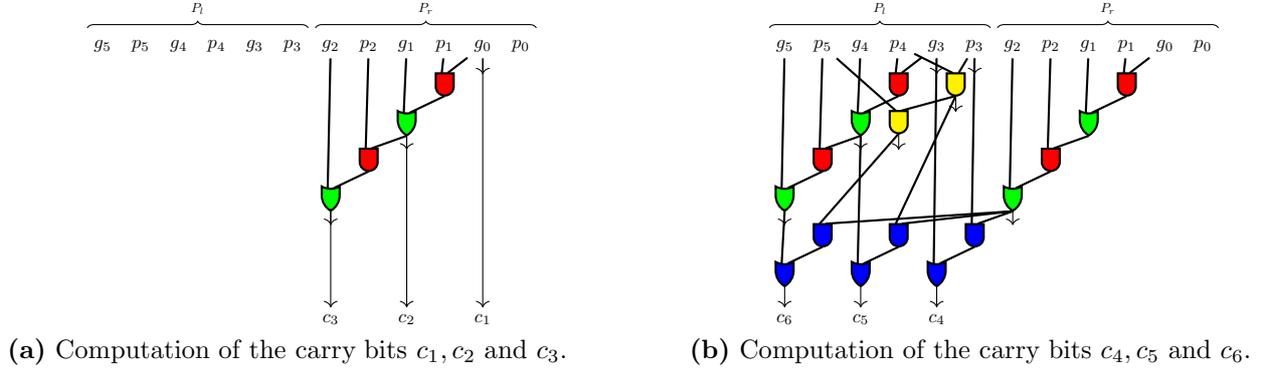
\begin{figure}[htb]
		\centering
		\begin{tikzpicture}[scale=0.5]
			\node[scale=0.7] at (0,0) (g5) {$ g_5 $};
			\node[scale=0.7] at (1,0) (p5) {$ p_5 $};
			\node[scale=0.7] at (2,0) (g4) {$ g_4 $};
			\node[scale=0.7] at (3,0) (p4) {$ p_4 $};
			\node[scale=0.7] at (4,0) (g3) {$ g_3 $};
			\node[scale=0.7] at (5,0) (p3) {$ p_3 $};
			\node[scale=0.7] at (6,0) (g2) {$ g_2 $};
			\node[scale=0.7] at (7,0) (p2) {$ p_2 $};
			\node[scale=0.7] at (8,0) (g1) {$ g_1 $};
			\node[scale=0.7] at (9,0) (p1) {$ p_1 $};
			\node[scale=0.7] at (10,0) (g0) {$ g_0 $};
			\node[scale=0.7] at (11,0) (p0) {$ p_0 $};

			\node[fill=red, and gate US, draw, logic gate inputs=nn, rotate=270, thick, scale=0.4] at (7,-3) (a3) {};
			\node[fill=green, or gate US, draw, logic gate inputs=nn, rotate=270, thick, scale=0.4] at (6,-4) (o3) {};
			\node[fill=red, and gate US, draw, logic gate inputs=nn, rotate=270, thick, scale=0.4] at (9,-1) (a4) {};
			\node[fill=green, or gate US, draw, logic gate inputs=nn, rotate=270, thick, scale=0.4] at (8,-2) (o4) {};
			
			\node[scale=0.7] at (6,-7.25) (c3) {$ c_3 $};
			\node[scale=0.7] at (8,-7.25) (c2) {$ c_2 $};
			\node[scale=0.7] at (10,-7.25) (c1) {$ c_1 $};
			\node at (6,-5) (l1) {};
			\node at (8,-3) (l2) {};						
			\node at (10,-1) (l3) {};
			
			\node[below=3.75cm, align=flush center] at (p3){\parbox{7.5cm}{\textbf{(a)} Computation of the carry bits $ c_1,c_2 $ and $ c_3 $.}};
			
			\node[scale=0.5] at (2.5,1) {$ P_l $};
			\node[scale=0.5] at (8.5,1) {$ P_r $};
			
			\draw[decorate,decoration={brace,raise=0.5ex}] (g5.north west) -- (p3.north east);
			\draw[decorate,decoration={brace,raise=0.5ex}] (g2.north west) -- (p0.north east);
			
			\draw[thick]  (g2)--(o3.input 2);
			\draw[thick]  (p2)--(a3.input 2);
			\draw[thick]  (g1)--(o4.input 2);
			\draw[thick]  (p1)--(a4.input 2);
			\draw[thick]  (g0)--(a4.input 1);
			\draw[thick]  (a4.output)--(o4.input 1);
			\draw[thick]  (o4.output)--(a3.input 1);
			\draw[thick]  (a3.output)--(o3.input 1);
			\draw[->]  (o3)--(l1);
			\draw[->]  (o4)--(l2);
			\draw[->]  (g0)--(l3);
			\draw[->]  (o3)--(c3);
			\draw[->]  (o4)--(c2);
			\draw[->]  (g0)--(c1);
\end{tikzpicture}
\hspace*{10mm}
\begin{tikzpicture}[scale=0.5]
			\node[scale=0.7] at (0,0) (g5) {$ g_5 $};
			\node[scale=0.7] at (1,0) (p5) {$ p_5 $};
			\node[scale=0.7] at (2,0) (g4) {$ g_4 $};
			\node[scale=0.7] at (3,0) (p4) {$ p_4 $};
			\node[scale=0.7] at (4,0) (g3) {$ g_3 $};
			\node[scale=0.7] at (5,0) (p3) {$ p_3 $};
			\node[scale=0.7] at (6,0) (g2) {$ g_2 $};
			\node[scale=0.7] at (7,0) (p2) {$ p_2 $};
			\node[scale=0.7] at (8,0) (g1) {$ g_1 $};
			\node[scale=0.7] at (9,0) (p1) {$ p_1 $};
			\node[scale=0.7] at (10,0) (g0) {$ g_0 $};
			\node[scale=0.7] at (11,0) (p0) {$ p_0 $};
			\node[fill=red, and gate US, draw, logic gate inputs=nn, rotate=270, thick, scale=0.4] at (1,-3) (a1) {};
			\node[fill=green, or gate US, draw, logic gate inputs=nn, rotate=270, thick, scale=0.4] at (0,-4) (o1) {};
			\node[fill=red, and gate US, draw, logic gate inputs=nn, rotate=270, thick, scale=0.4] at (3,-1) (a2) {};
			\node[fill=green, or gate US, draw, logic gate inputs=nn, rotate=270, thick, scale=0.4] at (2,-2) (o2) {};
			\node[fill=red, and gate US, draw, logic gate inputs=nn, rotate=270, thick, scale=0.4] at (7,-3) (a3) {};
			\node[fill=green, or gate US, draw, logic gate inputs=nn, rotate=270, thick, scale=0.4] at (6,-4) (o3) {};
			\node[fill=red, and gate US, draw, logic gate inputs=nn, rotate=270, thick, scale=0.4] at (9,-1) (a4) {};
			\node[fill=green, or gate US, draw, logic gate inputs=nn, rotate=270, thick, scale=0.4] at (8,-2) (o4) {};
			\node[fill=yellow, and gate US, draw, logic gate inputs=nn, rotate=270, thick, scale=0.4] at (4.5,-1) (a5) {};
			\node[fill=yellow, and gate US, draw, logic gate inputs=nn, rotate=270, thick, scale=0.4] at (3,-2) (a6) {};
			\node[fill=blue, and gate US, draw, logic gate inputs=nn, rotate=270, thick, scale=0.4] at (1,-5) (a7) {};
			\node[fill=blue, or gate US, draw, logic gate inputs=nn, rotate=270, thick, scale=0.4] at (0,-6) (o5) {};
			\node[fill=blue, and gate US, draw, logic gate inputs=nn, rotate=270, thick, scale=0.4] at (3,-5) (a8) {};
			\node[fill=blue, or gate US, draw, logic gate inputs=nn, rotate=270, thick, scale=0.4] at (2,-6) (o6) {};
			\node[fill=blue, and gate US, draw, logic gate inputs=nn, rotate=270, thick, scale=0.4] at (5,-5) (a9) {};
			\node[fill=blue, or gate US, draw, logic gate inputs=nn, rotate=270, thick, scale=0.4] at (4,-6) (o7) {};
			\node[scale=0.7] at (0,-7.25) (c6) {$ c_6 $};
			\node[scale=0.7] at (2,-7.25) (c5) {$ c_5 $};
			\node[scale=0.7] at (4,-7.25) (c4) {$ c_4 $};
			\node at (4,-1) (l1) {};
			\node at (5,-1) (l2) {};						
			\node at (4.5,-2) (l3) {};
			\node at (2,-3) (l4) {};
			\node at (3,-3) (l5) {};
			\node at (0,-5) (l6) {};
			\node at (6,-5) (l7) {};
			
			\node[below=3.75cm, align=flush center] at (p3){\parbox{7.5cm}{\textbf{(b)} Computation of the carry bits $ c_4,c_5 $ and $ c_6 $.}};
			
			\node[scale=0.5] at (2.5,1) {$ P_l $};
			\node[scale=0.5] at (8.5,1) {$ P_r $};
			
			\draw[decorate,decoration={brace,raise=0.5ex}] (g5.north west) -- (p3.north east);
			\draw[decorate,decoration={brace,raise=0.5ex}] (g2.north west) -- (p0.north east);
			
			\draw[thick]  (g5)--(o1.input 2);
			\draw[thick]  (p5)--(a1.input 2);
			\draw[thick]  (p5)--(a6.input 2);
			\draw[thick]  (g4)--(o2.input 2);
			\draw[thick]  (p4)--(a2.input 2);
			\draw[thick]  (p4)--(a5.input 2);
			\draw[thick]  (g3)--(a2.input 1);
			\draw[->]  (g3)--(l1);
			\draw[->]  (p3)--(l2);
			\draw[thick]  (p3)--(a5.input 1);
			\draw[thick]  (g2)--(o3.input 2);
			\draw[thick]  (p2)--(a3.input 2);
			\draw[thick]  (g1)--(o4.input 2);
			\draw[thick]  (p1)--(a4.input 2);
			\draw[thick]  (g0)--(a4.input 1);
			\draw[thick]  (a2.output)--(o2.input 1);
			\draw[thick]  (a5.output)--(a6.input 1);
			\draw[->]  (a5)--(l3);
			\draw[thick]  (a4.output)--(o4.input 1);
			\draw[thick]  (o2.output)--(a1.input 1);
			\draw[->]  (o2)--(l4);
			\draw[->]  (a6)--(l5);
			\draw[thick]  (a6.output)--(a7.input 2);
			\draw[thick]  (a5.output)--(a8.input 2);
			\draw[thick]  (p3)--(a9.input 2);
			\draw[thick]  (o4.output)--(a3.input 1);
			\draw[thick]  (a1.output)--(o1.input 1);
			\draw[thick]  (a3.output)--(o3.input 1);
			\draw[->]  (o1)--(l6);
			\draw[thick]  (o1.output)--(o5.input 2);
			\draw[thick]  (o2.output)--(o6.input 2);
			\draw[thick]  (g3)--(o7.input 2);
			\draw[thick]  (o3.output)--(a7.input 1);
			\draw[thick]  (o3.output)--(a8.input 1);
			\draw[thick]  (o3.output)--(a9.input 1);
			\draw[->]  (o3)--(l7);
			\draw[thick]  (a7.output)--(o5.input 1);
			\draw[thick]  (a8.output)--(o6.input 1);
			\draw[thick]  (a9.output)--(o7.input 1);
			\draw[->]  (o5)--(c6);
			\draw[->]  (o6)--(c5);
			\draw[->]  (o7)--(c4);
		\end{tikzpicture}
\caption[Visualization of the adder construction by Hermann on 6 input pairs.]{Visualization of the adder construction of Theorem~\ref{daddercirc} on 6 input pairs.}
		\label{figdaddercirc}
\end{figure}

	Figure~\ref{figdaddercirc} illustrates the adder construction of Theorem~\ref{daddercirc} for 6 input pairs. The recursively constructed adders on $ P_l $ and $ P_r $ as well as the \textsc{And-Or} path circuit from Theorem~\ref{thmdaop} on $ P_r $ are indicated with red (\textsc{And}) and green (\textsc{Or}) gates. The yellow gates represent the \textsc{And}-prefix circuit by Ladner and Fischer (Theorem~\ref{thmlf}). Each of these sub-circuits are indicated by a trivial implementation. The blue gates are the extra gates introduced in Equation \eqref{eqqadderconstr} (two gates for each carry bit $ c_4,c_5,c_6 $). 

The depth of the adder from Theorem~\ref{thmdaop} is optimal up to a constant 
by Theorem~\ref{thmlbaop} and it has a sub-quadratic size. To achieve linear size, 
Brenner and Silvanus \cite{BS24} developed a framework that linearizes given adders with a sub-quadratic size such that the depth does not increase much. 
Since we use this framework in our approach 
to linearize delay-optimizing adders, we will present it in more detail 
in Section~\ref{lpartframework}.
Using their linearization framework, 
Brenner and Silvanus \cite{BS24} constructed the following two linear size adder circuits,
for which the depth increased only by an additive term of size
$\log_2\log_2\log_2n + \text{const}$.
	

\begin{theorem}[Brenner and Silvanus \cite{BS24}] \label{daddercirclin0}
For any $ n\in \mathbb{N}_{\geq 4} $, one can construct 
in running time $ \mathcal{O}(n\log_2\log_2n) $
an adder circuit $ A_{n} $ on $ n $ input pairs with depth 
		\begin{align*}
			\mathrm{d}(A_{n})&\leq \log_2n + \log_2\log_2n + \log_2\log_2\log_2n+6.6 \intertext{and size} \mathrm{s}(A_n)&\leq 21.6n.
		\end{align*} 
\end{theorem} 

\begin{theorem}[Brenner and Silvanus \cite{BS24}] \label{daddercirclin}
For any $ n\in \mathbb{N}_{\geq 4} $, one can construct 
in running time $ \mathcal{O}(n\log_2\log_2n) $
an adder circuit $ A_{n} $ on $ n $ input pairs with depth 
   \begin{align*}
   \mathrm{d}(A_{n})&\leq \log_2n + \log_2\log_2n + \log_2\log_2\log_2n+7.6 \intertext{and size} \mathrm{s}(A_n)&\leq 16.7n.
   \end{align*} 
\end{theorem}

\subsubsection{Delay-Optimizing Algorithms}
	The asymptotically best delay bound of a linear size adder circuit was provided by Spirkl \cite{Sp14} (Corollary 3.38) who claimed a delay bound of order $ \lceil\log_2W\rceil$ $+\mathcal{O}(\sqrt{\log_2n}) $ and a linear size. 
	
	Using a similar adder construction as described in the previous section, we hope to achieve a significantly better delay bound and a sub-quadratic size making use of fast \textsc{And-Or} path circuits and \textsc{And}-prefix circuits. Brenner and Hermann \cite{BH19} constructed the \textsc{And-Or} path circuit with the best known delay. The size of their circuit was proven by Hermann \cite{Her20} to be in $ \mathcal{O}(n\log_2n) $.

\begin{theorem}[\cite{BH19} and \cite{Her20} (Thm. 4.3.4)] \label{andorcirc}
		Let $ m\in \mathbb{N} $ with $ m\geq 3 $, Boolean variables $ x=(x_0,\dots, x_{m-1}) $ and arrival times $ a(x_0),\dots, a(x_{m-1})\in \mathbb{N} $ be given. Let $ W= \sum_{i=0}^{m-1}2^{a(x_i)} $. 
One can construct in running time $ \mathcal{O}(m\log_2^2m) $
an \textsc{And-Or}-path circuit $ AOP $ on $ x $ with delay 
		\begin{align*}
			\mathrm{del}(AOP)\leq &\log_2W+\log_2\log_2m+\log_2\log_2\log_2m + 4.3, \intertext{size} 
			\mathrm{s}(AOP)\leq &m(\log_2m+\log_2\log_2m+\log_2\log_2\log_2m +3.3)-1 \intertext{and fanout}
			\mathrm{fanout}(AOP)\leq&\log_2m+\log_2\log_2m+\log_2\log_2\log_2m +3.3\,.
		\end{align*}
\end{theorem}
	
	\begin{remark}
Computing each carry bit separately with these \textsc{And-Or} path circuits as in 
Figure~\ref{figac}~(b) yields the best delay guarantee known of
$ \log_2W+\log_2\log_2(2n-1)+\log_2\log_2\log_2(2n-1) + 4.3 $ for an adder on $ n $ input pairs with weight $ W $. However, its size would be at least quadratic and, thus, it would be unsuitable in practice.
	\end{remark}

	The best \textsc{And-Or} path circuit regarding delay with a linear size was provided by Spirkl \cite{Sp14}. For given input pairs $ p_0,g_0,\dots, p_{n-1},g_{n-1} $ with arrival times $ a(p_0),a(g_0),\dots, a(p_{n-1}),a(g_{n-1})\in\mathbb{N} $ such that $ a(p_i)=a(g_i) $ for all $ i\in\{0,\dots, n\} $, Spirkl claimed that the carry bit $ c_{n} $ can be computed by a circuit with delay at most $ \log_2V+2\sqrt{2\log_2n}+6 $, size at most $ 6n-6 $ and fanout at most $ 2^{\sqrt{2\log_2n}}+1 $, where $ V=\sum_{i=0}^{n-1}2^{a(p_i)} $. Hence, the circuit computes an \textsc{And-Or} path of length $ 2n-1 $. For arbitrary inputs $ x_0,\dots, x_{m-1} $ with arbitrary arrival times $ a(x_0),\dots, a(x_{m-1}) $ and weight $ W=\sum_{i=0}^{m-1}2^{a(x_i)} $, we can use Spirkl's result with $ V=W $, $ n=\frac{m+1}{2} $ for odd $ m $ and $ n=\frac{m}{2}+1 $ for even $ m $ using an artificial input. We get the following result.
	
	\begin{theorem}[Spirkl \cite{Sp14} (Thm. 2.26)]
		\label{andorcircspir}
		Let $ m\in \mathbb{N} $ with $ m\geq 3 $, Boolean variables $ x=(x_0,\dots, x_{m-1}) $ and arrival times $ a(x_0),\dots, a(x_{m-1})\in \mathbb{N} $ be given. Let $ W= \sum_{i=0}^{m-1}2^{a(x_i)} $ be the weight of the inputs. We can construct in polynomial time
an \textsc{And-Or}-path circuit $ AOP $ on $ x $ with delay 
		\begin{align*}
			\mathrm{del}(AOP)\leq &\log_2W+2\sqrt{2\log_2m}+6, \intertext{size} 
			\mathrm{s}(AOP)\leq &3m \intertext{and fanout}
			\mathrm{fanout}(AOP)\leq&2^{\sqrt{2\log_2m}}+1.
		\end{align*}
	\end{theorem}

	We still need an \textsc{And}-prefix circuit guaranteeing a good delay in order to attempt an adder construction. The best known \textsc{And}-prefix circuit regarding the delay is given by Rautenbach, Szegedy and Werber \cite{RSW}. They addressed the \textsc{Parallel Prefix Problem}\index{parallel prefix problem@\textsc{Parallel Prefix Problem}}, which is the \textsc{Parallel And-Prefix Problem} for an arbitrary, associative, binary operator $ \circ:\{0,1\}\times\{0,1\}\to\{0,1\} $.

\begin{theorem}[Rautenbach, Szegedy, Werber \cite{RSW} (Thm. 2)] \label{andprecirc0}
Let $ \circ:\{0,1\}\times\{0,1\}\to\{0,1\} $ be an associative operator. 
Let $ m\in \mathbb{N} $ with $ m\geq 3 $, Boolean variables $ x_0,\dots, x_{m-1} $ and 
arrival times $ a(x_0),\dots, a(x_{m-1})\in \mathbb{N} $ be given and let 
$ W= \sum_{i=0}^{m-1}2^{a(x_i)} $. The \textsc{Parallel Prefix Problem} can be solved 
in polynomial runtime by a circuit $ S_n $ over the basis $ \{\circ\} $ with delay 
\begin{align*}
\mathrm{del}(S_n)\leq &\log_2(W)+3\log_2\log_2m + \mathcal{O}(1)
   \intertext{ and size}
\mathrm{s}(S_n)\in&\mathcal{O}(m\log_2\log_2m)\,.
\end{align*}
\end{theorem}

Thus, there are constants $ c_d, c_s\in\mathbb{R} $ such that the circuits 
from the theorem above  
have a delay of at most $ \log_2(W)+3\log_2\log_2m + c_d $ and a size of at most $ c_sm\log_2\log_2m $. However, these constants were not specified in~\cite{RSW}. 
Without concrete values for these constants we would not be able to compute concrete delay and size bounds for our adders. Hence, we will compute upper bounds for $ c_d $ and $ c_s $ in Section~\ref{analandprecirc}.

For some sub-circuits used during our adder constructions, we do not need the best possible asymptotic delay bounds but rather rely on smaller sizes. This may be the case for small instances or for our construction of a linear size adder in Section~\ref{linadder}. For these cases, we capture some delay bounds of depth-minimizing circuits presented in Section~\ref{depthoptalg}.
	


\begin{corollary} \label{cordaddercirclin}
Assume that we are given arrival times 
$ a(p_0),a(g_0),\dots, a(p_{n-1}),a(g_{n-1})$ $\in \mathbb{N} $ 
to our inputs. Let $ W = \sum_{i=0}^{n-1}\big(2^{a(p_{i})} + 2^{a(g_{i})}\big) $.
Then, the following statements hold.
\begin{itemize}
\item[(a)]
The circuit $ A_n $ from Theorem~\ref{daddercirc} has delay of at most\,
$
\log_2W + \log_2n + \log_2\log_2n + 2.65.
$
\item[(b)]
The circuit $ A_n $ from Theorem~\ref{daddercirclin0} has delay of at most\,
$
\log_2W+\log_2n + \log_2\log_2n + \log_2\log_2\log_2n+6.6.
$
\item[(c)]
The circuit $ A_n $ from Theorem~\ref{daddercirclin} has delay of at most\,
$
\log_2W+\log_2n + \log_2\log_2n + \log_2\log_2\log_2n+7.6.
$
\end{itemize}
\end{corollary}

\begin{proof}
In all cases, we can observe that 
\[
   \text{del}(A_n)\leq \max\{a(p_i), a(g_i); 0\leq i\leq n-1\} + \text{d}(A_n).
\] 
The results follow from the depth bound of $ A_n $, stated in 
Theorem~\ref{daddercirc}, Theorem~\ref{daddercirclin0}, and Theorem~\ref{daddercirclin},
and the fact that 
\[
\max\{a(p_i), a(g_i); 0\leq i\leq n-1\}\leq \log_2W. \tag*{\qedhere}
\]
\end{proof}


The results of the next corollary can be proven analogously.

\begin{corollary}\label{corthmall}
Assume that we are given arrival times $ a(x_0),\dots, a(x_{n-1})\in \mathbb{N} $ to our
inputs. Let $W = \sum_{i=0}^{n-1}2^{a(x_i)}$. Then, the following statements hold.
\begin{itemize}
\item[(a)]
The circuit $ S_n $ from Theorem~\ref{thmlf} has delay at most\,\,
$log_2W+\lceil\log_2n\rceil + f$.
 \item[(b)]
The circuit $AOP_m$ from Theorem~\ref{thmdaop} has delay at most\,\,
$\log_2W+\log_2m+\log_2\log_2m + 0.65$.
\end{itemize}
\end{corollary}



\section{Analysis of an \textsc{And}-Prefix Circuit}
	\label{analandprecirc}
	
In this section, we compute concrete constants for the delay and size bounds of the circuit constructed by Rautenbach, Szegedy and Werber \cite{RSW} for the \textsc{Parallel Prefix Problem} (see Theorem~\ref{andprecirc0}). For this we need to take a closer look at their circuit construction which we describe in our own notation in the following.
	
	Let $ S(t_0,\dots, t_{n-1}) $ denote a circuit over the basis $ \{\circ\} $ that solves the prefix problem on $ n $ inputs $ x=(x_0,x_1,\dots,x_{n-1}) $ with arrival times $ t_0,\dots, t_{n-1}$ $\in \mathbb{N} $ for some associative operator $ \circ:\{0,1\}\times\{0,1\}\to\{0,1\} $. The construction of $ S(t_0,\dots, t_{n-1}) $ is then done as follows:

For $ n=1 $ the circuit $ S(t_0) $ consists of the input vertex $ x_0 $.

For $ n\geq2 $ we apply the following steps.
\begin{enumerate}
\item Partition the set $ \{0,1,\dots,n-1\} $ into $ l:=\lceil\sqrt{n}\rceil $ sets 
{\allowdisplaybreaks
\begin{align*}
			P^{(1)}=&\{0,1,\dots,n_1-1\},\\
			P^{(2)}=&\{(n_1),(n_1+1),\dots,(n_1+n_2-1)\},\\&\dots,\\
			P^{(l)}=&\{(n_1+n_2+\dots+n_{l-1}),(n_1+n_2+\dots+n_{l-1}+1),
			 \dots,(n_1+n_2+\dots+n_l-1)\}
\end{align*}
}
such that $ n_1\geq n_2\geq \dots \geq n_l $ and $ n_1-n_l\leq 1 $.
\item For $ 1\leq i\leq l $ we use the following dynamic programming approach to construct a circuit $ C_i $ over the basis $ \{\circ\} $ calculating $ y_i:=\circ_{j\in P^{(i)}}x_j $: In what follows, $ C_{j_1,j_2} $ will denote a circuit calculating \[\overset{j_2}{\underset{j=j_1}{\circ}}x_{(n_1+n_2+\dots+n_{i-1}+j)}\] for $ 0\leq j_1\leq j_2\leq n_i-1 $. If $ j_1=j_2 $, then $ C_{j_1,j_2} $ consists only of the corresponding input vertex.
		
If $ j_1<j_2 $, we recursively construct $ C_{j_1,j_2} $ using one $ \circ $-gate joining 
the outputs of two circuits $ C_{j_1,k} $ and $ C_{k+1,j_2} $ such that 
$
   \max\{\text{del}(C_{j_1,k}),\text{del}(C_{k+1,j_2})\}
$
is minimized.
		
Let $ C_i=C_{0,n_i-1} $.
\item For $ 1\leq i\leq l $, we recursively construct \[S(t_{(n_1+n_2+\dots+n_{i-1})},\dots,t_{(n_1+n_2+\dots+n_i-2)})\] and use these circuits to calculate all $ n_i-1$ prefixes on the inputs $ x_j $ for $ j\in P^{(i)}\backslash \{n_1+n_2+\dots+n_i-1\} $.
\item We construct $ S(t(y_1),t(y_2),\dots,t(y_{l-1})) $ to calculate all $ (l-1) $ prefixes on the inputs $ y_j $ with their arrival times $ t(y_j) $ for $ 1\leq j\leq l-1 $ calculated by the circuits constructed in Step 2.
\item For $ 2\leq i\leq l $ and $ j\in P^{(i)}\backslash\{n_1+n_2+\dots +n_i-1\} $ 
we join the output $ (y_1\circ y_2\circ \dots \circ y_{i-1}) $ of the circuit constructed 
in Step 4 with the output $ \circ_{\substack{k\in P^{(i)}\\k\leq j}}x_k $ of the circuit constructed 
in Step 3 using one $ \circ $-gate\vspace{-4pt} calculating 
\[
\overset{j-1}{\underset{k=0}{\circ}}x_k=(y_1\circ y_2\circ\dots\circ y_{i-1})\circ \Bigg(\underset{\substack{k\in P^{(i)}\\k< j}}{\circ}x_k\Bigg).
\]
\item Finally, we join the output $ (y_1\circ y_2\circ\dots\circ y_{l-1}) $ of the circuit constructed in Step 4 with the output $ y_l $ of the circuit constructed in Step 2 using one $ \circ $-gate which calculates $ \circ_{i=0}^{n-1}x_i $.
\end{enumerate}

	Rautenbach, Szegedy and Werber derived recursion formulas for the delay and size of the circuits resulting from this algorithm. For $ n\in\mathbb{N}_{\geq 1} $ let $ \text{s}(n) $ denote the maximum size and $ \text{del}(W,n) $ denote the maximum delay of a circuit $ S(t_0,\dots,t_{n-1}) $ for arrival times $ t_0,\dots, t_{n-1}\in \mathbb{N} $ and $ W=\sum_{i=0}^{n-1}2^{t_i} $.
	
	\begin{lemma}[Rautenbach, Szegedy, Werber \cite{RSW} (Lem. 2)]
		\label{lemrec}
		For $ n\in\mathbb{N}_{\geq 3} $ we have the recursions 
		\begin{align*}
			\mathrm{s}(n)&\leq (\lceil\sqrt{n}\rceil+1)\hspace{2pt}\mathrm{s}(\lceil\sqrt{n}\rceil-1) + 2(n-\lceil\sqrt{n}\rceil)
			\intertext{and}
			\mathrm{del}(W,n)&\leq \mathrm{del}(4W,\lceil\sqrt{n}\rceil-1)+1.\tag*{\qed}
		\end{align*}
	\end{lemma}

Using these recursion formulas, Rautenbach, Szegedy and Werber showed that there are constants 
$ c_d, c_s\in\mathbb{R} $ such that 
$\text{del}(W,n)\leq \log_2W+ 3\log_2\log_2n +c_d $ and $\text{s}(n)\leq c_sn\log_2\log_2n$.

Note that the fanout of the constructed circuit, in particular of the nodes $ y_1\circ \dots \circ y_{i} $ for $ 1\leq i\leq l-1 $, is in $ \mathcal{O}(\sqrt{n}) $ due to Step 5 of the construction. 
Rautenbach, Szegedy and Werber \cite{RSW} also developed a modified circuit with a fanout of 2 and a delay of at most $ \log_2W+\log_2n+7\log_2\log_2n+\mathcal{O}(1) $. Since our main objective is the delay-optimization combined with the size reduction of adders, we only focus on the delay-optimizing construction given above and disregard the high fanout.

\begin{figure}[htb]
\centering
   \begin{tikzpicture}[scale=0.75]
      \node[outer sep=0pt] (i1) at (0, 4){$x_0$};
      \node[outer sep=0pt] (i2) at (1, 4){$x_1$};
      \node[outer sep=0pt] (i3) at (2, 4){$x_2$};
      \node[outer sep=0pt] (i4) at (3, 4){$x_3$};

      \node[fill=blue, and gate US, draw, logic gate inputs=nn, rotate=270, thick, scale=0.8] at (1,3) (gates1){};
      \node[and gate US, draw, rotate=270, thick, scale=1] at (1,3) {};
      \node[fill=blue, and gate US, draw, logic gate inputs=nn, rotate=270, thick, scale=0.8] at (3,3) (gates2){};
      \node[and gate US, draw, rotate=270, thick, scale=1] at (3,3) {};
      \node[fill=red, and gate US, draw, logic gate inputs=nn, rotate=270, thick, scale=0.6] at (2,2) (circ1){};
      \node[fill=yellow, and gate US, draw, logic gate inputs=nn, rotate=270, thick, scale=0.6] at (3,2) (circ2){};
      \node at (0,0) (o1){};
      \node at (1,0) (o2){};
      \node at (2,0) (o3){};
      \node at (3,0) (o4){};

      \node[below=3.3cm, align=flush center] at (i3){\parbox{4.5cm}{\textbf{(a)} Circuit $ S(t_0,\dots, t_3) $.}};

      \draw[thick] (i1)--(gates1.input 2);
      \draw[thick] (i2)--(gates1.input 1);
      \draw[thick] (i3)--(gates2.input 2);
      \draw[thick] (i4)--(gates2.input 1);
      \draw[thick] (gates1.output)--(circ1.input 2);
      \draw[thick] (i3)--(circ1.input 1);
      \draw[thick] (gates1.output)--(circ2.input 2);
      \draw[thick] (gates2.output)--(circ2.input 1);
      \draw[->] (i1)--(o1);
      \draw[->] (gates1.output)--(o2);
      \draw[->] (circ1.output)--(o3);
      \draw[->] (circ2.output)--(o4);
   \end{tikzpicture} \hspace*{-0.5cm}
   \begin{tikzpicture}[scale=0.75]
      \node[outer sep=0pt] (i1) at (0, 4){$x_0$};
      \node[outer sep=0pt] (i2) at (1, 4){$x_1$};
      \node[outer sep=0pt] (i3) at (2, 4){$x_2$};
      \node[outer sep=0pt] (i4) at (3, 4){$x_3$};
      \node[outer sep=0pt] (i5) at (4, 4){$x_4$};
      \node[outer sep=0pt] (i6) at (5, 4){$x_5$};
      
      \node[fill=blue, and gate US, draw, logic gate inputs=nn, rotate=270, thick, scale=0.8] at (1,3) (gates1){};
      \node[and gate US, draw, rotate=270, thick, scale=1] at (1,3) {};
      \node[fill=blue, and gate US, draw, logic gate inputs=nn, rotate=270, thick, scale=0.8] at (3,3) (gates2){};
      \node[and gate US, draw, rotate=270, thick, scale=1] at (3,3) {};
      \node[fill=blue, and gate US, draw, logic gate inputs=nn, rotate=270, thick, scale=0.8] at (5,3) (gates3){};
      \node[and gate US, draw, rotate=270, thick, scale=1] at (5,3) {};
      \node[fill=red, and gate US, draw, logic gate inputs=nn, rotate=270, thick, scale=0.6] at (2,2) (circ1){};
      \node[fill=green, and gate US, draw, logic gate inputs=nn, rotate=270, thick, scale=0.6] at (3,2) (circ2){};
      \node[fill=red, and gate US, draw, logic gate inputs=nn, rotate=270, thick, scale=0.6] at (4,1) (circ3){};
      \node[fill=yellow, and gate US, draw, logic gate inputs=nn, rotate=270, thick, scale=0.6] at (5,1) (circ4){};
      \node at (0,0) (o1){};
      \node at (1,0) (o2){};
      \node at (2,0) (o3){};
      \node at (3,0) (o4){};
      \node at (4,0) (o5){};
      \node at (5,0) (o6){};
      
      \node[below=3.3cm, align=flush center] at (i4){\parbox{3.8cm}{\textbf{(b)} Circuit $ S(t_0,\dots, t_5) $.}};
      
      \draw[thick] (i1)--(gates1.input 2);
      \draw[thick] (i2)--(gates1.input 1);
      \draw[thick] (i3)--(gates2.input 2);
      \draw[thick] (i4)--(gates2.input 1);
      \draw[thick] (i5)--(gates3.input 2);
      \draw[thick] (i6)--(gates3.input 1);
      \draw[thick] (gates1.output)--(circ1.input 2);
      \draw[thick] (i3)--(circ1.input 1);
      \draw[thick] (gates1.output)--(circ2.input 2);
      \draw[thick] (gates2.output)--(circ2.input 1);
      \draw[thick] (circ2.output)--(circ3.input 2);
      \draw[thick] (i5)--(circ3.input 1);
      \draw[thick] (circ2.output)--(circ4.input 2);
      \draw[thick] (gates3.output)--(circ4.input 1);
      \draw[->] (i1)--(o1);
      \draw[->] (gates1.output)--(o2);
      \draw[->] (circ1.output)--(o3);
      \draw[->] (circ2.output)--(o4);
      \draw[->] (circ3.output)--(o5);
      \draw[->] (circ4.output)--(o6);
   \end{tikzpicture}\hspace*{0.5cm}
   \begin{tikzpicture}[scale=0.75]
      \node[outer sep=0pt] (i1) at (0, 4){$x_0$};
      \node[outer sep=0pt] (i2) at (1, 4){$x_1$};
      \node[outer sep=0pt] (i3) at (2, 4){$x_2$};
      \node[outer sep=0pt] (i4) at (3, 4){$x_3$};
      \node[outer sep=0pt] (i5) at (4, 4){$x_4$};
      \node[outer sep=0pt] (i6) at (5, 4){$x_5$};
      \node[outer sep=0pt] (i7) at (6, 4){$x_6$};
      \node[outer sep=0pt] (i8) at (7, 4){$x_7$};
      \node[outer sep=0pt] (i9) at (8, 4){$x_8$};

      \node[fill=blue, and gate US, draw, logic gate inputs=nnn, rotate=270, thick, scale=0.6] at (2,3) (gates1){};
      \node[and gate US, draw, rotate=270, thick, scale=1] at (2,3) {};
      \node[fill=blue, and gate US, draw, logic gate inputs=nnn, rotate=270, thick, scale=0.6] at (5,3) (gates2){};
      \node[and gate US, draw, rotate=270, thick, scale=1] at (5,3) {};
      \node[fill=blue, and gate US, draw, logic gate inputs=nnn, rotate=270, thick, scale=0.6] at (8,3) (gates3){};
      \node[and gate US, draw, rotate=270, thick, scale=1] at (8,3) {};
      \node[fill=orange, and gate US, draw, logic gate inputs=nn, rotate=270, thick, scale=0.6] at (1,3) (circ1){};
      \node[fill=orange, and gate US, draw, logic gate inputs=nn, rotate=270, thick, scale=0.6] at (4,3) (circ2){};
      \node[fill=orange, and gate US, draw, logic gate inputs=nn, rotate=270, thick, scale=0.6] at (7,3) (circ3){};
      \node[fill=red, and gate US, draw, logic gate inputs=nn, rotate=270, thick, scale=0.6] at (3,2) (circ4){};
      \node[fill=red, and gate US, draw, logic gate inputs=nn, rotate=270, thick, scale=0.6] at (4,2) (circ5){};
      \node[fill=green, and gate US, draw, logic gate inputs=nn, rotate=270, thick, scale=0.6] at (5,2) (circ6){};
      \node[fill=red, and gate US, draw, logic gate inputs=nn, rotate=270, thick, scale=0.6] at (6,1) (circ7){};
      \node[fill=red, and gate US, draw, logic gate inputs=nn, rotate=270, thick, scale=0.6] at (7,1) (circ8){};
      \node[fill=yellow, and gate US, draw, logic gate inputs=nn, rotate=270, thick, scale=0.6] at (8,1) (circ9){};
      \node at (0,0) (o1){};
      \node at (1,0) (o2){};
      \node at (2,0) (o3){};
      \node at (3,0) (o4){};
      \node at (4,0) (o5){};
      \node at (5,0) (o6){};
      \node at (6,0) (o7){};
      \node at (7,0) (o8){};
      \node at (8,0) (o9){};

      \node[below=3.3cm, align=flush center] at (i5){\parbox{3.8cm}{\textbf{(c)} Circuit $ S(t_0,\dots, t_8) $.}};

      \draw[thick] (i1)--(gates1.input 3);
      \draw[thick] (i2)--(gates1.input 2);
      \draw[thick] (i3)--(gates1.input 1);
      \draw[thick] (i4)--(gates2.input 3);
      \draw[thick] (i5)--(gates2.input 2);
      \draw[thick] (i6)--(gates2.input 1);
      \draw[thick] (i7)--(gates3.input 3);
      \draw[thick] (i8)--(gates3.input 2);
      \draw[thick] (i9)--(gates3.input 1);
      \draw[thick] (i1)--(circ1.input 2);
      \draw[thick] (i2)--(circ1.input 1);
      \draw[thick] (i4)--(circ2.input 2);
      \draw[thick] (i5)--(circ2.input 1);
      \draw[thick] (i7)--(circ3.input 2);
      \draw[thick] (i8)--(circ3.input 1);
      \draw[thick] (gates1.output)--(circ4.input 2);
      \draw[thick] (i4)--(circ4.input 1);
      \draw[thick] (gates1.output)--(circ5.input 2);
      \draw[thick] (circ2)--(circ5.input 1);
      \draw[thick] (gates1.output)--(circ6.input 2);
      \draw[thick] (gates2.output)--(circ6.input 1);
      \draw[thick] (circ6.output)--(circ7.input 2);
      \draw[thick] (i7)--(circ7.input 1);
      \draw[thick] (circ6.output)--(circ8.input 2);
      \draw[thick] (circ3)--(circ8.input 1);
      \draw[thick] (circ6.output)--(circ9.input 2);
      \draw[thick] (gates3.output)--(circ9.input 1);
      \draw (i1)--(o1);
      \draw[->] (circ1.output)--(o2);
      \draw[->] (gates1.output)--(o3);
      \draw[->] (circ4.output)--(o4);
      \draw[->] (circ5.output)--(o5);
      \draw[->] (circ6.output)--(o6);
      \draw[->] (circ7.output)--(o7);
      \draw[->] (circ8.output)--(o8);
      \draw[->] (circ9.output)--(o9);
   \end{tikzpicture}
\caption[Prefix circuit by Rautenbach, Szegedy and Werber on 4, 6 and 9 inputs.]{$ S(t_0,\dots,t_{n-1}) $ for $ n=4,6 $ and $ 9 $.}
\label{figAP}
\end{figure}
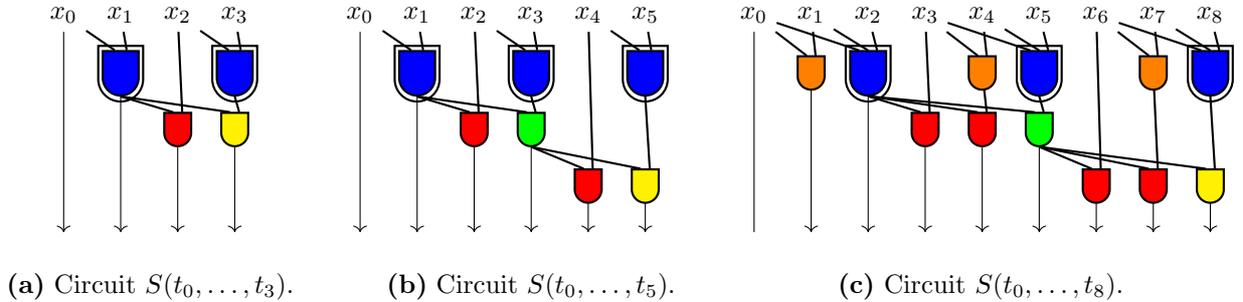

In order to derive concrete bounds for the constants $ c_d $ and $ c_s $,
we first need to analyse the algorithm on small instances.
Figure~\ref{figAP} shows the prefix circuits $ S(t_0,\dots,t_{n-1}) $ computed via the construction by Rautenbach, Szegedy and Werber for $ n=4,6 $ and $ 9 $. The blue symbols represent the circuits $ C_i $ computed in Step 2 of the construction. In the cases $ n=4 $ and $ n=6 $ they just consist of one $ \circ $-gate. In the case $ n=9 $ they are circuits of size and depth 2. The orange symbols represent the gates constructed during the recursive calls at Step 3 of the algorithm. The green symbols are the gates introduced during Step 4 and the red symbols are the gates coming from Step 5. Lastly, the yellow symbols represent the one gate added in Step 6 of the construction.

	\begin{lemma}
		\label{lem3-9}
		For $ n\in\{3,\dots,9\} $, inputs $ x=(x_0,\dots, x_{n-1}) $ with arrival times $ t_0,\dots, t_{n-1} $ and weight $ W=\sum_{i=0}^{n-1}2^{t_i} $, we have 
		\begin{align*}
      \mathrm{del}(W,n)&\leq \log_2W+3\log_2\log_2n+0.007
			\intertext{and}
			\mathrm{s}(n)&\leq 3.114n\log_2\log_2n.
		\end{align*}
	\end{lemma}

	\begin{proof}
		First, assume that $ n\in \{3,4\} $. In Figure~\ref{figAP} we see that $ S(t_0,\dots,t_3) $ has depth 2 and size 4. Since $ S(t_0,\dots,t_2) $ is smaller or equal to $ S(t_0,\dots,t_3) $ in depth and size, we have 
		\begin{align*} 
			\text{del}(W,n)&\leq\log_2W+2\leq \log_2W+3\log_2\log_23 + 0.007
			\leq \log_2W+3\log_2\log_2n + 0.007
			\intertext{and}
			\text{s}(n)&\leq4\leq 3.114*3\log_2\log_23 \leq 3.114n\log_2\log_2n.
		\end{align*}
	
		Next, assume that $ n\in\{5,6\} $. $ S(t_0,\dots,t_5) $ has depth 3 and size 7 as seen in Figure~\ref{figAP}. As before, $ S(t_0,\dots,t_4) $ is smaller than $ S(t_0,\dots,t_5) $ in depth and size due to the construction of these circuits. We get 
		\begin{align*}
			\text{del}(W,n)&\leq\log_2W +3\leq \log_2W+3\log_2\log_25 + 0.007
			\leq \log_2W+3\log_2\log_2n + 0.007
			\intertext{and}
			\text{s}(n)&\leq7\leq 3.114*5\log_2\log_25 \leq 3.114n\log_2\log_2n.
		\end{align*}
	
Finally, assume that $ n\in\{7,8,9\} $. Figure~\ref{figAP}
implies a depth of 4 and a size of 15 for $S(t_0,\dots, t_8)$.
Again, the circuits for $ n=7 $ and $ n=8 $ are by
construction smaller  than $ S(t_0,\dots, t_8) $ in depth and size.
Therefore, we have 
   \begin{align*}
	\text{del}(W,n)&\leq\log_2W +4\leq \log_2W+3\log_2\log_27 + 0.007
	\leq \log_2W+3\log_2\log_2n + 0.007
	\intertext{and}
	\text{s}(n)&\leq15\leq 3.114*7\log_2\log_27 \leq 3.114n\log_2\log_2n.\tag*{\qedhere}
   \end{align*}
\end{proof}

Before we formulate and prove the main result of this section, we need a numerical lemma.
	
\begin{lemma}
	\label{lemnum0}
	Assume $ n\in\mathbb{N} $ with $ n\geq 2 $. Then \[6.228\log_2\log_2n-1.114\sqrt{n}-8.228\leq 0.\]
\end{lemma}
	
\begin{proof}
Consider the function $f:\mathbb{R}_{\geq2}\to\mathbb{R}$ with
\[
   f(x) = 6.228\log_2\log_2x-1.114\sqrt{x}-8.228\,.
\]
Its derivative is 
\[
   f'(x) = \frac{6.228}{(\ln(2))^2 x \log_2(x)} - \frac{0.557}{\sqrt{x}},
\]
which is positive for $x \leq 25$ (as 
$\frac{6.228}{(\ln(2))^2 \sqrt{25} \log_2(25)} - 0.557 > 0$)
and negative for $x \geq 26$ (as
$\frac{6.228}{(\ln(2))^2 \sqrt{26} \log_2(26)} - 0.557 < 0$).
Since $f(25) < 0$ and $f(26) < 0$, this proves the lemma.
\end{proof}
\linebox{
\begin{theorem} \label{andprecirc}
Let $ n\in \mathbb{N} $ with $ n\geq 3 $. The prefix problem for an associative operator 
$ \circ:\{0,1\}\times\{0,1\}\to\{0,1\} $ on inputs $ x_0,\dots, x_{n-1} $ with arrival times 
$ a(x_0),\dots, a(x_{n-1})\in \mathbb{N} $ and weight $ W=\sum_{i=0}^{n-1}2^{a(x_i)} $ 
can be solved in polynomial time by a circuit $ S_n $ over the basis $ \{\circ\} $ with delay
\begin{align*}
   \mathrm{del}(S_n)\leq &\log_2W+3\log_2\log_2n + 0.007 \intertext{and size}
   \mathrm{s}(S_n)\leq &3.114n\log_2\log_2n\,.
\end{align*} 
\end{theorem}
}
	\begin{proof}
		We use the construction by Rautenbach, Szegedy and Werber described above. The polynomial runtime is already implied by Theorem~\ref{andprecirc0}. We prove the delay and size bounds by induction on $ n $.
		
		For $ n\in\{3,\dots,9\} $ the statement follows due to Lemma~\ref{lem3-9}.
		
		Assume now that $ n\geq 10 $ and the statement holds true for all $ n'<n $. In this case, we use the recursion formulas from Lemma~\ref{lemrec}. We note that $ \lceil\sqrt{n}\rceil-1\geq 3 $ for $ n\geq 10 $. Therefore, we can apply the induction hypothesis in the following calculations. 
For the delay we get
		\setlength{\mywidth}{\widthof{$\stackrel{Lem. \ref{lemrec}..}{\leq}$}}
		\begin{align*}
			\text{del}(W,n)\klgleich[Lem. \ref{lemrec}]& \text{del}(4W,\lceil\sqrt{n}\rceil-1)+1
			\stackrel{(IH)}{=} \log_2(4W)+3\log_2\log_2(\lceil\sqrt{n}\rceil-1) + 1.007\\
			\klgleich& \log_2(W)+3\log_2\log_2(\sqrt{n}) + 3.007 
			\, = \, \log_2(W)+3\log_2\log_2n + 0.007.
		\end{align*}
For the size we calculate
{\allowdisplaybreaks
\begin{align*}
			\text{s}(n)\klgleich[Lem. \ref{lemrec}]& (\lceil\sqrt{n}\rceil+1)\hspace{2pt}\text{s}(\lceil\sqrt{n}\rceil-1) + 2(n-\lceil\sqrt{n}\rceil)\\
			\klgleich[(IH)]&3.114(\lceil\sqrt{n}\rceil+1)(\lceil\sqrt{n}\rceil-1)\log_2\log_2(\lceil\sqrt{n}\rceil-1) + 2(n-\lceil\sqrt{n}\rceil)\\
			\klgleich& 3.114(\sqrt{n}+2)\sqrt{n} \log_2\log_2\sqrt{n} + 2(n-\sqrt{n})\\
			\gleich& 3.114(n+2\sqrt{n}) (\log_2\log_2n-1) + 2(n-\sqrt{n})\\
			\gleich&3.114n\log_2\log_2n + 6.228\sqrt{n}\log_2\log_2n-1.114n-8.228\sqrt{n}\\
			\gleich&3.114n\log_2\log_2n + \sqrt{n}(6.228\log_2\log_2n - 1.114\sqrt{n}-8.228)\\
			\klgleich[Lem. \ref{lemnum0}]&3.114n\log_2\log_2n.\tag*{\qedhere}
\end{align*}
}
	\end{proof}

\section{Delay-Optimizing Adder Circuit}
\label{deloptadder}
	
In this section, we will construct an adder circuit whose delay is close to 
the best known delay bound from Theorem~\ref{andorcirc},
while maintaining a sub-quadratic size. In Section~\ref{algorithm}, we will describe our 
general algorithm, and in Section~\ref{adderconstruction}, we will apply it to compute an 
adder with a delay of at most $\log_2W+3\log_2\log_2n + \text{const}$ and a size of 
$\mathcal{O}(n\log_2^2n$) on $ n $ input pairs with weight $ W $.
	
\subsection{Algorithm} \label{algorithm}
Our idea is to adapt the depth-optimizing construction that was used in 
\cite{BS24} to prove Theorem~\ref{daddercirc}
to the case of delay-optimization. Hence, we want to use the alternating split 
(Theorem~\ref{thmaltsplit}) to recursively construct an adder with the help of a 
given family of \textsc{And-Or} path circuits and a given family of 
\textsc{And}-prefix circuits. 
In contrast to the setting of Theorem~\ref{daddercirc},
we have to work with different arrival times, so 
we have to be careful with the choice of our sub-instances and cannot apply the 
approach from \cite{BS24} directly.
	
Assume, a partition of an instance $ p_0,g_0,$ $\dots, p_{n-1},g_{n-1} $ with arrival times $ a(p_i),a(g_i)\in\mathbb{N} $ and weight $W$ 
into two consecutive parts $ P_l $ and $ P_r $ is given. 
Assume that we constructed an adder as seen in Figure~\ref{figdaddercirc} with recursively computed adders $ A_r $ (respectively, $ A_l $) on $ P_r $ (respectively, $ P_l $), a given \text{And-Or} path circuit $ AOP $ on $ P_r $, and a given \textsc{And}-prefix circuit $ S $ on the propagate signals of $ P_l $. This adder would then have a delay of at most 
	\begin{equation}
		\max\{\text{del}(A_r), \text{del}(A_l)+1, \text{del}(AOP)+2, \text{del}(S)+2\} \label{eqdel}
	\end{equation} 
as seen in Figure~\ref{figdaddercirc}. In order to prove a delay bound inductively, we have to make sure that the quantity \eqref{eqdel} is smaller than our claimed delay bound. Therefore, the delay $ \text{del}(A_l) $ of the recursively computed adder on $ P_l $ needs to be at least by 1 smaller than our claimed total delay bound. 
This is the case if the weight $ W_{P_l} $ of the inputs in $ P_l $ is at most half the weight of the whole instance 
(since then $\log_2  W_{P_l} \leq \log_2 \left(\frac{W}{2}\right) = \log_2 W - \frac{1}{2}$).
For the size computation we also have to make sure that our sub-instances are not to large.
	
	Combining these two considerations, our adder construction, given by Algorithm \ref{alg1}, works as follows. We define the sub-instance $ P_l $ to contain the leftmost $ k_l $ input pairs, where \[k_l = \min\Bigg\{\max\bigg\{m\in\mathbb{N}:\sum_{i=1}^{m}\big(2^{a(p_{n-i})} + 2^{a(g_{n-i})}\big)\leq \frac{W}{2}\bigg\}, \Big\lfloor \frac{n}{2}\Big\rfloor \Bigg\}.\]
	
	Then we make a case distinction. If $ k_l=\lfloor\frac{n}{2}\rfloor $, we can define $ P_r $ to contain the rest of the instance and construct our adder just like described above and depicted in Figure~\ref{figdaddercirc}.
	
\begin{figure}[htb]
		\centering
		\begin{tikzpicture}[scale=0.5]
			\node[scale=0.7] at (0,0) (g11) {$ g_{11} $};
			\node[scale=0.7] at (1,0) (p11) {$ p_{11} $};
			\node[scale=0.7] at (2,0) (g10) {$ g_{10} $};
			\node[scale=0.7] at (3,0) (p10) {$ p_{10} $};
			\node[scale=0.7] at (4,0) (g9) {$ g_9 $};
			\node[scale=0.7] at (5,0) (p9) {$ p_9 $};
			\node[scale=0.7] at (6,0) (g8) {$ g_8 $};
			\node[scale=0.7] at (7,0) (p8) {$ p_8 $};
			\node[scale=0.7] at (8,0) (g7) {$ g_7 $};
			\node[scale=0.7] at (9,0) (p7) {$ p_7 $};
			\node[scale=0.7] at (10,0) (g6) {$ g_6 $};
			\node[scale=0.7] at (11,0) (p6) {$ p_6 $};
			\node[scale=0.7] at (12,0) (g5) {$ g_5 $};
			\node[scale=0.7] at (13,0) (p5) {$ p_5 $};
			\node[scale=0.7] at (14,0) (g4) {$ g_4 $};
			\node[scale=0.7] at (15,0) (p4) {$ p_4 $};
			\node[scale=0.7] at (16,0) (g3) {$ g_3 $};
			\node[scale=0.7] at (17,0) (p3) {$ p_3 $};
			\node[scale=0.7] at (18,0) (g2) {$ g_2 $};
			\node[scale=0.7] at (19,0) (p2) {$ p_2 $};
			\node[scale=0.7] at (20,0) (g1) {$ g_1 $};
			\node[scale=0.7] at (21,0) (p1) {$ p_1 $};
			\node[scale=0.7] at (22,0) (g0) {$ g_0 $};
			\node[scale=0.7] at (23,0) (p0) {$ p_0 $};
			
			\node[scale=0.7] at (2.5,1) {$ P_l $};
			\node[scale=0.7] at (9.5,1) {$ P_m $};
			\node[scale=0.7] at (17.5,1) {$ P_r $};
			
			\draw[decorate,decoration={brace,raise=0.5ex}] (g11.north west) -- (p9.north east);
			\draw[decorate,decoration={brace,raise=0.5ex}] (g7.north west) -- (p6.north east);
			\draw[decorate,decoration={brace,raise=0.5ex}] (g5.north west) -- (p0.north east);
			
			\draw[fill=red] (-0.4,-1) rectangle (5.4,-2);
			\draw[fill=red] (7.6,-1) rectangle (11.4,-2);
			\draw[fill=red] (11.6,-1) rectangle (23.4,-2);
			
			\draw[fill=yellow] (-0.4,-3) rectangle (5.4,-4);
			\draw[fill=yellow] (7.6,-3) rectangle (11.4,-4);
			
			\draw[fill=green] (7.6,-5) rectangle (11.4,-6);
			\draw[fill=green] (11.6,-5) rectangle (23.4,-6);
			
			\node[scale=0.7] at (2.5,-1.5) {$ A_{k_l}^l $};
			\node[scale=0.7] at (9.5,-1.5) {$ A_{k_m}^m $};
			\node[scale=0.7] at (17.5,-1.5) {$ A_{k_r}^r $};
			\node[scale=0.7] at (2.5,-3.5) {$ S_{k_l}^l $};
			\node[scale=0.7] at (9.5,-3.5) {$ S_{k_m}^m $};
			\node[scale=0.7] at (9.5,-5.5) {$ AOP_{k_m}^m $};
			\node[scale=0.7] at (17.5,-5.5) {$ AOP_{k_r}^r $};
			
			\draw[->]  (0,-2)--(0,-2.5);
			\draw[->]  (2,-2)--(2,-2.5);
			\draw[->]  (4,-2)--(4,-2.5);
			
			\draw[->]  (8,-2)--(8,-2.5);
			\draw[->]  (10,-2)--(10,-2.5);
			
			\draw[->]  (12,-2)--(12,-2.5);
			\draw[->]  (14,-2)--(14,-2.5);
			\draw[->]  (16,-2)--(16,-2.5);
			\draw[->]  (18,-2)--(18,-2.5);
			\draw[->]  (20,-2)--(20,-2.5);
			\draw[->]  (22,-2)--(22,-2.5);
			
			\draw[->]  (1,-4)--(1,-4.5);
			\draw[->]  (3,-4)--(3,-4.5);
			\draw[->]  (5,-4)--(5,-4.5);
			
			\draw[->]  (9,-4)--(9,-4.5);
			\draw[->]  (11,-4)--(11,-4.5);
			
			\draw[->]  (8,-6)--(8,-6.5);
			\draw[->]  (12,-6)--(12,-6.5);
		\end{tikzpicture}
		\caption[Visualization of the computed sub-circuits during Algorithm~\ref{alg1}.]{Visualization of the computed sub-circuits during Algorithm~\ref{alg1} for $ n=12 $ and $ k_l=3 $.}
		\label{figpartition}
	\end{figure}
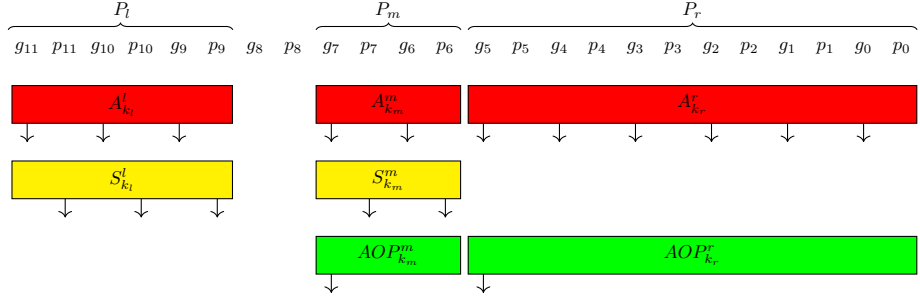

	However, if $ k_l<\lfloor\frac{n}{2}\rfloor $ (i.e., if more than half of the total weight lies in the left half of the instance), we divide the instance into four parts as seen in Figure~\ref{figpartition}. We define $ P_r $ to contain the rightmost $ k_r=\lfloor\frac{n}{2}\rfloor $ input pairs and $ P_m $ to contain the remaining $ k_m $ input pairs, except for the pair to the right of $ P_l $ which we will consider separately. Hence, $ n=k_l+k_m+k_r+1 $. Note that $ P_l $ or $ P_m $ may contain no input pairs if $ k_l=0 $, or if $ k_l=\frac{n}{2}-1 $ and $ n $ is even, respectively. By the definition of $ k_l $ we know that $ P_l $ and the pair to the right of $ P_l $ combine more that half of the total weight of the instance. Consequently, both $ P_r $ and $ P_m $ contain less weight than $ \frac{W}{2} $. Moreover, each sub-instance contains at most $ \lfloor\frac{n}{2}\rfloor $ input pairs.
	
Then, we compute adders $ A_{k_l}^l $, $ A_{k_m}^m $ and $ A_{k_r}^r $ on $ P_l $, $ P_m $ and $ P_r $, \textsc{And}-prefix circuits $ S_{k_l} $ and $ S_{k_m} $ on the propagate signals of $ P_l $ and $ P_m $ as well as \textsc{And-Or} path circuits $ AOP_{k_m} $ and $ AOP_{k_r} $ on $ P_m $ and $ P_r $, as seen in Figure~\ref{figpartition}. 
	
	\begin{algorithm}[htb]
		\caption{Delay-optimizing adder construction framework}
		\label{alg1}
		\begin{algorithmic}[1]
			\Require $ n\in \mathbb{N}, n\geq 3 $, $ n $ input pairs $ p_0, g_0, \dots, p_{n-1},g_{n-1} $ with arrival times $ a(p_{i}), a(g_{i})\in \mathbb{N} $ for $ i\in \{0,\dots, n-1\} $, adder circuits $ (A_k)_{k<n} $, \textsc{And-Or}-path circuits $ (AOP_k)_{k<n} $, \textsc{And}-prefix circuits $ (S_k)_{k<n} $.
			
			\Ensure An adder circuit $ C_n $ on $ p_0, g_0, \dots, p_{n-1},g_{n-1} $.
			
			\State $ k_l \gets \min\{\max\{m\in\mathbb{N}:\sum_{i=1}^{m}\big(2^{a(p_{n-i})} + 2^{a(g_{n-i})}\big)\leq \frac{W}{2}\}, \lfloor \frac{n}{2}\rfloor \} $.
			\If{$ k_l = \lfloor \frac{n}{2}\rfloor $}
			\State $ k_r \gets n-k_l $.
			\State $ P_r \gets (p_0,g_0,\dots ,p_{k_r-1},g_{k_r-1}) $.
			\State $ P_l \gets (p_{k_r},g_{k_r},\dots ,p_{n-1},g_{n-1}) $.
			\State Compute adder circuits $ A_{k_r}^r, A_{k_l}^l $ on $ P_r, P_l $, respectively.
			\State Compute \textsc{And-Or}-path circuit $ AOP_{k_r}^r $ on $ P_r $.
			\State Compute \textsc{And}-prefix circuit $ S_{k_l}^l $ on inputs $ p_i $ of $ P_{k_l} $ ($ p_{k_r}, \dots, p_{n-1} $).
			\For{$ i=1 $ to $ k_r $}
			\State $ \text{out}_i(C_n):=\text{out}_i(A_{k_r}^r) $.
			\EndFor
			\For{$ i=1 $ to $ k_l $}
			\State $ \text{out}_{k_r+i}(C_n):=\text{out}_i(A_{k_l}^l)\lor \big(\text{out}_i(S_{k_l}^l)\land \text{out}(AOP_{k_r}^r)\big) $.
			\EndFor 
			\State \Return $ C_n $.
			\Else
			\State $ k_r \gets \lfloor \frac{n}{2}\rfloor $.
			\State $ k_m \gets n-k_l-k_r-1 = \lceil \frac{n}{2}\rceil -k_l-1 $.
			\State $ P_r \gets (p_0,g_0,\dots ,p_{k_r-1},g_{k_r-1}) $.
			\State $ P_m \gets (p_{k_r},g_{k_r},\dots ,p_{k_r+k_m-1},g_{k_r+k_m-1}) $.
			\State $ P_l \gets (p_{k_r+k_m+1},g_{k_r+k_m+1},\dots ,p_{n-1},g_{n-1}) $.
			\State Compute adder circuits $ A_{k_r}^r, A_{k_m}^m, A_{k_l}^l $ on $ P_r, P_m, P_l $, respectively.
			\State Compute \textsc{And-Or}-path circuits $ AOP_{k_m}^m, AOP_{k_r}^r $ on $ P_m,P_r $, respectively.
			\State Compute \textsc{And}-prefix circuits $ S_{k_l}^l, S_{k_m}^m $ on inputs $ p_i $ of $ P_{k_l}, P_{k_m} $, respectively.
			\For{$ i=1 $ to $ k_r $}
			\State $ \text{out}_i(C_n):=\text{out}_i(A_{k_r}^r) $.
			\EndFor
			\For{$ i=1 $ to $ k_m $}
			\State $ \text{out}_{k_r+i}(C_n):=\text{out}_i(A_{k_m}^m)\lor 	\big(\text{out}_i(S_{k_m}^m)\land \text{out}(AOP_{k_r}^r)\big) $.
			\EndFor 
			\State $ \text{out}_{k_r+k_m+1}(C_n):=g_{k_r+k_m}\lor \bigg(p_{k_r+k_m} \land \Big( \text{out}(AOP_{k_m}^m)\lor \big(\text{out}_{k_m}(S_{k_m}^m)\land \text{out}(AOP_{k_r}^r)\big)\Big)\bigg) $.
			\For{$ i=1 $ to $ k_l $}
			\State {\small $ \text{out}_{k_r+k_m+1+i}(C_n):=$\\ \hspace*{2.8cm}$\text{out}_i(A_{k_l}^l)\lor \Bigg(\text{out}_i(S_{k_l}^l)\land \bigg(g_{k_r+k_m}\lor \Big(p_{k_r+k_m} \land \big(\text{out}(AOP_{k_m}^m)\lor (\text{out}_{k_m}(S_{k_m}^m)\land \text{out}(AOP_{k_r}^r))\big)\Big)\bigg)\Bigg) $.}
			\EndFor
			\State \Return $ C_n $.
			\EndIf
		\end{algorithmic}
	\end{algorithm}
	
	The carry bits are then computed in four steps:
	\begin{itemize}
		\item The first $ k_r $ carry bits are given directly by $ A_{k_r}^r $.
		\item The next $ k_m $ carry bits are given by the expression \[ \text{out}_i(A_{k_m}^m)\lor \big(\text{out}_i(S_{k_m}^m)\land \text{out}(AOP_{k_r}^r)\big) \] for $ 1\leq i \leq k_m $ via the alternating split (Theorem~\ref{thmaltsplit}).
		\item The $ (k_r+k_m+1) $th carry bit is computed via \[ g_{k_r+k_m}\lor \bigg(p_{k_r+k_m} \land \Big( \text{out}(AOP_{k_m}^m)\lor \big(\text{out}_{k_m}(S_{k_m}^m)\land \text{out}(AOP_{k_r}^r)\big)\Big)\bigg).\]		
		\item The last $ k_l $ carry bits are given by 
		\begin{align*}
			\text{out}_i(A_{k_l}^l)\lor \Bigg(\text{out}_i(S_{k_l}^l)\land \bigg(&g_{k_r+k_m}\lor \Big(p_{k_r+k_m} \land  \big(\text{out}(AOP_{k_m}^m)\lor  (\text{out}_{k_m}(S_{k_m}^m)\land \text{out}(AOP_{k_r}^r))\big)\Big)\bigg)\Bigg)
		\end{align*} 
		for $ 1\leq i \leq k_m $ again via the alternating split.
	\end{itemize}

	Note that the formula \[ g_{k_r+k_m}\lor \bigg(p_{k_r+k_m} \land \Big( \text{out}(AOP_{k_m}^m)\lor \big(\text{out}_{k_m}(S_{k_m}^m)\land \text{out}(AOP_{k_r}^r)\big)\Big)\bigg) \] indeed computes the $ (k_r+k_m+1) $th carry bit, i.e., the function $ g^*((g_{k_r+k_m},$ $p_{k_r+k_m},\dots, g_1,p_1,g_0)) $, because of the definition of $ g^* $ and the alternating split (Theorem~\ref{thmaltsplit}).

	The next lemma bounds the delay and the size of the adder constructed in Algorithm \ref{alg1} in terms of the delay and size of the computed sub-circuits.

	\begin{lemma}
		\label{lemalg1}
		Let $ n\in \mathbb{N} $ with $ n\geq 3 $, input pairs $ p_0, g_0, \dots, p_{n-1},g_{n-1} $ with arrival times $ a(p_{i}), a(g_{i}) $ for $ i\in \{0,\dots, n-1\} $, adder circuits $ (A_k)_{k<n} $, \textsc{And-Or}-path circuits $ (AOP_k)_{k<n} $ and \textsc{And}-prefix circuits $ (S_k)_{k<n} $ be given and let \[k_l=\min\Bigg\{\max\bigg\{m\in\mathbb{N}:\sum_{i=1}^{m}\big(2^{a(p_{n-i})} + 2^{a(g_{n-i})}\big)\leq \frac{W}{2}\bigg\}, \Big\lfloor \frac{n}{2}\Big\rfloor \Bigg\}.\] Algorithm $ \ref{alg1} $ computes an adder circuit $ C_n $ with 
		\begin{itemize}
			\item $ \begin{aligned}[t]
					\mathrm{del}(C_n)&\leq \max\{\mathrm{del}(A_{k_r}^r), \mathrm{del}(A_{k_l}^l)+1, \mathrm{del}(AOP_{k_r}^r)+2, \mathrm{del}(S_{k_l}^l)+2\} \\
					\alignedintertext{and} 
					\mathrm{s}(C_n)&\leq \mathrm{s}(A_{k_r}^r) + \mathrm{s}(A_{k_l}^l) + \mathrm{s}(AOP_{k_r}^r) + \mathrm{s}(S_{k_l}^l) + 2k_l,
				\end{aligned} $ \vspace{0pt}\\
				if $ k_l= \lfloor \frac{n}{2} \rfloor $.
			\item $ \begin{aligned}[t]
				\mathrm{del}(C_n)\leq& \max\{\mathrm{del}(A_{k_r}^r), \mathrm{del}(A_{k_m}^m) + 1,\mathrm{del}(A_{k_l}^l)+1 , a(g_{k_r+k_m}) + 3,\\
				&a(p_{k_r+k_m}) + 4,\, \mathrm{del}(AOP_{k_m}^m)+5,\, \mathrm{del}(AOP_{k_r}^r)+6,\, \mathrm{del}(S_{k_l}^l)+2, \mathrm{del}(S_{k_m}^m)+6\} \\
				\alignedintertext{and} 
				\mathrm{s}(C_n)\leq& \mathrm{s}(A_{k_r}^r) + \mathrm{s}(A_{k_m}^m) + \mathrm{s}(A_{k_l}^l) + \mathrm{s}(AOP_{k_r}^r) + \mathrm{s}(AOP_{k_m}^m) + 
				\mathrm{s}(S_{k_m}^m) + \mathrm{s}(S_{k_l}^l) + 2k_m + 2k_l + 3,
			\end{aligned} $ \vspace{0pt}\\
			if $ k_l< \lfloor \frac{n}{2} \rfloor $.
		\end{itemize}
	\end{lemma}

	\begin{proof}
		First, assume that $ k_l= \lfloor \frac{n}{2} \rfloor $.
		The delay of $ C_n $ is the maximum delay of any of the carry bits computed by $ C_n $.
		The first $ k_r $ carry bits (computed in line 10 of Algorithm \ref{alg1}) have a delay of at most $ \text{del}(A_{k_r}) $. The last $ k_l $ carry bits (computed in line 12) have a delay of at most \[ \max\{\text{del}(A_{k_l}^l)+1, \text{del}(AOP_{k_r}^r)+2, \text{del}(S_{k_l}^l)+2\}. \] Thus, the total delay bound follows as claimed.
		
		The size bound arises from adding up the sizes of the sub-circuits computed in lines 6 to 8 of the algorithm together with some additional gates which are needed to compute all of the carry bits. In this case, we need two additional gates to compute each of the $ k_l $ carry bits $ c_{k_r+1},\dots c_{n+1} $ resulting in $ 2k_l $ additional gates.
		
		Now, assume that $ k_l< \lfloor \frac{n}{2} \rfloor $.
		Again, we look at the delay of the individual carry bits. The first ones (computed in line 24) have a delay of at most $ \text{del}(A_{k_r}^r) $. The carry bits computed in line 26 have a delay of at most \[ \max\left\{\text{del}(A_{k_m}^m)+1, \text{del}(AOP_{k_r}^r)+2, \text{del}(S_{k_m}^m)+2\right\}. \] The ($ k_r+k_m+1 $)th carry bit (computed in line 27) has a delay of at most 
		\begin{align*} 
			\max&\left\{a(g_{k_r+k_m})+1, a(p_{k_r+k_m})+2,\, \text{del}(AOP_{k_m}^m)+3,\, \text{del}(AOP_{k_r}^r)+4, \text{del}(S_{k_m}^m)+4\right\}.
		\end{align*} 
		The last $ k_r $ carry bits (computed in line 29) have a delay of at most 
\[
\max\big\{\text{del}(A_{k_l}^l)+1,\, \text{del}(S_{k_l}^l)+2,\, a(g_{k_r+k_m})+3, a(p_{k_r+k_m})+4,\, \text{del}(AOP_{k_m}^m)+5,\, \text{del}(AOP_{k_r}^r)+6,\, \text{del}(S_{k_m}^m)+6\big\}.
\] 
		Altogether, the delay bound follows.
		
		For the size bound we add the sizes of all sub-circuits computed in lines 20 to 22 of the algorithm. In this case, we have to add two gates for each carry bit computed in line 26, three gates for the carry bit computed in line 27 and two gates for each carry bit computed in line 29. Note that in line 27 we can reuse the gate between $ \text{out}_{k_m}(S_{k_m}^m) $ and $ \text{out}(AOP_{k_r}^r) $ which is already constructed in line 26. In line 29 we only need to add the first two gates and can reuse the remaining ones. Hence, we need $ 2k_m + 2k_l + 3 $ additional gates.
	\end{proof}

\subsection{Adder Construction}
	\label{adderconstruction}
We will apply Algorithm~\ref{alg1} using the fastest known \textsc{And-Or} path circuits
(Theorem~\ref{andorcirc}), and the fastest known \textsc{And}-prefix circuits
(Theorem~\ref{andprecirc0}, with the specific bounds from in Theorem~\ref{andprecirc}),
as sub-circuits. 
We receive an adder with a sub-quadratic size whose delay exceeds the delay of the \textsc{And}-prefix circuits only by a constant.
	
Before we can show this, we need a series of numerical lemmas.

	\begin{lemma}
		\label{numlem1}
		The function \[f(x)=\frac{\log_2\log_2x}{\log_2x}\] has its maximum at $ x=2^e $ with $ f(2^e)\leq 0.531 $ and is decreasing for $ x\geq 2^e $.
	\end{lemma}

	\begin{proof}
We rewrite $ f $ as \[f(x)=\frac{\ln(\ln x)-\ln(\ln2)}{\ln x}\] and calculate its derivative \[f'(x)=\frac{1-\ln(\ln x)+\ln(\ln2)}{x\ln^2x}=\frac{1-\ln(\log_2x)}{x\ln^2x}.\] 
Note that $ f'(x)=0 $ if and only if $ x=2^e $ and $ f'(x) $ is negative for all $ x>2^e $ which implies the lemma.
	\end{proof}

	\begin{lemma}
		\label{lemnum}
		Assume $ n\in \mathbb{N} $ with $ 6\leq n\leq 374 $. Then
		\begin{enumerate}[(i)]
			\item $ \log_2n\leq 2\log_2\log_2n + 2.357 $,
			\item $ 6.2\leq 2.399\log_2n $.
		\end{enumerate}
	\end{lemma}

\begin{proof}
The first statement is verified easily for $n=6$, so assume $n \geq 7$. Then, we have
$n > 2^e$, so by \Cref{numlem1} $n \mapsto \frac{\log_2 \log_2 n}{\log_2 n}$
is decreasing. Therefore, we get
\begin{align*}
   2\log_2\log_2n + 2.357 
     \geq \left( 2 \cdot \frac{\log_2 \log_2(374)}{\log_2(374)}
         + \frac{2.357}{\log_2(374)}\right) \log_2 n 
   \geq (0.7243 + 0.2757) \log_2 n = \log_2 n
\end{align*}
		
		The second statement holds since the logarithm is increasing and 
$\log_26\geq \frac{6.2}{2.399}$.
\end{proof}

	\begin{lemma}
		\label{numlem}
		Assume $ n\in \mathbb{N} $ with $ n\geq 375 $. Then we have the following inequalities:
\begin{enumerate}[(i)]
   \item $ \log_2\lceil\frac{n}{2}\rceil \leq \log_2n-0.99 $,
   \item $ \log_2\log_2\log_2n\leq 2\log_2\log_2n - 4.561 $,
   \item $ \log_2\log_2n + \log_2\log_2\log_2n+3.3 \leq 0.939 \log_2n $,
   \item $ 3.114\log_2\log_2n\leq 1.128\log_2n $,
   \item $ 2\log_2n-1\geq 1.883\log_2n $.
\end{enumerate}
	\end{lemma}

\begin{proof}
For $x\geq 375$, we have $x > \log_2(x) > 7 > 2^e$. Thus, by \Cref{numlem1},
both $x\mapsto \frac{\log2 \log2 x}{\log_2 x}$ and
$x\mapsto \frac{\log_2 \log_2 \log_2 x}{\log_2 \log_2 x}$ are then decreasing.
Since $n\geq 375$, we can apply \Cref{numlem1} several times in the following proofs.
\begin{enumerate}[(i)]
\item		For the first statement we calculate 
		\setlength{\mywidth}{\widthof{$\stackrel{n\geq 151..}{\leq}$}}
		\begin{align*}
			\log_2\Big\lceil\frac{n}{2}\Big\rceil & \,\leq\, \log_2\Big(\frac{n+1}{2}\Big)=\log_2\left(n\frac{n+1}{2n}\right)= \log_2n-1+\log_2\left(1+\frac{1}{n}\right)
			\klgleich[$ n\geq 375 $]\log_2n-0.99.
		\end{align*}
	
\item
We apply \Cref{numlem1} and get
\begin{align*}
   \log_2\log_2\log_2n + 4.561 
       & \leq \left(\frac{\log_2 \log_2\log_2 (375)}{\log_2 \log_2(375)} 
                   + \frac{4.561}{\log_2 \log_2(375)} \right) \log_2 \log_2 n\\
       & \leq (0.52662 + 1.4732) \log_2 \log_2 n \quad \leq \quad 2 \log_2 \log_2 n
\end{align*}
\item
Using \Cref{numlem1}, we calculate
\begin{align*}
   \log_2\log_2n + \log_2\log_2\log_2n+3.3
    & \leq \frac{\log_2 \log_2 (375)}{\log_2 (375)} \log n + 
          \frac{\log_2 \log_2 \log_2 (375)}{\log_2 \log_2 (375)} \log_2\log_2n +
           \frac{3.3}{\log(375)} \log n\\
    &  \leq \frac{\log_2 \log_2 (375)}{\log_2 (375)} \log_2 n + 
          \frac{\log_2 \log_2 \log_2 (375)}{ \log_2 (375)} \log_2n +
           \frac{3.3}{\log(375)} \log n\\
    & \leq (0.3621 + 0.1907 + 0.386) \log_2n  
    \quad \leq \quad 0.939 \log_2n
\end{align*}

		
\item
For the fourth statement, we use Lemma~\ref{numlem1} and calculate
		\[3.114\log_2\log_2n\stackrel{\substack{\text{Lem. \ref{numlem1}},\\ n\geq 375}}{\leq}3.114\frac{\log_2\log_2375}{\log_2375}\log_2n\leq 1.128\log_2n.\]
		
\item
The last statement holds true since $ 0.117\log_2(375)\geq 1 $ 
and the logarithm is monotonely increasing.\qedhere
\end{enumerate}
	\end{proof}

	Now, we can finally state and prove the main theorem of this section.

\linebox{
	\begin{theorem}
		\label{mainthm1}
		Let $ n\in \mathbb{N} $ with $ n\geq 3 $ and input pairs $ p_0, g_0, \dots, p_{n-1},g_{n-1} $ with arrival times $ a:\{p_0, g_0\dots p_{n-1}, g_{n-1}\}\to \mathbb{N} $ be given and let $ W = \sum_{i=0}^{n-1}\big(2^{a(p_{i})} + 2^{a(g_{i})}\big) $ be the weight of the input pairs. We can construct an adder circuit $ C_{n} $ on these input pairs with delay 
		\begin{align*}
			\mathrm{del}(C_{n})&\leq \log_2W + 3\log_2\log_2n + c_1 \intertext{and size} \mathrm{s}(C_n)&\leq c_2n\log_2^2n
		\end{align*} 
		in polynomial time for $ c_1=5.007 $ and $ c_2=2.422$.
	\end{theorem}
}

	\begin{proof}
		We prove the delay and size bounds by induction on $ n $.
		
		\textbf{Case 1:} Assume that $ 3\leq n\leq 5 $. In this case we use the ripple carry adder circuit which has depth and size $ 2n-2 $ (see Example \ref{extrem_adders}). This circuit fulfills our desired delay and size bounds for $ c_1\geq 5 $ and $ c_2\geq 1 $ since 
		\begin{align*}
			2n-2&\leq 3\log_2\log_2n+5
			\intertext{and}
			2n-2&\leq n\log_2^2n
		\end{align*}
		for $ n\in \{3,4,5\} $.
		
\textbf{Case 2:} Assume that $ 6\leq n\leq 374 $. 
Here, we use the adder circuit constructed by Brenner and Silvanus \cite{BS24}
from Theorem~\ref{daddercirc}, which we gave a delay bound for
in Corollary~\ref{cordaddercirclin}~(a), and call it $ C_n $. It has delay 
{\allowdisplaybreaks
		\setlength{\mywidth}{\widthof{$\stackrel{Lem. \ref{lemnum} (ii)..}{\leq}$}}
\begin{align*}
\text{del}(C_n)&\klgleich[Cor. \ref{cordaddercirclin}~(a)] \log_2W + \log_2n + \log_2\log_2n + 2.65 \klgleich[Lem. \ref{lemnum} (i)] \log_2W + 3\log_2\log_2n + c_1 
\intertext{and size} \text{s}(C_n)
&\klgleich[Thm. \ref{daddercirc}] 6.2n\log_2n \klgleich[Lem. \ref{lemnum} (ii)] c_2n\log_2^2n
\end{align*}
} 
		for $ c_1\geq 5.007 $ and $ c_2\geq 2.399 $.
 	
 		\textbf{Case 3:} Assume that $ n\geq 375 $. Here, we apply our adder construction framework (Algorithm \ref{alg1}). Inductively, we can assume that adder circuits $ C_i $ on $ i $ input pairs with the claimed delay and size bounds can be constructed for all $ 3\leq i<n $.
 	
 		\textbf{Case 3.1:} Assume that $ k_l=\lfloor\frac{n}{2}\rfloor $. Then $ k_r=\lceil\frac{n}{2}\rceil $ and $ W_{P_{k_l}}\leq \frac{W}{2} $. The following sub-circuits are used in our algorithm:
 		\begin{itemize}
 			\item We recursively compute adders $ A_{k_r}^r $ and $ A_{k_l}^l $ on $ P_r $ and $ P_l $, respectively. By induction, they fulfill
 			\setlength{\mywidth}{\widthof{$\stackrel{W_{P_l}\leq \frac{W}{2}, }{\leq}$}}
 			{\allowdisplaybreaks
 			\begin{align}
 				\text{del}(A_{k_r}^r)&\klgleich[(IH)] \log_2W_{P_r} + 3\log_2\log_2k_r + c_1 \nonumber\\ 
 				&\klgleich[$\substack{W_{P_r}\leq W,\\{k_r\leq n}}$] \log_2W +3\log_2\log_2n + c_1, \label{eq1}\\
 				\text{del}(A_{k_l}^l)&\klgleich[(IH)] \log_2W_{P_l} + 3\log_2\log_2k_l + c_1 \nonumber\\ 
 				&\klgleich[$\substack{W_{P_l}\leq \frac{W}{2},\\k_l\leq n}$] \log_2W +3\log_2\log_2n + c_1 -1, \label{eq2}\\
 				\text{s}(A_{k_r}^r)&\klgleich[(IH)]c_2k_r\log_2^2k_r, \label{eq3} \\
 				\text{s}(A_{k_l}^l)&\klgleich[(IH)]c_2k_l\log_2^2k_l. \label{eq4}
 			\end{align}}\noindent
\item For the \textsc{And-Or}-path circuit $ AOP_{k_r} $, we use the circuit 
from Theorem~\ref{andorcirc}. $ AOP_{k_r} $ has $ 2k_r-1 \leq n $ inputs since $ k_r\leq \frac{n+1}{2} $. Therefore, we have
 			\setlength{\mywidth}{\widthof{$\stackrel{Lem. \ref{numlem} (iii)..}{\leq}$}}
{\allowdisplaybreaks
 			\begin{align}
 				\text{del}(AOP_{k_r}^r)\klgleich& \log_2W_{P_{k_r}}+\log_2\log_2n+\log_2\log_2\log_2n + 4.3 \nonumber \\
 				\klgleich[$ W_{P_{k_r}}\leq W $]& \log_2W+\log_2\log_2n+\log_2\log_2\log_2n + 4.3 \nonumber \\
 				\klgleich[Lem. \ref{numlem} (ii)]& \log_2W + 3\log_2\log_2n - 0.261, \label{eq5} \\
 				\text{s}(AOP_{k_r}^r)\klgleich& n\log_2n+n\log_2\log_2n+n\log_2\log_2\log_2n+ 
 				3.3n-1 \nonumber\\
 				\klgleich[Lem. \ref{numlem} (iii)]& 1.939n\log_2n. \label{eq6}
 			\end{align}
}
 			\item For the \textsc{And}-prefix circuit $ S_{k_l} $ on the propagate signals of $ P_{k_l} $, we use the circuit of Rautenbach, Szegedy and Werber, which we proved specific delay and size bounds for in Theorem~\ref{andprecirc}, guaranteeing the bounds
 			\setlength{\mywidth}{\widthof{$\stackrel{ \sum_{i=0}^{k_l-1}2^{a(p_{k_r+i})}\leq \frac{W}{2}, }{\leq}$}}
{\allowdisplaybreaks
 			\begin{align}
 				\text{del}(S_{k_l})&\klgleich \log_2\Bigg(\sum_{i=0}^{k_l-1}2^{a(p_{k_r+i})}\Bigg)+3\log_2\log_2k_l + 0.007 \nonumber \\
 				&\klgleich[$ \substack{\sum_{i=0}^{k_l-1}2^{a(p_{k_r+i})}\leq \frac{W}{2}, \\k_l\leq n} $] \log_2W+3\log_2\log_2n - 0.993, \label{eq7} \\
 				\text{s}(S_{k_l})&\klgleich 3.114k_l\log_2\log_2k_l \nonumber \\
 				&\klgleich[$ k_l\leq \frac{n}{2} $] 1.557n \log_2\log_2n \nonumber \\
 				&\klgleich[Lem. \ref{numlem} (iv)] 0.564n\log_2n. \label{eq8}
 			\end{align}
}
 		\end{itemize} 
 		
 		Altogether, these bounds combined with Lemma~\ref{lemalg1} imply the delay bound 
{\allowdisplaybreaks
 		\setlength{\mywidth}{\widthof{$\stackrel{Lem... \ref{lemalg1}}{\leq}$}}
 		\begin{align*}
 			\text{del}(C_n)\klgleich& \max\{\text{del}(A_{k_r}), \text{del}(A_{k_l})+1, \text{del}(AOP_{k_r})+2, \text{del}(S_{k_l})+2\}\\
 			\klgleich[$ \substack{\eqref{eq1}, \eqref{eq2}, \\ \eqref{eq5}, \eqref{eq7}} $]& \max\{\log_2W +3\log_2\log_2n + c_1,\, \log_2W + 3\log_2\log_2n + 1.739,\\
 &\phantom{\max\{} \log_2W+3\log_2\log_2n+ 1.007\}\\
 			\klgleich& \log_2W +3\log_2\log_2n + c_1
 		\end{align*}
}
 		for $ c_1\geq 1.739 $.
 		
 		For the size bound we calculate
 		\setlength{\mywidth}{\widthof{$\stackrel{Lem. \ref{numlem} (i),..}{\leq}$}}
 		\begin{align}
 			\text{s}(A_{k_r})+\text{s}(A_{k_l})\klgleich[$ \eqref{eq3},\eqref{eq4} $]&c_2k_r\log_2^2k_r + c_2k_l\log_2^2k_l\nonumber\\
 			\klgleich[$ \substack{\text{Lem. \ref{numlem} (i),}\\ k_l\leq \frac{n}{2}} $]& c_2\big(k_r(\log_2n-0.99)^2+k_l(\log_2n-1)^2\big)\nonumber\\
 			\klgleich& c_2\big(k_r(\log_2^2n-1.98\log_2n+0.981)+
 			k_l(\log_2n^2-2\log_2n+1)\big)\nonumber\\
 			\gleich[$ k_r+k_l=n $]& c_2\big(n\log_2^2n-\log_2n(1.98k_r+2k_l)+
 			0.981k_r+k_l\big).\label{eq10}
 		\end{align}
 		Furthermore, we have 
 		\begin{equation*}
 			1.98k_r+2k_l\geq 1.98\frac{n+1}{2}+2\frac{n-1}{2}=1.99n-0.01
 		\end{equation*}
 		and 
 		\begin{equation*}
 			0.981k_r+k_l\leq 0.981\frac{n}{2}+\frac{n}{2}\leq 0.991n. 
 		\end{equation*}
 		Now, Equation \eqref{eq10} can be simplified to 
 		\begin{equation}
 			\text{s}(A_{k_r})+\text{s}(A_{k_l})\leq c_2(n\log_2^2n-1.99n\log_2n+0.01\log_2n +0.991n).\label{eq11}
 		\end{equation}
 		With these calculations we can now use Lemma~\ref{lemalg1} to calculate the total size bound
 		\setlength{\mywidth}{\widthof{$\stackrel{1.99=0.137+1.853..}{\leq}$}}
{\allowdisplaybreaks
 		\begin{align}
 			\text{s}(C_n)\klgleich&\text{s}(A_{k_r}) + \text{s}(A_{k_l}) + \text{s}(AOP_{k_r}) + \text{s}(S_{k_l}) + 2k_l\nonumber\\
 			\klgleich[$ \substack{\eqref{eq11},\eqref{eq6},\\ \eqref{eq8}, 2k_l\leq n} $]& c_2(n\log_2^2n-1.99n\log_2n+0.01\log_2n+0.991n) +
 			1.939n\log_2n + 0.564n\log_2n + n\nonumber\\
 			\gleich[$ 1.99=0.116+1.874 $]& c_2n\log_2^2n - c_2(0.116n\log_2n-0.991n-0.01\log_2n)
 			-n\log_2n\Big(1.874c_2-2.503-\frac{1}{\log_2n}\Big)\nonumber\\
   \gleich & c_2n\log_2^2n - c_2 n \log_2 n 
             \left(0.116 - \frac{0.991}{\log_2 n} - \frac{0.01}{n} \right) 
   -n\log_2n\left(1.874c_2-2.503-\frac{1}{\log_2n}\right) \nonumber\\
   \klgleich[$n \geq 375$] & c_2n\log_2^2n - c_2 n \log_2 n 
             \left(0.116 - \frac{0.991}{\log_2 375} - \frac{0.01}{375} \right) 
   -n\log_2n\left(1.874c_2-2.503-\frac{1}{\log_2n}\right) \nonumber\\
   \klgleich & c_2n\log_2^2n - c_2 n \log_2 n 
             \left(0.116 - 0.1159 - 0.00003 \right) 
   -n\log_2n\left(1.874c_2-2.503-\frac{1}{\log_2n}\right) \nonumber\\
   \klgleich & c_2n\log_2^2n 
    -n\log_2n\left(1.874c_2-2.503-\frac{1}{\log_2 375}\right) \nonumber\\
 			\klgleich& c_2n\log_2^2n,\nonumber
 		\end{align}
}
where the last inequality holds if \[c_2\geq \frac{2.503+\frac{1}{\log_2375}}{1.874},\] 
which is satisfied if $c_2\geq 1.399$.

 		
 		\textbf{Case 3.2:} Assume that $ k_l<\lfloor\frac{n}{2}\rfloor $. Then $ k_r=\lfloor\frac{n}{2}\rfloor $ and $ k_m = \lceil \frac{n}{2}\rceil -k_l-1 $. In this case, we have $ k_r,k_m,k_l\leq \frac{n}{2} $ and $ W_{P_r},W_{P_m},W_{P_l}\leq \frac{W}{2} $. Note that $ k_l $ or $ k_m $ may be 0. We use the same type of sub-circuits in this part of our algorithm:
  		{\allowdisplaybreaks
 		\begin{itemize}
 			\item If $ k_l,k_m\geq 3 $, we recursively compute adders $ A_{k_r}^r,A_{k_m}^m $ and $ A_{k_l}^l $ on $ P_r,P_m $ and $ P_l $, respectively. By induction, they fulfill
{\allowdisplaybreaks
 			\setlength{\mywidth}{\widthof{$\stackrel{W_{P_m}\leq \frac{W}{2}, }{\leq}$}}
 			\begin{align}
 				\text{del}(A_{k_r}^r)&\klgleich[(IH)] \log_2W_{P_r} + 3\log_2\log_2k_r + c_1 \nonumber\\ 
 				&\klgleich[$\substack{W_{P_r}\leq \frac{W}{2},\\{k_r\leq n}}$] \log_2W +3\log_2\log_2n + c_1-1, \label{eq13}\\
 				\text{del}(A_{k_m}^m)&\klgleich[(IH)] \log_2W_{P_m} + 3\log_2\log_2k_m + c_1 \nonumber\\ 
 				&\klgleich[$\substack{W_{P_m}\leq \frac{W}{2},\\k_l\leq n}$] \log_2W +3\log_2\log_2n + c_1 -1, \label{eq14}\\
 				\text{del}(A_{k_l}^l)&\klgleich[(IH)] \log_2W_{P_l} + 3\log_2\log_2k_l + c_1 \nonumber\\ 
 				&\klgleich[$\substack{W_{P_l}\leq \frac{W}{2},\\k_l\leq n}$] \log_2W +3\log_2\log_2n + c_1 -1, \label{eq15}\\
 				\text{s}(A_{k_r}^r)&\klgleich[(IH)]c_2k_r\log_2^2k_r\nonumber\\
 				&\klgleich[$ k_r\leq \frac{n}{2} $]c_2k_r(\log_2n-1)^2, \label{eq16} \\
 				\text{s}(A_{k_m}^m)&\klgleich[(IH)]c_2k_m\log_2^2k_m\nonumber\\
 				&\klgleich[$ k_m\leq \frac{n}{2} $]c_2k_m(\log_2n-1)^2, \label{eq17} \\
 				\text{s}(A_{k_l}^l)&\klgleich[(IH)]c_2k_l\log_2^2k_l\nonumber\\
 				&\klgleich[$ k_l\leq \frac{n}{2} $]c_2k_l(\log_2n-1)^2. \label{eq18}
 			\end{align}
}
 			If $ k_l\leq2 $ or $ k_m\leq2 $, we just use the trivial adder circuit on $ P_l $ or $ P_m $. In this case, the inequalities above are still satisfied for $ c_1\geq 0 $ and $ c_2\geq 1 $.
 			
 			\item For the \textsc{And-Or}-path circuits $ AOP_{k_r}^r $ and $ AOP_{k_m}^m $, assuming that $ k_m\geq 3 $, we again use the circuit 
from Theorem~\ref{andorcirc}. Since $ k_r,k_m\leq \frac{n}{2} $, $ AOP_{k_r}^r $ and $ AOP_{k_m}^m $ have less than $ n $ inputs. Therefore, we have
 			\setlength{\mywidth}{\widthof{$\stackrel{Lem. \ref{numlem} (iii)..}{\leq}$}}
 			\begin{align}
 				\text{del}(AOP_{k_r}^r)\klgleich& \log_2W_{P_r}+\log_2\log_2n+\log_2\log_2\log_2n + 4.3 \nonumber \\
 				\klgleich[$ W_{P_r}\leq \frac{W}{2} $]& \log_2W+\log_2\log_2n+\log_2\log_2\log_2n + 3.3 \nonumber \\
 				\klgleich[Lem. \ref{numlem} (ii)]& \log_2W + 3\log_2\log_2n - 1.261, \label{eq19} \\
 				\text{del}(AOP_{k_m}^m)\klgleich& \log_2W_{P_m}+\log_2\log_2n+\log_2\log_2\log_2n + 4.3 \nonumber \\
 				\klgleich[$ W_{P_m}\leq \frac{W}{2} $]& \log_2W+\log_2\log_2n+\log_2\log_2\log_2n + 3.3 \nonumber \\
 				\klgleich[Lem. \ref{numlem} (ii)]& \log_2W + 3\log_2\log_2n - 1.261, \label{eq20} \\
 				\text{s}(AOP_{k_r}^r)\klgleich& n\log_2n+n\log_2\log_2n+n\log_2\log_2\log_2n+ 
 				3.3n-1 \nonumber \\
 				\klgleich[Lem. \ref{numlem} (iii)]& 1.939n\log_2n, \label{eq21} \\
 				\text{s}(AOP_{k_m}^m)\klgleich& n\log_2n+n\log_2\log_2n+n\log_2\log_2\log_2n+ 
 				 3.3n-1 \nonumber \\
 				\klgleich[Lem. \ref{numlem} (iii)]& 1.939n\log_2n. \label{eq22}
 			\end{align}
 			If $ k_m\leq 2 $, we use the trivial \textsc{And-Or}-path circuit as $ AOP_{k_m}^m $ which also satisfies the above inequalities.
 			
 			\item For the \textsc{And}-prefix circuits $ S_{k_m}^m $ and $ S_{k_l}^l $ on the propagate signals of $ P_m $ and $ P_l $, respectively, we again use the circuit of Rautenbach, Szegedy and Werber (Theorem~\ref{andprecirc}). Here, we have
 			\setlength{\mywidth}{\widthof{$\stackrel{ \sum_{i=1}^{k_l}2^{a(p_{k_r+k_m+i})}\leq \frac{W}{2},}{\leq}$}}
 			\begin{align}
 				\text{del}(S_{k_m}^m)\klgleich& \log_2\Bigg(\sum_{i=0}^{k_m-1}2^{a(p_{k_r+i})}\Bigg)+3\log_2\log_2k_m+ 0.007 \nonumber \\
 				\klgleich[$ \substack{\sum_{i=0}^{k_m-1}2^{a(p_{k_r+i})}\leq \frac{W}{2}, \\k_m\leq n} $]& \log_2W+3\log_2\log_2n - 0.993, \label{eq23} \\
 				\text{del}(S_{k_l}^l)\klgleich& \log_2\Bigg(\sum_{i=1}^{k_l}2^{a(p_{k_r+k_m+i})}\Bigg)+3\log_2\log_2k_l + 0.007 \nonumber \\
 				\klgleich[$ \substack{\sum_{i=1}^{k_l}2^{a(p_{k_r+k_m+i})}\leq \frac{W}{2}, \\k_l\leq n} $]& \log_2W+3\log_2\log_2n - 0.993, \label{eq24} \\
 				\text{s}(S_{k_m}^m)\klgleich& 3.114k_m\log_2\log_2k_m \nonumber \\
 				\klgleich[$ k_m\leq n $]& 3.114k_m \log_2\log_2n \nonumber \\
 				\klgleich[Lem. \ref{numlem} (iv)]& 1.128k_m\log_2n, \label{eq25} \\
 				\text{s}(S_{k_l}^l)\klgleich& 3.114k_l\log_2\log_2k_l \nonumber \\
 				\klgleich[$ k_l\leq n $]& 3.114k_l \log_2\log_2n \nonumber \\
 				\klgleich[Lem. \ref{numlem} (iv)]& 1.128k_l\log_2n. \label{eq26}
 			\end{align}
 		\end{itemize}}\noindent
 	
 		Now, we again use Lemma~\ref{lemalg1} to prove the desired delay and size bounds. For the delay, we have 
 		\setlength{\mywidth}{\widthof{$\stackrel{ \eqref{eq13},\eqref{eq14},\eqref{eq15}, }{\leq}$}}
{\allowdisplaybreaks
\begin{align*}
\text{del}(C_n)\klgleich 
& \max\big\{\text{del}(A_{k_r}^r), \text{del}(A_{k_m}^m) + 1,\text{del}(A_{k_l}^l)+1 , a(g_{k_r+k_m})+3, a(p_{k_r+k_m}) + 4,\\
&\phantom{\max\big\{}\text{del}(AOP_{k_m}^m)+5, \text{del}(AOP_{k_r}^r)+6, \text{del}(S_{k_l}^l)+2, \text{del}(S_{k_m}^m)+6\big\} \\
\klgleich[$ \substack{\eqref{eq13},\eqref{eq14},\eqref{eq15},\\ \eqref{eq19},\eqref{eq20},\\ \eqref{eq23},\eqref{eq24}} $]& \max\big\{\log_2W +3\log_2\log_2n + c_1, \log_2W + 3\log_2\log_2n + 4.739,\\ 
&\phantom{\max\big\{} \log_2W+3\log_2\log_2n+ 5.007\big\}\\
 			\klgleich& \log_2W +3\log_2\log_2n + c_1
\end{align*}}\noindent
 		for $ c_1\geq 5.007 $. Here we used that $ a(g_{k_r+k_m}) $ and $ a(p_{k_r+k_m}) $ are less than $ \log_2W $.
 		
 		For the size bound, we calculate 
 		\setlength{\mywidth}{\widthof{$\stackrel{ k_r+k_m+k_l=n-1..}{\leq}$}}
{\allowdisplaybreaks
 		\begin{align}
 			\text{s}(A_{k_r}^r)+\text{s}(A_{k_m}^m)+\text{s}(A_{k_l}^l)\klgleich[$ \eqref{eq16},\eqref{eq17},\eqref{eq18} $]&c_2(k_r+k_m+k_l)(\log_2n-1)^2\nonumber\\
 			\klgleich[$ k_r+k_m+k_l=n-1 $]&c_2n(\log_2^2n-2\log_2n+1)\nonumber\\
 			\klgleich[Lem. \ref{numlem} (v)]&c_2n(\log_2^2n-1.883\log_2n).\label{eq28}
 		\end{align}
}
Moreover, inequalities $ \eqref{eq25}$ and $\eqref{eq26} $ and the fact that 
$ k_m+k_l=\Big\lceil\frac{n}{2}\Big\rceil-1\leq \frac{n-1}{2}\leq \frac{n}{2}$  give us
 		\begin{align}
 			\text{s}(S_{k_m}^m)+\text{s}(S_{k_l}^l)&\leq1.128\log_2n(k_m+k_l)
 			\leq 0.564n\log_2n. \label{eq27}
 		\end{align}
 		Lemma~\ref{lemalg1} together with our size bounds for the computed sub-circuits results in the total size bound
 		\setlength{\mywidth}{\widthof{$\stackrel{k_m+k_l\leq \frac{n-1}{2}..}{\leq}$}}
 {\allowdisplaybreaks\begin{align*}
 			\text{s}(C_n)\klgleich&\text{s}(A_{k_r}^r) + \text{s}(A_{k_m}^m) + \text{s}(A_{k_l}^l) + \text{s}(AOP_{k_r}^r) + \text{s}(AOP_{k_m}^m) + 
 			\text{s}(S_{k_m}^m) + \text{s}(S_{k_l}^l) + 2k_m + 2k_l + 3\\
 			\klgleich[$ \substack{\eqref{eq28},\eqref{eq21},\\ \eqref{eq22}, \eqref{eq27}} $]&c_2n(\log_2^2n-1.883\log_2n) + 3.878n\log_2n + 
 			0.564n\log_2n + 2k_m + 2k_l + 3\\
 			\klgleich[$ k_m+k_l\leq \frac{n-1}{2} $]&c_2n\log_2^2n-n\log_2n(1.883c_2-4.442)+n+2\\
 			\gleich&c_2n\log_2^2n-n\log_2n\left(1.883c_2-4.442-\frac{1}{\log_2n}-\frac{2}{n\log_2n}\right)\\
 			\klgleich[$ n\geq 375 $]&c_2n\log_2^2n-n\log_2n\left(1.883c_2-4.442-
 			\frac{1}{\log_2375}-\frac{2}{375\log_2375}\right)\\
 			\klgleich& c_2n\log_2^2n-n\log_2n(1.883c_2-4.56)\\
 			\klgleich& c_2n\log_2^2n
 		\end{align*}}
 		for $ c_2\geq \frac{4.56}{1.883} $, which is the case if $c_2\geq 2.422$.
 		
Thus, we get the delay and size bounds for $ C_n $ with constants
$c_1 \geq 5.007$ and $c_2 \geq 2.422$.
 	
For proving the runtime, we show by induction on $ n $ that there exists a constant $ \alpha $ such that the total number of steps needed for computing $ C_n $ is bounded by $ n^{\alpha} $. It suffices to consider the case where $ n\geq 151 $. 
 		
 		\textbf{Case 1:} Assume that $ k_l=\lfloor\frac{n}{2}\rfloor $. Then $ k_r\leq \frac{n+1}{2}\leq \frac{76}{151}n $ since $ n\geq~151 $. By the induction hypothesis we can construct the sub-circuits $ A_{k_r}^r $ and $ A_{k_l}^l $ in at most $ k_r^{\alpha} $ and $ k_l^{\alpha} $ steps, respectively. The rest of the algorithm, including the computation of the \textsc{And-Or}-path circuit $ AOP_{k_r}^r $ and the \textsc{And}-prefix circuit $ S_{k_l}^l $ (see Theorem~\ref{andorcirc} and Theorem~\ref{andprecirc}), can be performed in polynomial time. For this part, we assume that we need at most $ n^{\beta} $ steps for some fixed $ \beta >0 $. Then, the total number of computation steps is at most
 		\[k_r^{\alpha}+k_l^{\alpha}+n^{\beta}\leq \Big(\frac{76}{151}n\Big)^{\alpha}+ \Big(\frac{n}{2}\Big)^{\alpha}+n^{\beta}\leq \Big(\Big(\frac{76}{151}\Big)^{\alpha}+ \frac{1}{2^{\alpha}}+n^{\beta -\alpha}\Big)n^{\alpha}\leq n^{\alpha}\]
 		for a sufficiently large $ \alpha $.
 		
 		\textbf{Case 2:} Assume that $ k_l<\lfloor\frac{n}{2}\rfloor $. Then $ k_r\leq \frac{n}{2} $ and $ k_l\leq\frac{n}{2} $. Again, by the induction hypothesis we can construct the sub-circuits $ A_{k_r}^r, A_{k_m}^m $ and $ A_{k_l}^l $ in at most $ k_r^{\alpha}, k_m^{\alpha} $ and $ k_l^{\alpha} $ steps, respectively. The rest of the algorithm can again be performed in a total amount of at most $ n^{\gamma} $ computation steps for some fixed $ \gamma>0 $. Combined, the number of steps does not exceed 
 		\[k_r^{\alpha}+k_m^{\alpha}+k_l^{\alpha}+n^{\gamma}\leq 3\Big(\frac{n}{2}\Big)^{\alpha}+n^{\gamma}\leq \Big(\frac{3}{2^{\alpha}}+n^{\gamma -\alpha}\Big)n^{\alpha}\leq n^{\alpha}\] for a sufficiently large $ \alpha $. 
This finishes the proof of the polynomial runtime and also the proof of the theorem.
\end{proof}



\begin{remark}
The fanout of the adder from \Cref{mainthm1} is in $ \mathcal{O}(\sqrt{n}) $ since the fanout of the \textsc{And}-prefix circuit by Rautenbach, Szegedy and Werber from Theorem~\ref{andprecirc}, which we used as sub-circuits in Algorithm \ref{alg1}, is also in $\mathcal{O}(\sqrt{n})$ (see the remark in Section~\ref{analandprecirc}).
\end{remark}

\section{Size Reduction of Adder Circuits}
	\label{sizereduction}
	
In this section, we are going to use another adder construction framework, 
presented in \cite{BS24},
in order to reduce the size of our adder circuit from Theorem~\ref{mainthm1},
at the cost of a relatively small delay increase. 
In Section~\ref{lpartframework}, we will present the adder construction framework
from \cite{BS24}  which we will then 
use to compute three additional adder circuits.
The first adder, shown in Section~\ref{smalleradder}, has a size in $ \mathcal{O}(n\log_2n) $, whereas the second and the third adder, both presented in Section~\ref{linadder}, have a size in $ \mathcal{O}(n\log_2\log_2n) $ and $ \mathcal{O}(n) $, respectively. The delay of each of these circuits exceeds the delay of the circuit from Theorem~\ref{mainthm1} by a summand of order $ \mathcal{O}(\log_2\log_2\log_2n) $. 

\subsection{$ l $-Part Adder Construction Framework}
	\label{lpartframework}
	
	\begin{algorithm}[htb]
\caption[l-part adder construction framework]{l-part adder construction framework \cite{BS24} (Alg. 4.2)}
		\label{alg2}
		\begin{algorithmic}[1]
			\Require $ n\in \mathbb{N}, n\geq 2 $, $ n $ input pairs $ p_0, g_0, \dots, p_{n-1},g_{n-1} $, a family of adder circuits $ (A_k)_{k\in\mathbb{N}} $, a family of \textsc{And-Or}-path circuits $ (AOP_k)_{k\in\mathbb{N}} $, a family of \textsc{And}-prefix circuits $ (S_k)_{k\in\mathbb{N}} $.
			
			\Ensure An adder circuit $ C_n $ on $ p_0, g_0, \dots, p_{n-1},g_{n-1} $.
			
			\State Choose $ k\in\mathbb{N}_{>0} $ and $ l:=\lceil n/k\rceil $.
			\For{$ j = 0 $ to $ l-2 $}
			\State $ P^{(j)}:=(p_jk,g_jk, \dots ,p_{(j+1)k-1},g_{(j+1)k-1}) $.
			\EndFor
			\State $ P^{(l-1)} := (p_{(l-1)k},g_{(l-1)k},\dots ,p_{n-1},g_{n-1}) $.
			\For{$ j=0 $ to $ l-1 $}
			\State $ n_j\gets | P^{(j)}|, N_j \gets n_0+\dots+n_{j-1} $.
			\EndFor
			\State Construct an adder circuit $ A_{n_0}^{0} $ on $ P^{(0)} $.
			\For{$ i=1 $ to $ n_0 $}
			\State $ out_i(C_n):=out_i(A_{n_0}^{(0)}) $.
			\EndFor
			\For{$ j=1 $ to $ l-1 $}
			\State Construct an adder circuit $ A_{n_j}^{j} $ on $ P^{(j)} $.
			\State Construct an \textsc{And}-prefix circuit $ S_{n_j}^{j} $ on the inputs $ p_i $ of $ P^{(j)} $.
			\State Construct an \textsc{And-Or}-path circuit $ AOP_{N_j}^{j} $ on the $ N_j $ input pairs $ P^{(j-1)}\dplus\dots\dplus P^{(0)} $.
			\For{$ i=1 $ to $ n_j $}
			\State $ \text{out}_{N_j+i}(C_n):=\text{out}_i(A_{n_j}^{(j)})\lor \big(\text{out}_i(S_{n_j}^{(j)})\land \text{out}(AOP_{N_j}^{(j)})\big) $.
			\EndFor
			\EndFor \\
			\Return $ C_n $.
		\end{algorithmic}
	\end{algorithm}
	
	Algorithm \ref{alg2} provides a procedure to construct an adder circuit on $ n $ input pairs given another family of adder circuits $ (A_k)_{k\in\mathbb{N}} $, a family of \textsc{And-Or}-path circuits $ (AOP_k)_{k\in\mathbb{N}} $ and a family of \textsc{And}-prefix circuits $ (S_k)_{k\in\mathbb{N}} $. For a given $ k\in\mathbb{N} $, the instance is partitioned into $ l:=\lceil\frac{n}{k}\rceil $ consecutive parts $ P^{(0)},\dots, P^{(l-1)} $ of $ n_j\leq k $ input pairs for each $ 0\leq j \leq l-1 $. The carry bits are again computed using the alternating split (Theorem~\ref{thmaltsplit}). Hence, we compute an adder circuit $ A_{n_j}^j $ on $ P^{(j)} $ for all $ 0\leq j\leq l-1 $, an \textsc{And}-prefix circuit $ S_{n_j}^j $ on the propagate signals of $ P^{(j)} $ and an \textsc{And-Or} path circuit $ AOP_{N_j}^j $ on the $ N_j:=n_0+\dots+n_{j-1} $ input pairs $ P^{(j-1)}\dplus\dots\dplus P^{(0)} $ for all $ 1\leq j\leq l-1 $, where $ P^{(j-1)}\dplus\dots\dplus P^{(0)} $ is the concatenation of the input parts $ P^{(j-1)},\dots, P^{(0)} $.
	
	The first $ n_0 $ carry bits can be directly computed by $ A_{n_0}^0 $. The $ (N_j+i) $th carry bit is given by $ \text{out}_i(A_{n_j}^{(j)})\lor \big(\text{out}_i(S_{n_j}^{(j)})\land \text{out}(AOP_{N_j}^{(j)})\big) $ for all $ 1\leq i\leq n_j $ and $ 1\leq j\leq l-1 $ due to Theorem~\ref{thmaltsplit}. Hence, Algorithm \ref{alg2} computes all carry bits correctly.

	The main difficulty in constructing a fast and small adder using this $ l $-part adder construction framework is the construction of the \textsc{And-Or} path circuits $ AOP_{N_j}^j $ for $ 1\leq j\leq l-1 $ in line 13 of Algorithm \ref{alg2}. If each of these $ l-1 $ circuits is computed separately with a linear size each, their total size would be in 
$ \Theta(l*n)=\Theta(n^2/k) $, which would be undesirable. 
The next lemma allows to compute the circuits $ AOP_{N_j}^j $ more efficiently.

	\begin{lemma}[\cite{BS24} (Cor. 4.5)]
		\label{lemandor}
		Let $ n $ input pairs $ p_0,g_0,\dots,$ $ p_{n-1},g_{n-1} $ and a partition $ P^{(0)},\dots, P^{(l-1)} $ of the inputs pairs in $ l\in\mathbb{N} $ consecutive parts be given, where each part $ P^{(j)} $ contains $ n_j $ input pairs for $ 0\leq j\leq l-1 $. Then, we have
		\begin{align*}
			g^*(g_{n-1},p_{n-1},\dots, g_1,p_1,g_0)
			=&g^*\big(h^*(P^{(l-1)}),a(P^{(l-1)}),\dots, h^*(P^{(1)}),a(P^{(1)}),h^*(P^{(0)})\big),
		\end{align*}
		where 
		\begin{align*}
			h^*(P^{(j)}):=&g^*(g_{n_0+\dots+n_j-1},p_{n_0+\dots+n_j-1},\dots, g_{n_0+\dots+n_{j-1}}) \intertext{and}
			a(P^{(j)}):=&\bigwedge((p_{n_0+\dots+n_{j-1}},\dots, p_{n_0+\dots+n_j-1}))
		\end{align*} 
		for all $ 0\leq j\leq l-1 $.
	\end{lemma}

	\begin{proof}
		Iteratively applying the alternating split (Theorem~\ref{thmaltsplit}) gives us the desired equation 
{\allowdisplaybreaks
		\begin{align*}
			 & g^*(g_{n-1},p_{n-1},\dots, g_1,p_1,g_0)\\
			=\, &h^*(P^{(l-1)})\lor\bigg(a(P^{(l-1)})\land g^*\Big(P^{(l-2)}\dplus\dots\dplus P^{(0)}\Big)\bigg)\\
			=\, &h^*(P^{(l-1)})\lor\Bigg(a(P^{(l-1)})\land \bigg(h^*(P^{(l-2)})\lor \Big(a(P^{(l-2)})\land \big(\dots
			\lor (a(P^{(1)})\land h^*(P^{(0)}))\dots\big)\Big)\bigg)\Bigg)\\
			=\, &g^*\big(h^*(P^{(l-1)}),a(P^{(l-1)}),\dots, h^*(P^{(1)}),a(P^{(1)}),h^*(P^{(0)})\big),
		\end{align*}
}
		where $ P^{(l-2)}\dplus\dots\dplus P^{(0)} $ is the concatenation of $ P^{(l-2)},\dots, P^{(0)} $.
	\end{proof}

\begin{figure}[htb]
		\centering
		\begin{tikzpicture}
			\node[scale=0.7] at (0,0) (g11) {$ g_{11} $};
			\node[scale=0.7] at (0.5,0) (p11) {$ p_{11} $};
			\node[scale=0.7] at (1,0) (g10) {$ g_{10} $};
			\node[scale=0.7] at (1.5,0) (p10) {$ p_{10} $};
			\node[scale=0.7] at (2,0) (g9) {$ g_9 $};
			\node[scale=0.7] at (2.5,0) (p9) {$ p_9 $};
			\node[scale=0.7] at (3,0) (g8) {$ g_8 $};
			\node[scale=0.7] at (3.5,0) (p8) {$ p_8 $};
			\node[scale=0.7] at (4,0) (g7) {$ g_7 $};
			\node[scale=0.7] at (4.5,0) (p7) {$ p_7 $};
			\node[scale=0.7] at (5,0) (g6) {$ g_6 $};
			\node[scale=0.7] at (5.5,0) (p6) {$ p_6 $};
			\node[scale=0.7] at (6,0) (g5) {$ g_5 $};
			\node[scale=0.7] at (6.5,0) (p5) {$ p_5 $};
			\node[scale=0.7] at (7,0) (g4) {$ g_4 $};
			\node[scale=0.7] at (7.5,0) (p4) {$ p_4 $};
			\node[scale=0.7] at (8,0) (g3) {$ g_3 $};
			\node[scale=0.7] at (8.5,0) (p3) {$ p_3 $};
			\node[scale=0.7] at (9,0) (g2) {$ g_2 $};
			\node[scale=0.7] at (9.5,0) (p2) {$ p_2 $};
			\node[scale=0.7] at (10,0) (g1) {$ g_1 $};
			\node[scale=0.7] at (10.5,0) (p1) {$ p_1 $};
			\node[scale=0.7] at (11,0) (g0) {$ g_0 $};
			\node[scale=0.7] at (11.5,0) (p0) {$ p_0 $};
			
			\node[fill=red, and gate US, draw, logic gate inputs=nn, rotate=270, thick, 	scale=0.4] at (0.5,-1.5) (a1) {};
			\node[fill=green, or gate US, draw, logic gate inputs=nn, rotate=270, thick, 		scale=0.4] at (0,-2) (o1) {};
			\node[fill=red, and gate US, draw, logic gate inputs=nn, rotate=270, thick, 		scale=0.4] at (1.5,-0.5) (a2) {};
			\node[fill=green, or gate US, draw, logic gate inputs=nn, rotate=270, thick, 		scale=0.4] at (1,-1) (o2) {};
			\node[fill=red, and gate US, draw, logic gate inputs=nn, rotate=270, thick, 		scale=0.4] at (3.5,-1.5) (a3) {};
			\node[fill=green, or gate US, draw, logic gate inputs=nn, rotate=270, thick, 		scale=0.4] at (3,-2) (o3) {};
			\node[fill=red, and gate US, draw, logic gate inputs=nn, rotate=270, thick, 		scale=0.4] at (4.5,-0.5) (a4) {};
			\node[fill=green, or gate US, draw, logic gate inputs=nn, rotate=270, thick, 		scale=0.4] at (4,-1) (o4) {};
			\node[fill=red, and gate US, draw, logic gate inputs=nn, rotate=270, thick, 		scale=0.4] at (6.5,-1.5) (a5) {};
			\node[fill=green, or gate US, draw, logic gate inputs=nn, rotate=270, thick, 		scale=0.4] at (6,-2) (o5) {};
			\node[fill=red, and gate US, draw, logic gate inputs=nn, rotate=270, thick, 		scale=0.4] at (7.5,-0.5) (a6) {};
			\node[fill=green, or gate US, draw, logic gate inputs=nn, rotate=270, thick, 		scale=0.4] at (7,-1) (o6) {};
			\node[fill=red, and gate US, draw, logic gate inputs=nn, rotate=270, thick, 		scale=0.4] at (9.5,-1.5) (a7) {};
			\node[fill=green, or gate US, draw, logic gate inputs=nn, rotate=270, thick, 		scale=0.4] at (9,-2) (o7) {};
			\node[fill=red, and gate US, draw, logic gate inputs=nn, rotate=270, thick, 		scale=0.4] at (10.5,-0.5) (a8) {};
			\node[fill=green, or gate US, draw, logic gate inputs=nn, rotate=270, thick, 		scale=0.4] at (10,-1) (o8) {};
			
			\node[fill=yellow, and gate US, draw, logic gate inputs=nn, rotate=270, 	thick, 	scale=0.4] at (1.5,-1) (y1) {};
			\node[fill=yellow, and gate US, draw, logic gate inputs=nn, rotate=270, 	thick, 	scale=0.4] at (2.25,-0.5) (y2) {};
			\node[fill=yellow, and gate US, draw, logic gate inputs=nn, rotate=270, 	thick, 	scale=0.4] at (4.5,-1) (y3) {};
			\node[fill=yellow, and gate US, draw, logic gate inputs=nn, rotate=270, 	thick, 	scale=0.4] at (5.25,-0.5) (y4) {};
			\node[fill=yellow, and gate US, draw, logic gate inputs=nn, rotate=270, 	thick, 	scale=0.4] at (7.5,-1) (y5) {};
			\node[fill=yellow, and gate US, draw, logic gate inputs=nn, rotate=270, 	thick, 	scale=0.4] at (8.25,-0.5) (y6) {};
			
			\node[fill=blue, or gate US, draw, logic gate inputs=nn, rotate=270, thick, 		scale=0.4] at (3.5,-5) (b1) {};
			\node[fill=blue, and gate US, draw, logic gate inputs=nn, rotate=270, thick, 		scale=0.4] at (4.5,-4.5) (b2) {};
			\node[fill=blue, or gate US, draw, logic gate inputs=nn, rotate=270, thick, 		scale=0.4] at (5.5,-4) (b3) {};
			\node[fill=blue, and gate US, draw, logic gate inputs=nn, rotate=270, thick, 		scale=0.4] at (6.5,-3.5) (b4) {};
			\node[fill=blue, or gate US, draw, logic gate inputs=nn, rotate=270, thick, 		scale=0.4] at (7.5,-3) (b5) {};
			\node[fill=blue, and gate US, draw, logic gate inputs=nn, rotate=270, thick, 		scale=0.4] at (8.5,-2.5) (b6) {};
			
			\node at (0,-2.5) (l1) {};
			\node at (3,-2.5) (l2) {};						
			\node at (6,-2.5) (l3) {};
			\node at (9,-2.5) (l4) {};
			\node at (1.5,-1.5) (l5) {};
			\node at (2.25,-1) (l6) {};
			\node at (2.5,-0.5) (l7) {};
			\node at (4.5,-1.5) (l8) {};
			\node at (5.25,-1) (l9) {};
			\node at (5.5,-0.5) (l10) {};						
			\node at (7.5,-1.5) (l11) {};
			\node at (8.25,-1) (l12) {};
			\node at (8.5,-0.5) (l13) {};
			\node at (3.5,-5.5) (l14) {};
			\node at (5.5,-4.5) (l15) {};
			\node at (7.5,-3.5) (l16) {};
			
			\node[scale=0.7] at (1.25,0.5) {$ P^{(3)} $};
			\node[scale=0.7] at (4.25,0.5) {$ P^{(2)} $};
			\node[scale=0.7] at (7.25,0.5) {$ P^{(1)} $};
			\node[scale=0.7] at (10.25,0.5) {$ P^{(0)} $};
			
			\draw[decorate,decoration={brace,raise=0.5ex}] (g11.north west) -- (p9.north east);
			\draw[decorate,decoration={brace,raise=0.5ex}] (g8.north west) -- (p6.north east);
			\draw[decorate,decoration={brace,raise=0.5ex}] (g5.north west) -- (p3.north east);
			\draw[decorate,decoration={brace,raise=0.5ex}] (g2.north west) -- (p0.north east);
			
			\draw[thick]  (g11)--(o1.input 2);
			\draw[thick]  (p11)--(a1.input 2);
			\draw[thick]  (p11)--(y1.input 2);
			\draw[thick]  (g10)--(o2.input 2);
			\draw[thick]  (p10)--(a2.input 2);
			\draw[thick]  (p10)--(y2.input 2);
			\draw[thick]  (g9)--(a2.input 1);
			\draw[thick]  (p9)--(y2.input 1);
			\draw[->]  (p9)--(l7);
			\draw[thick]  (g8)--(o3.input 2);
			\draw[thick]  (p8)--(a3.input 2);
			\draw[thick]  (p8)--(y3.input 2);
			\draw[thick]  (g7)--(o4.input 2);
			\draw[thick]  (p7)--(a4.input 2);
			\draw[thick]  (p7)--(y4.input 2);
			\draw[thick]  (g6)--(a4.input 1);
			\draw[thick]  (p6)--(y4.input 1);
			\draw[->]  (p6)--(l10);
			\draw[thick]  (g5)--(o5.input 2);
			\draw[thick]  (p5)--(a5.input 2);
			\draw[thick]  (p5)--(y5.input 2);
			\draw[thick]  (g4)--(o6.input 2);
			\draw[thick]  (p4)--(a6.input 2);
			\draw[thick]  (p4)--(y6.input 2);
			\draw[thick]  (g3)--(a6.input 1);
			\draw[thick]  (p3)--(y6.input 1);
			\draw[->]  (p3)--(l13);
			\draw[thick]  (g2)--(o7.input 2);
			\draw[thick]  (p2)--(a7.input 2);
			\draw[thick]  (g1)--(o8.input 2);
			\draw[thick]  (p1)--(a8.input 2);
			\draw[thick]  (g0)--(a8.input 1);
			
			\draw[thick]  (a2.output)--(o2.input 1);
			\draw[thick]  (y2.output)--(y1.input 1);
			\draw[->]  (y2)--(l6);
			\draw[thick]  (a4.output)--(o4.input 1);
			\draw[thick]  (y4.output)--(y3.input 1);
			\draw[->]  (y4)--(l9);
			\draw[thick]  (a6.output)--(o6.input 1);
			\draw[thick]  (y6.output)--(y5.input 1);
			\draw[->]  (y6)--(l12);
			\draw[thick]  (a8.output)--(o8.input 1);
			
			\draw[thick]  (o2.output)--(a1.input 1);
			\draw[->]  (y1)--(l5);
			\draw[thick]  (y1.output)--(b2.input 2);
			\draw[thick]  (o4.output)--(a3.input 1);
			\draw[->]  (y3)--(l8);
			\draw[thick]  (y3.output)--(b4.input 2);
			\draw[thick]  (o6.output)--(a5.input 1);
			\draw[->]  (y5)--(l11);
			\draw[thick]  (y5.output)--(b6.input 2);
			\draw[thick]  (o8.output)--(a7.input 1);
			
			\draw[thick]  (a1.output)--(o1.input 1);
			\draw[thick]  (a3.output)--(o3.input 1);
			\draw[thick]  (a5.output)--(o5.input 1);
			\draw[thick]  (a7.output)--(o7.input 1);
			
			\draw[->]  (o1)--(l1);
			\draw[thick]  (o1.output)--(b1.input 2);
			\draw[->]  (o3)--(l2);
			\draw[thick]  (o3.output)--(b3.input 2);
			\draw[->]  (o5)--(l3);
			\draw[thick]  (o5.output)--(b5.input 2);
			\draw[->]  (o7)--(l4);
			\draw[thick]  (o7.output)--(b6.input 1);
			
			\draw[thick]  (b6.output)--(b5.input 1);
			\draw[thick]  (b5.output)--(b4.input 1);
			\draw[->]  (b5)--(l16);
			\draw[thick]  (b4.output)--(b3.input 1);
			\draw[thick]  (b3.output)--(b2.input 1);
			\draw[->]  (b3)--(l15);
			\draw[thick]  (b2.output)--(b1.input 1);
			\draw[->]  (b1)--(l14);
		\end{tikzpicture}
		\caption[Visualization of the \textsc{And-Or} path circuit construction in Algorithm \ref{alg2}.]{Illustration of Lemma~\ref{AOPpathlem} for 4 input parts with 3 input pairs each.}
		\label{figlemandor}
	\end{figure}
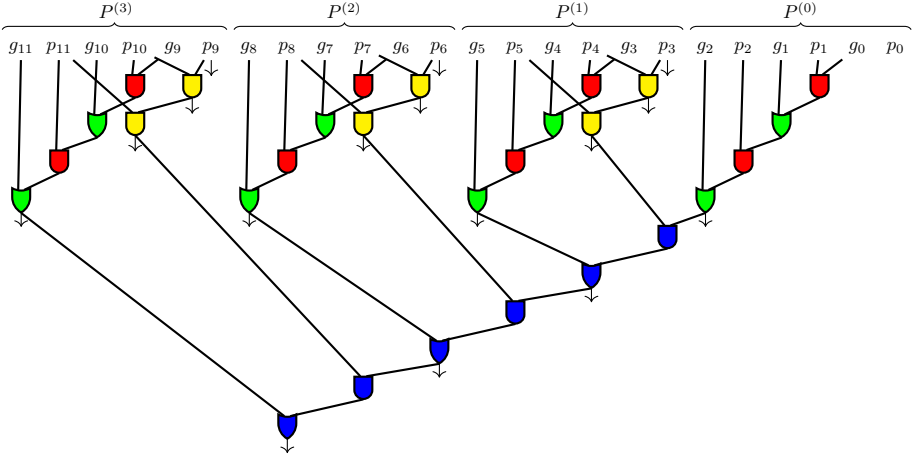

	Figure~\ref{figlemandor} shows a circuit resulting from Lemma~\ref{lemandor} for $ l=4 $ input parts of 3 input pairs each. The red and green gates represent the \textsc{And-Or} path circuits computed on each input part. The yellow gates depict the \textsc{And} function on the propagate signals computed on the parts $ P^{(1)},\dots,P^{(l-1)} $. The blue gates represent the $ \textsc{And-Or} $ path circuit whose inputs are the outputs of the other sub-circuits. 
	
	We can see that each \textsc{And-Or} path on the input pairs $ P^{(j-1)}\dplus\dots\dplus$ $ P^{(0)} $ for $ 1\leq j\leq l $ can be computed by taking the output after each blue \textsc{Or} gate. Hence, the \textsc{And-Or} path circuits $ AOP_{N_j} $ from Algorithm \ref{alg2} can be realized by computing an adder circuit on the input pairs
	\[\text{out}(AOP_{n_0}^{0}),\text{out}_{n_1}(S_{n_1}^{1}), \text{out}(AOP_{n_1}^{1}),\dots,\text{out}_{n_{l-2}}(S_{n_{l-2}}^{l-2}), \text{out}(AOP_{n_{l-2}}^{l-2}), \] where $ AOP_{n_j}^j $ is an \textsc{And-Or} path circuit computed on the input part $ P^{(j)} $ for $ 0\leq j\leq l-1 $. This idea is summarized in the next lemma.

\begin{lemma}[\cite{BS24} (Cor 4.6)] \label{AOPpathlem}
In addition to the input of Algorithm \ref{alg2}, let a family $ (B_k)_{k\in\mathbb{N}} $ of adder circuits be given. For each $ j\in\{0,\dots,l-1\} $, compute an \textsc{And-Or}-path circuit $ AOP_{n_j}^{j} $ on the input part $ P^{(j)} $. Compute an adder circuit $ B_l $ on the $ l $ input pairs 
		\[\mathrm{out}(AOP_{n_0}^{0}),\mathrm{out}_{n_1}(S_{n_1}^{1}), \mathrm{out}(AOP_{n_1}^{1}),\dots,\mathrm{out}_{n_{l-1}}(S_{n_{l-1}}^{l-1}), \mathrm{out}(AOP_{n_{l-1}}^{l-1}). \]
		Then, we have $ \mathrm{out}(AOP_{N_j}^{j})=\mathrm{out}_j(B_l) $ for all $ j\in\{1,\dots,l-1\} $.
	\end{lemma}
	
	\begin{proof}
		This follows directly from Lemma~\ref{lemandor}.
	\end{proof}

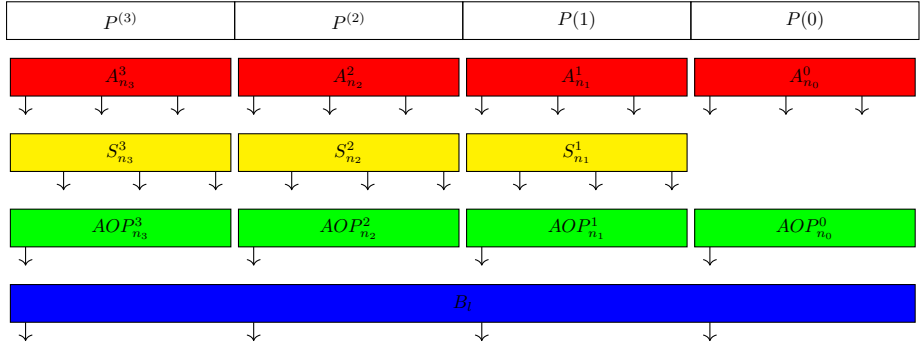
\begin{figure}[htb]
		\centering
		\begin{tikzpicture}[scale=0.5]
			
			\draw (-0.5,0.5) rectangle (5.5,-0.5);
			\draw (5.5,0.5) rectangle (11.5,-0.5);
			\draw (11.5,0.5) rectangle (17.5,-0.5);
			\draw (17.5,0.5) rectangle (23.5,-0.5);
			
			\draw[fill=red] (-0.4,-1) rectangle (5.4,-2);
			\draw[fill=red] (5.6,-1) rectangle (11.4,-2);
			\draw[fill=red] (11.6,-1) rectangle (17.4,-2);
			\draw[fill=red] (17.6,-1) rectangle (23.4,-2);
			
			\draw[fill=yellow] (-0.4,-3) rectangle (5.4,-4);
			\draw[fill=yellow] (5.6,-3) rectangle (11.4,-4);
			\draw[fill=yellow] (11.6,-3) rectangle (17.4,-4);
			
			\draw[fill=green] (-0.4,-5) rectangle (5.4,-6);
			\draw[fill=green] (5.6,-5) rectangle (11.4,-6);
			\draw[fill=green] (11.6,-5) rectangle (17.4,-6);
			\draw[fill=green] (17.6,-5) rectangle (23.4,-6);
			
			\draw[fill=blue] (-0.4,-7) rectangle (23.4,-8);
			
			\node[scale=0.7] at (2.5,0) {$ P^{(3)} $};
			\node[scale=0.7] at (8.5,0) {$ P^{(2)} $};
			\node[scale=0.7] at (14.5,0) {$ P{(1)} $};
			\node[scale=0.7] at (20.5,0) {$ P{(0)} $};
			
			\node[scale=0.7] at (2.5,-1.5) {$ A_{n_3}^3 $};
			\node[scale=0.7] at (8.5,-1.5) {$ A_{n_2}^2 $};
			\node[scale=0.7] at (14.5,-1.5) {$ A_{n_1}^1 $};
			\node[scale=0.7] at (20.5,-1.5) {$ A_{n_0}^0 $};
			
			\node[scale=0.7] at (2.5,-3.5) {$ S_{n_3}^3 $};
			\node[scale=0.7] at (8.5,-3.5) {$ S_{n_2}^2 $};
			\node[scale=0.7] at (14.5,-3.5) {$ S_{n_1}^1 $};
			
			\node[scale=0.7] at (2.5,-5.5) {$ AOP_{n_3}^3 $};
			\node[scale=0.7] at (8.5,-5.5) {$ AOP_{n_2}^2 $};
			\node[scale=0.7] at (14.5,-5.5) {$ AOP_{n_1}^1 $};
			\node[scale=0.7] at (20.5,-5.5) {$ AOP_{n_0}^0 $};
			
			\node[scale=0.7] at (11.5,-7.5) {$ B_l $};
			
			\draw[->]  (0,-2)--(0,-2.5);
			\draw[->]  (2,-2)--(2,-2.5);
			\draw[->]  (4,-2)--(4,-2.5);
			\draw[->]  (6,-2)--(6,-2.5);
			\draw[->]  (8,-2)--(8,-2.5);
			\draw[->]  (10,-2)--(10,-2.5);			
			\draw[->]  (12,-2)--(12,-2.5);
			\draw[->]  (14,-2)--(14,-2.5);
			\draw[->]  (16,-2)--(16,-2.5);
			\draw[->]  (18,-2)--(18,-2.5);
			\draw[->]  (20,-2)--(20,-2.5);
			\draw[->]  (22,-2)--(22,-2.5);
			
			\draw[->]  (1,-4)--(1,-4.5);
			\draw[->]  (3,-4)--(3,-4.5);
			\draw[->]  (5,-4)--(5,-4.5);
			\draw[->]  (7,-4)--(7,-4.5);
			\draw[->]  (9,-4)--(9,-4.5);
			\draw[->]  (11,-4)--(11,-4.5);
			\draw[->]  (13,-4)--(13,-4.5);
			\draw[->]  (15,-4)--(15,-4.5);
			\draw[->]  (17,-4)--(17,-4.5);
			
			\draw[->]  (0,-6)--(0,-6.5);
			\draw[->]  (6,-6)--(6,-6.5);
			\draw[->]  (12,-6)--(12,-6.5);
			\draw[->]  (18,-6)--(18,-6.5);
			
			\draw[->]  (0,-8)--(0,-8.5);
			\draw[->]  (6,-8)--(6,-8.5);
			\draw[->]  (12,-8)--(12,-8.5);
			\draw[->]  (18,-8)--(18,-8.5);
		\end{tikzpicture}
		\caption[Visualization of the computed sub-circuits during Algorithm~\ref{alg2}.]{Visualization of the computed sub-circuits during Algorithm \ref{alg2} combined with Lemma~\ref{AOPpathlem} for 4 input parts.}
		\label{figalg2}
	\end{figure}

	From now on, we will always realize the \text{And-Or} path circuits $ AOP_{N_j} $ from line 13 of Algorithm \ref{alg2} as indicated by Lemma~\ref{AOPpathlem} which proves to be way more efficient than computing them separately. Hence, instead of computing sub-circuits $ AOP_{N_j}^j $ for all $ 1\leq j\leq l-1 $, we will compute\vspace{-3pt} sub-circuits $ AOP_{n_j}^j $ on each input part $ P^{(j)} $ as well as an additional adder $ B_l $. This is the adder whose size we are going to reduce. Figure~\ref{figalg2} shows all sub-circuits computed during Algorithm \ref{alg2} using this subroutine.
	
	\begin{remark}
		We actually do not need to compute the circuit $ AOP_{n_{l-1}}^{(l-1)} $ in this procedure. In order to realize the circuits $ AOP_{N_j}^j $ for $ 1\leq j\leq l-1 $, it would suffice to compute an adder circuit $ B_{l-1} $ on the $ l-1 $ input pairs \[\text{out}(AOP_{n_0}^{0}),\text{out}_{n_1}(S_{n_1}^{1}), \text{out}(AOP_{n_1}^{1}),\dots,\text{out}_{n_{l-2}}(S_{n_{l-2}}^{l-2}), \text{out}(AOP_{n_{l-2}}^{l-2}). \]
		We include the sub-circuit $ AOP_{n_{l-1}}^{(l-1)} $ since it does not harm our analysis and simplifies notation.
	\end{remark}
	
	The next theorem formulates the delay and size bounds for the circuit resulting from Algorithm \ref{alg2} combined with Lemma~\ref{AOPpathlem} in terms of the delay and size of the computed sub-circuits.

	\begin{theorem}
		\label{thmalg2}
		Let $ n\in\mathbb{N} $ with $ n\geq 2 $, two families of adder circuits $ (A_k)_{k\in\mathbb{N}} $ and $ (B_l)_{l\in\mathbb{N}} $, a family of \textsc{And-Or}-path circuits $ (AOP_k)_{k\in\mathbb{N}} $ and a family of \textsc{And}-prefix circuits $ (S_k)_{k\in\mathbb{N}} $ be given. Using Lemma~\ref{AOPpathlem} for the computation of the \textsc{And-Or}-paths in line 13, Algorithm \ref{alg2} computes an adder circuit $ C_n $ on $ n $ input pairs with given arrival times with delay 
		\begin{align*}
			\mathrm{del}(C_n)\leq& \max\Big\{\mathrm{del}(A_{n_0}^{0}),\, \widetilde{\mathrm{del}}(B_l)+2,\, \max_{j\in\{1,\dots,l-1\}}\{\mathrm{del}(A_{n_j}^{j})+1, \mathrm{del}(S_{n_j}^{j})+2\}\Big\}
			\intertext{and size}
			\mathrm{s}(C_n)\leq& \,\mathrm{s}(A_{n_0}^{0})+\mathrm{s}(AOP_{n_{0}}^{0})+
			\sum_{j=1}^{l-1}\big(\mathrm{s}(A_{n_j}^{j})+\mathrm{s}(AOP_{n_{j}}^{j})+\mathrm{s}(S_{n_j}^{j})\big) + \mathrm{s}(B_l)+2n,
		\end{align*} 
		where $ \widetilde{\mathrm{del}}(B_l) $ is the delay of $ B_l $ computed on the input pairs \[\mathrm{out}(AOP_{n_0}^{0}),\mathrm{out}_{n_1}(S_{n_1}^{1}), \mathrm{out}(AOP_{n_1}^{1}),\dots,\mathrm{out}_{n_{l-1}}(S_{n_{l-1}}^{l-1}), \mathrm{out}(AOP_{n_{l-1}}^{l-1}).\]
	\end{theorem}

	\begin{proof}
The delay of $ C_n $ can be bounded by the maximum delay of any of the output signals. Therefore, we have 
		\begin{align*}
			\text{del}(C_n)\leq& \max\left\{\text{del}(A_{n_0}^{0}),\, \max_{j\in\{1,\dots,l-1\}}\left\{\text{del}(A_{n_j}^{j})+1, \text{del}(S_{n_j}^{j})+2, \text{del}(AOP_{N_j}^{j})+2\right\}\right\}.
		\end{align*} 
	 	Since we use Lemma~\ref{AOPpathlem} for the computation of the output of the circuits $ AOP_{N_j}^{j} $, we can substitute the quantities $ \text{del}(AOP_{N_j}^{j}) $ with $ \widetilde{\text{del}}(B_l) $ for all $ j\in\{1,\dots,l-1\} $ and achieve the desired delay bound.
	 	
	 	For the size bound, we add the sizes of all sub-circuits needed in our algorithm, including the circuits $ AOP_{n_j}^{j} $ and $ B_l $ introduced in Lemma~\ref{AOPpathlem}. 
Moreover, we need $\sum_{j=1}^{l-1}2n_j\leq 2n $ additional gates arising from line 15 of Algorithm \ref{alg2}. Combined we achieve the desired size bound.
	\end{proof}

	In the following, we explain why this construction can help to reduce the size of any adder circuit, in particular the circuit from Theorem~\ref{mainthm1}, while increasing the delay only by a small amount. As hinted at before, the circuit $ B_l $, introduced in Lemma~\ref{AOPpathlem}, is the adder whose size is going to be reduced. Note that $ B_l $ gets $ l=\lceil\frac{n}{k}\rceil $ input pairs whereas the other sub-circuits get $ k $ inputs or input pairs. If the size of $ B_l $ is bounded by $ l\sigma(l) $ for some sub-linear, increasing function $ \sigma $, then it would be linear in $ n $ if we chose $ k=\lceil\sigma(n)\rceil $ because in this case we would have $ l\sigma(l)\in\mathcal{O}(\frac{n}{k}\sigma(l))=\mathcal{O}(n) $. If we chose the other sub-circuits to be linear as well, then Theorem~\ref{thmalg2} implies that we would get a linear size circuit in the end. Hence, we get a general linearization framework for any adder circuit with sub-quadratic size, which we will describe more precisely in Section~\ref{linadder}. But first, we will reduce the size of our circuit from Theorem~\ref{mainthm1} by a factor of $ \log_2n $ to a size in $ \mathcal{O}(n\log_2n) $ in the next section.

\subsection{Smaller Adder Circuits}
	\label{smalleradder}
	
In this section, we are going to construct an adder circuit with a delay of at most 
$ \log_2W + 3\log_2\log_2n + 3\log_2\log_2\log_2n + \text{const} $ and a size of $\mathcal{O}(n\log_2n)$ 
on $ n $ input pairs with weight $ W $. 
This is done by using the $ l $-part adder construction framework described in the previous section with the adder from Theorem~\ref{mainthm1} as the adder $ B_l $ from Lemma~\ref{AOPpathlem}. In order to reduce the size of this adder by a factor of $ \log_2n $, we choose $ k=\lceil\log_2\rceil $ as the size of the input parts.
Before we specifically compute the smaller adder in Section~\ref{sizenlogn}, we provide a general construction which reduces the size of any circuit with size $ n\log_2n\sigma(n) $ by the factor $ \sigma(n) $ while increasing its delay by a summand of order $ \mathcal{O}(\log_2\log_2\sigma(n)) $ for some sub-linear, increasing function $ \sigma $.
\subsubsection{General Construction}
	\label{genconstr1}
\linebox{
	\begin{theorem}
		\label{mainthm2}
		Let $ l\in\mathbb{N} $ with $ l\geq 2 $. Assume that for any given input pairs $ p_0, g_0, \dots, p_{l-1},g_{l-1} $ with arrival times $ a:\{p_0, g_0\dots p_{l-1}, g_{l-1}\}\to \mathbb{N} $ and weight $ W_l $, we can construct an adder circuit $ B_l $ with delay 
		\begin{align*}
			\mathrm{del}(B_l)&\in [\log_2W_l+\delta(l)-1,\log_2W_l+\delta(l)] \intertext{and size} 
			\mathrm{s}(B_l)&\leq l\log_2(l)\sigma(l)
		\end{align*} for all $ l\in\mathbb{N} $, where $ \delta,\sigma:\mathbb{N}\to \mathbb{R}_{\geq 0} $ are monotonely increasing functions. Then, for any $ n\in\mathbb{N} $ and input pairs $ p_0, g_0, \dots, p_{n-1},g_{n-1} $ with arrival times $ a:\{p_0, g_0\dots p_{n-1}, g_{n-1}\}\to \mathbb{N} $ and weight $ W $, there is an adder circuit $ C_n $ with a size in $ \mathcal{O}(n\log_2n) $ and delay
		\begin{align*}
			\mathrm{del}(C_n)\leq&\max\{\log_2W+\log_2\sigma(n)+\log_2\log_2\sigma(n),\, \mathrm{del}(B_n)+3\log_2\log_2\sigma(n)\}+ const.
		\end{align*}
	\end{theorem}
}

	\begin{remark}
		It is reasonable to assume that there is a non-negative and monotonely increasing function $ \delta $ such that $ \text{del}(B_l)\in [\log_2W_l+\delta(l)-1,\log_2W_l+\delta(l)] $ since $ \text{del}(B_l)\geq \log_2W_l-1 $ for any given arrival times due to Proposition~\ref{proplb}. Furthermore, the delay of a circuit can artificially be increased by adding unnecessary gates.
	\end{remark}

	\begin{proof}
		We apply Algorithm \ref{alg2} combined with Lemma~\ref{AOPpathlem} to our inputs with the given adder family $ (B_l)_{l\in\mathbb{N}} $ and the following other families of sub-circuits:
\begin{itemize}
   \item For the family of adder circuits $ (A_k)_{k\in\mathbb{N}} $, we use the 
depth-optimizing circuits from Theorem~\ref{daddercirc}, with bounds
			\setlength{\mywidth}{\widthof{$\stackrel{Thm. \ref{cordaddercirclin}~(a)..}{\leq}$}}
			\begin{align}
				\text{del}(A_k)\klgleich[Cor. \ref{cordaddercirclin}~(a)]&\log_2W + \log_2k + \log_2\log_2k + 2.65,\label{eq29}\\
				\text{s}(A_k)\klgleich&6.2k\log_2k.\nonumber
			\end{align}
\item We use the \textsc{And-Or}-path circuits $ (AOP_k)_{k\in\mathbb{N}} $ 
from Theorem~\ref{andorcirc}. On $ k $ input pairs the circuit has less than $ 2k $ inputs which leads to the delay and size bounds
\begin{align}
	\text{del}(AOP_k)\leq\,&\log_2W+\log_2\log_2(2k)+\log_2\log_2\log_2(2k) + 4.3,\label{eq30}\\
	\text{s}(AOP_k)\leq\,& 2k\log_2(2k)+2k\log_2\log_2(2k)+ 2k\log_2\log_2\log_2(2k) + 6.6k-1.\nonumber
\end{align}
			\item For the family of \textsc{And}-prefix circuits $ (S_k)_{k\in\mathbb{N}} $ we use the circuits by Rautenbach, Szegedy and Werber (Theorem~\ref{andprecirc}) which guarantee the delay and size bounds 
			\begin{align}
				\text{del}(S_k)&\leq \log_2W+3\log_2\log_2k+ 0.007,\label{eq31}\\
				\text{s}(S_k)&\leq 3.114k\log_2\log_2k. \nonumber
			\end{align}
		\end{itemize}
		We choose $ k:=\lceil \sigma(n)\rceil $. Then, $ l=\lceil \frac{n}{k}\rceil\leq \lceil \frac{n}{\sigma(n)}\rceil $ which implies that
		\[\text{s}(B_l)\leq l\log_2(l)\sigma(l)\leq \Big\lceil \frac{n}{\sigma(n)}\Big\rceil \log_2(n) \sigma(n)\in\mathcal{O}(n\log_2n).\]
		Since the sizes of the sub-circuits $ A_k, AOP_k $ and $ S_k $ are in $ \mathcal{O}(k\log_2k) $, we have
		\setlength{\mywidth}{\widthof{$\stackrel{Thm. \ref{thmalg2}..}{\leq}$}} 
		\begin{align*}
			\text{s}(C_n)\klgleich[Thm. \ref{thmalg2}]&\text{s}(A_{n_0}^{0})+\text{s}(AOP_{n_{0}}^{0})+
			\sum_{j=1}^{l-1}\big(\text{s}(A_{n_j}^{j})+\text{s}(AOP_{n_{j}}^{j})+\text{s}(S_{n_j}^{j})\big) + \text{s}(B_l)+2n\\
			\inn[$ n_j\leq k $]& \mathcal{O}(l*k\log_2k+n\log_2n+n)\\
			\gleich &\mathcal{O}(n\log_2n).
		\end{align*}
	
		In order to bound the delay of $ C_n $, we first have to examine the weight $ \widetilde{W} $ of the input pairs \[\text{out}(AOP_{n_0}^{0}),\text{out}_{n_1}(S_{n_1}^{1}), \text{out}(AOP_{n_1}^{1}),\dots,\text{out}_{n_{l-2}}(S_{n_{l-2}}^{l-2}), \text{out}(AOP_{n_{l-2}}^{l-2}) \] which are the inputs of our given circuit $ B_l $. Their arrival times are bounded by the delay of the corresponding circuits. For the propagate signal with index $ 0 $ we can implicitly assume an arrival time of $ 0 $.
		
		\noindent We calculate
		\setlength{\mywidth}{\widthof{$\stackrel{k=\lceil \sigma(n)\rceil...}{\leq}$}} 
{\allowdisplaybreaks
\begin{align*}
   \widetilde{W}\klgleich& \sum_{j=1}^{l-1}\Big(2^{\text{del}(AOP_{n_j}^j)}+2^{\text{del}(S_{n_j}^j)}\Big)+2^{\text{del}(AOP_{n_0}^0)}+2^0\\
   \klgleich[\eqref{eq30},\eqref{eq31}]&\sum_{j=0}^{l-1}\left(2^{\log_2W_{P^{(j)}}+\log_2\log_2(2n_j)+\log_2\log_2\log_2(2n_j) + 4.3}\right)+
   \sum_{j=1}^{l-1}\left(2^{\log_2W_{P^{(j)}}+3\log_2\log_2n_j+ 0.007}\right) + 1\\
   \klgleich[$ n_j\leq k $]&\sum_{j=0}^{l-1}\left(W_{P^{(j)}}\log_2(2k)\log_2\log_2(2k) * 2^{4.3}\right) + \sum_{j=1}^{l-1}\left(W_{P^{(j)}}\log_2^3k* 2^{0.007}\right) + 1\\
   \klgleich& \sum_{j=0}^{l-1}\left(W_{P^{(j)}}\log_2(2k)\log_2\log_2(2k) * 2^{4.3}+ W_{P^{(j)}}\log_2^3k* 2^{0.007}\right)\\
   \klgleich[$ (*) $]& \sum_{j=0}^{l-1} const*W_{P^{(j)}}\log_2^3k\\
   \klgleich[$ k=\lceil \sigma(n)\rceil $]& const*W\log_2^3\sigma(n),
\end{align*}}\noindent
		where we used at $ (*) $ that $ \log_2^3k $ grows faster in $ k $ than $ \log_2(2k)\log_2\log_2(2k) $. For the delay of $ B_l $ we now get 
		\setlength{\mywidth}{\widthof{$\stackrel{l\leq n..}{\leq}$}} 
		\begin{align}
			\widetilde{\text{del}}(B_l)\klgleich &\log_2\widetilde{W}+\delta(l)\nonumber\\
			\klgleich &const+\log_2W+3\log_2\log_2\sigma(n)+\delta(l)\nonumber\\
			\klgleich[$ l\leq n $]&\log_2W+\delta(n)+3\log_2\log_2\sigma(n)+const\nonumber\\
			\klgleich& \text{del}(B_n)+3\log_2\log_2\sigma(n)+const. \label{eq32}
		\end{align}
		Finally, we can combine our delay bounds for the sub-circuits to get the overall delay bound
		\setlength{\mywidth}{\widthof{$\stackrel{del(B_n)\geq \log_2W....}{\leq}$}} 
{\allowdisplaybreaks
\begin{align*}
\text{del}(C_n)\klgleich[Thm. \ref{thmalg2}]
  &\max\Big\{\text{del}(A_{n_0}^{0}),\, \widetilde{\text{del}}(B_l)+2,\,  
  \max_{j\in\{1,\dots,l-1\}}\{\text{del}(A_{n_j}^{j})+1,\, \text{del}(S_{n_j}^{j})+2\}\Big\}\\
  \klgleich[$ \eqref{eq29},\eqref{eq31},\eqref{eq32} $]&\max\Big\{\log_2W_{P^{(0)}} + \log_2n_0 + \log_2\log_2n_0 + 2.65,\\ 
  &\phantom{\max\Big\{}\text{del}(B_n)+3\log_2\log_2\sigma(n)+const+2, \\
  &\phantom{\max\Big\{}\max_{j\in\{1,\dots,l-1\}}\{\log_2W_{P^{(j)}} + \log_2n_j + \log_2\log_2n_j + 3.65,\\
  &\phantom{\max\Big\{}\log_2W_{P^{(j)}}+3\log_2\log_2n_j+ 2.007\}\Big\}\\
  \klgleich[$ \substack{n_j\leq k,\\ W_{P^{(j)}}\leq W} $]&\max\{\log_2W + \log_2k + \log_2\log_2k,\, \text{del}(B_n)+3\log_2\log_2\sigma(n),\\
 & \phantom{\max\{}
 \log_2W+3\log_2\log_2k\}+const\\
  \klgleich[$ \substack{k=\lceil\sigma(n)\rceil,\\\text{del}(B_n)\geq \log_2W} $]&\max\{\log_2W + \log_2\sigma(n) + \log_2\log_2\sigma(n), \text{del}(B_n)+3\log_2\log_2\sigma(n)\}+const.
\end{align*}}\noindent
This concludes the proof of the theorem.
\end{proof}

\subsubsection{Size $ \mathcal{O}(n\log_2n) $ Adder Construction}
	\label{sizenlogn}
	
In this section, we will construct a new adder using  Theorem~\ref{mainthm2} and the adder from Theorem~\ref{mainthm1} as $ B_l $. Hence, we have $ \sigma(n)=2.422\log_2n $ and $ \delta(n)=3\log_2\log_2n+5.007 $ in Theorem~\ref{mainthm2}. Thus, we expect a delay increase of $ 3\log_2\log_2\log_2n $ plus some constant, compared to the adder from Theorem~\ref{mainthm1}, and a size of $ \mathcal{O}(n\log_2n) $. 
	
	Before we can prove the specific delay and size bounds, we first need some numerical lemmas. The first lemma is required for smaller instances for which we will not use the $ l $-part adder construction framework.
	
	\begin{lemma}
		\label{numlem3}
		Assume that $ n\in \mathbb{N} $ with $ 3\leq n\leq 2^{21} $. Then
		\[\log_2n\leq 2\log_2\log_2n + 3\log_2\log_2\log_2n + 5.811.\]
	\end{lemma}
	
	\begin{proof}
The statement can be easily verified for $n=3$. If $n \geq 4$ then both $\log_2\log_2n$
and $\log_2\log_2\log_2n$ are non-negative, so the statement is clear for $\log_2n \leq 5.811$.
If $5.811 < \log_2 n \leq 2^e \leq 6.6$ then the statement follows from
$2\log_2 \log_2(n) > 2 \log_2 (5.811) > 2$. Hence, assume $\log_2 n \geq 2^e$. 
Then, by \Cref{numlem1}, both $n \mapsto \frac{\log_2 \log_2 n}{\log_2 n}$ and 
$n \mapsto \frac{\log_2 \log_2 \log_2 n}{\log_2 \log_2 n}$ are decreasing.
Thus, we can conclude:
\begin{align*}
   2\log_2\log_2n + 3\log_2\log_2\log_2n 
      & \geq 2  \frac{\log_2 \log_2 (2^{21})}{\log_2(2^{21})} \log_2 n
           + 3  \frac{\log_2 \log_2 \log_2 (2^{21})}{\log_2 \log_2(2^{21})} \frac{\log_2 \log_2 (2^{21})}{\log_2(2^{21})} \log_2 n\\
      & = 2  \frac{\log_2 \log_2 (2^{21})}{\log_2(2^{21})} \log_2 n
           + 3  \frac{\log_2 \log_2 \log_2 (2^{21})}{\log_2(2^{21})} \log_2 n\\
      & \geq (2\cdot 0.20915 + 3\cdot 0.101666) \log_2 n
      \, \geq \, 0.72329 \log_2 n\\
      & = \log_2 n - 0.27671 \log_2 n
      \, \geq \, \log_2 n - 0.27671\cdot 21
        \geq \log_2 n -5.811 \tag*{\qedhere}
\end{align*}
	\end{proof}

	\begin{lemma}
		\label{numlem4}
		Assume that $ n,k\in \mathbb{N} $ with $ n>2^{21} $ and $ k=\lceil\log_2n\rceil $. Then, the following inequalities hold:
		\begin{enumerate}[(i)]
			\item $ 2^{4.3}\log_2(2k)\log_2\log_2(2k)\leq 2.97\log_2^3k $,
			\item $ \log_2k\leq 1.016\log_2\log_2n $,
			\item $ \log_2\log_2k\leq \log_2\log_2\log_2n+0.023 $,
			\item $ \log_2\log_2k\leq 0.492\log_2\log_2n $,
			\item $ \log_2(2k)+\log_2\log_2(2k)+\log_2\log_2\log_2(2k)+3.3\leq 2.851 \log_2\log_2n$.
		\end{enumerate}
	\end{lemma}

\begin{proof}
First, note that $ k\geq 22 $. 
\begin{enumerate}[(i)]
\item Using $ k\geq 22 $, we get
\setlength{\mywidth}{\widthof{$\stackrel{\text{Lem.\ }\ref{numlem1}}{\leq}$}} 
{\allowdisplaybreaks
\begin{align*}
\log_2(2k)\log_2\log_2(2k)&\stackrel{\text{Lem.\ }\ref{numlem1}}{\leq}
     \log_2(2k)\frac{\log_2 \log_2(2\cdot 22)}{\log_2(2\cdot 22)} \log_2(2k)
   \, = \, \frac{\log_2 \log_2(2\cdot22)}{\log_2(2\cdot22)}(\log_2(k)+1)^2\\
   & \klgleich \frac{\log_2 \log_2(2\cdot22)}{\log_2(2\cdot22)} \left(1+\frac{1}{\log_2(22)}\right)^2\log^2_2(k)\\
   & \klgleich \frac{\log_2 \log_2(2\cdot22)}{\log_2(2\cdot22)\log_2(22)}
            \left(1+\frac{1}{\log_2(22)}\right)^2\log^3_2(k) 
   \, \leq \, 0.15075 \log^3_2(k),
\end{align*}
}
and therefore 
$2^{4.3}\log_2(2k)\log_2\log_2(2k) \leq 2^{4.3} 0.15075 \log^3_2(k)\leq 2.97 \log^3_2(k)$

		
\item
For the second statement, we calculate 
\begin{align*}
\log_2k&\leq\log_2(\log_2n+1)\stackrel{\log_2n\geq 21}{\leq} \log_2\Big(\Big(1+\frac{1}{21}\Big)\log_2n\Big)
      \,\leq\, 0.068 + \log_2\log_2n\\
      &\leq \,\Big(1+\frac{0.068}{\log_221}\Big)\log_2\log_2n \,\leq\, 1.016\log_2\log_2n.
\end{align*}
	
\item
We use statement (ii) to prove the third statement. We have
\begin{equation*}
   \log_2\log_2k\stackrel{(ii)}{\leq}\log_2(1.016\log_2\log_2n)\leq \log_2\log_2\log_2n + 0.023.
\end{equation*}
	
\item		For the next statement, it suffices to show that $ \log_2\log_2k\leq 0.484\log_2k $. The statement then follows from (ii). Indeed, we have
		\begin{equation*}
			\log_2\log_2k\stackrel{\substack{\text{Lem. \ref{numlem1}}, \\k\geq 22}}{\leq}\frac{\log_2\log_222}{\log_222}\log_2k\leq 0.484\log_2k.
		\end{equation*}
		
\item		We prove the last statement by examining the expression term by term. We get
		\setlength{\mywidth}{\widthof{$\stackrel{Lem. \ref{numlem1}..}{\leq}$}} 
{\allowdisplaybreaks
		\begin{align}
			\log_2(2k)\klgleich[(ii)]&1.016\log_2\log_2n+1,\label{eqq35}\\
			\log_2\log_2(2k)\klgleich[$\substack{\text{Lem. \ref{numlem1},} \\ k\geq 22 }$]& \frac{\log_2\log_244}{\log_244} \log_2(2k)
			\stackrel{\eqref{eqq35}}{\leq} 0.456\log_2\log_2n+0.449, \label{eqq36}\\
			\log_2\log_2\log_2(2k)\klgleich[Lem. \ref{numlem1}]& 0.531\log_2\log_2(2k)
			\stackrel{\eqref{eqq36}}{\leq} 0.243\log_2\log_2n+0.239 \nonumber
		\end{align}
}
		Summing up the above inequalities results in the desired bound
		\setlength{\mywidth}{\widthof{$\stackrel{n\geq 2^{20}..}{\leq}$}} 
		\begin{align*}
			\log_2(2k)+\log_2\log_2(2k)+\log_2\log_2\log_2(2k)+3.3
			\klgleich& 1.715\log_2\log_2n + 4.988\\
			\klgleich[$ n\geq 2^{21} $]&\Big(1.715+\frac{4.988}{\log_221}\Big)\log_2\log_2n\\
			\klgleich& 2.851\log_2\log_2n.\tag*{\qedhere}
		\end{align*}
\end{enumerate}
	\end{proof}

	\begin{lemma}
		\label{numlem5}
		Assume that $ n,k,l\in\mathbb{N} $ with $ n>2^{21} $, $ k=\lceil\log_2n\rceil $ and $ l=\lceil\frac{n}{k}\rceil $. Then, we have
		\[2.422l\log_2^2l\leq 2.422n\log_2n-4.336n\log_2\log_2n.\]
	\end{lemma}

	\begin{proof}
		We start this proof by bounding $ l $ and $ \log_2l $ in terms of $ n $. We calculate
		\setlength{\mywidth}{\widthof{$\stackrel{n\geq 2^{20}..}{\leq}$}}
{\allowdisplaybreaks
		\begin{align*}
			l\klgleich&\frac{n}{k}+1\leq \frac{n}{\log_2n}+1\stackrel{n\geq 2^{21}}{\leq}\Big(1+\frac{21}{2^{21}}\Big)\frac{n}{\log_2n}\leq 1.0001\frac{n}{\log_2n},\\
			\log_2l\klgleich&\log_2n-\log_2\log_2n+\log_21.0001
			\quad \leq \quad \log_2n-\log_2\log_2n+0.0002\\
			\klgleich[$ n\geq 2^{21} $]&\log_2n-\left(1-\frac{0.0002}{\log_221}\right)\log_2\log_2n \quad \leq \quad
			 \log_2n-0.9999\log_2\log_2n.
		\end{align*}
}
		Now, we use these inequalities to calculate
		\setlength{\mywidth}{\widthof{$\stackrel{Lem. \ref{numlem1},.}{\leq}$}}
		{\allowdisplaybreaks
		\begin{align*}
			2.422l\log_2^2l 
			\klgleich& 2.422\frac{n}{\log_2n}(\log_2n-0.9999\log_2\log_2n)^2\\
			\klgleich& 2.422\frac{n}{\log_2n}(\log_2^2n-1.9998\log_2n\log_2\log_2n+0.9999\log_2^2\log_2n)\\
			\klgleich[$ \substack{\text{Lem. \ref{numlem1},}\\ n\geq 2^{21}} $]&2.422\frac{n}{\log_2n}\left(\log_2^2n-\log_2n\log_2\log_2n\left(1.9998-\frac{0.9999\log_221}{21}\right)\right)\\
			\klgleich& 2.422\frac{n}{\log_2n}(\log_2^2n-1.7906\log_2n\log_2\log_2n)\\
			\klgleich& 2.422n\log_2n - 4.336n\log_2\log_2n.\tag*{\qedhere}
		\end{align*}}\noindent
	\end{proof}

	Now, we are ready to prove specific delay and size bounds for our second adder construction. 

\linebox{
	\begin{theorem}
		\label{mainthm3}
		Let $ n\in \mathbb{N} $ with $ n\geq 3 $ and input pairs $ p_0, g_0, \dots, p_{n-1},g_{n-1} $ with arrival times $ a:\{p_0, g_0,\dots ,p_{n-1}, g_{n-1}\}\to \mathbb{N} $ be given and let $ W = \big(\sum_{i=0}^{n-1}2^{a(p_{i})} + 2^{a(g_{i})}\big) $ be the weight of the input pairs. We can construct an adder circuit $ C_{n} $ on these input pairs with delay 
		\begin{align*}
			\mathrm{del}(C_{n})&\leq \log_2W + 3\log_2\log_2n + 3\log_2\log_2\log_2n + d_1  \intertext{and size} \mathrm{s}(C_n)&\leq d_2n\log_2n
		\end{align*} 
		in polynomial time where
$ d_1= 
   \begin{cases}
      8.461 &\text{if } n\leq 2^{21},\\
      9.067 &\text{otherwise,}
   \end{cases}$
and 
$ d_2= 
   \begin{cases}
      6.2 &\text{if } n\leq 2^{21},\\
      4.442 &\text{otherwise.}
\end{cases}$
	\end{theorem}
}

	\begin{proof}
First, assume that $ 3\leq n\leq 2097152=2^{21} $. In this case we use the depth-optimizing adder circuit from Theorem~\ref{daddercirc} as $ C_n $. It fulfills the desired bounds
		\setlength{\mywidth}{\widthof{$\stackrel{Thm. \ref{cordaddercirclin}~(a)..}{\leq}$}}
\begin{align*}
   \text{del}(C_n)\klgleich[Cor. \ref{cordaddercirclin}~(a)]&\log_2W + \log_2n + \log_2\log_2n + 2.65 \stackrel{\text{Lem. } \ref{numlem3}}{\leq}\log_2W + 3\log_2\log_2n + 3\log_2\log_2\log_2n + d_1
\end{align*}
and
\begin{align*}
			\text{s}(C_n)\quad \leq \quad &d_2n\log_2n
\end{align*}
	 	for $ d_1\geq 8.461 $ and $ d_2\geq 6.2 $.
	 	
	 	Hence, assume that $ n> 2^{21} $. Here, we apply Algorithm \ref{alg2} combined with Lemma~\ref{AOPpathlem} with the same families of sub-circuits used in the proof of Theorem~\ref{mainthm2}:
\begin{itemize}
   \item For the family of adder circuits $ (A_k)_{k\in\mathbb{N}} $ we again use the depth-optimizing circuits from Theorem~\ref{daddercirc}, with the delay and size bounds
	 		\setlength{\mywidth}{\widthof{$\stackrel{Thm. \ref{cordaddercirclin}~(a)..}{\leq}$}}
	 		\begin{align}
	 			\text{del}(A_k)\klgleich[Cor. \ref{cordaddercirclin}~(a)]&\log_2W + \log_2k + \log_2\log_2k + 2.65,\label{eq34}\\
	 			\text{s}(A_k)\klgleich&6.2k\log_2k.\label{eq35}
	 		\end{align}
\item The \textsc{And-Or}-path circuits from Theorem~\ref{andorcirc} 
are again used as the circuits $ (AOP_k)_{k\in\mathbb{N}} $ which gives us the delay and size bounds
	 		\begin{align}
	 			\text{del}(AOP_k)\leq\,&\log_2W+\log_2\log_2(2k)+\log_2\log_2\log_2(2k) + 4.3,\label{eq36}\\
	 			\text{s}(AOP_k)\leq\,& 2k\log_2(2k)+2k\log_2\log_2(2k)+
	 			2k\log_2\log_2\log_2(2k)+ 6.6k-1. \label{eq37}
	 		\end{align}
	 		\item For the family of \textsc{And}-prefix circuits $ (S_k)_{k\in\mathbb{N}} $ we again use the circuits by Rautenbach, Szegedy and Werber (Theorem~\ref{andprecirc}) which guarantee the delay and size bounds 
	 		\begin{align}
	 			\text{del}(S_k)&\leq \log_2W+3\log_2\log_2k+ 0.007,\label{eq38}\\
	 			\text{s}(S_k)&\leq 3.114k\log_2\log_2k. \label{eq39}
	 		\end{align}
 			\item For the family of adder circuits $ (B_l)_{l\in\mathbb{N}} $, introduced in Lemma~\ref{AOPpathlem}, we now use the adder circuits constructed in Theorem~\ref{mainthm1} with the respective constants given in Theorem~\ref{mainthm1}. We have the delay and size bounds
 			\begin{align}
 				\text{del}(B_l)&\leq \log_2W + 3\log_2\log_2l + 5.007,\label{eq40}\\ 
 				\text{s}(B_l)&\leq 2.422l\log_2^2l.\label{eq41}
 			\end{align}
	 	\end{itemize}
 		We choose $ k:=\lceil\log_2n\rceil $. Then $ k\geq 22 $ and $ l=\lceil\frac{n}{k}\rceil\leq\lceil\frac{n}{\log_2n}\rceil  $.
 		
 		First, we bound the delay of $ C_n $. We start by bounding the weight $ \widetilde{W} $ of the input pairs \[\text{out}(AOP_{n_0}^{0}),\text{out}_{n_1}(S_{n_1}^{1}), \text{out}(AOP_{n_1}^{1}),\dots,\text{out}_{n_{l-2}}(S_{n_{l-2}}^{l-2}), \text{out}(AOP_{n_{l-2}}^{l-2}) \] via a similar calculation as in the proof of Theorem~\ref{mainthm2}. We have 
 		\setlength{\mywidth}{\widthof{$\stackrel{Lem. \ref{numlem4} (i)..}{\leq}$}} 
{\allowdisplaybreaks
\begin{align}
   \widetilde{W}\klgleich& \sum_{j=1}^{l-1}\left(2^{\text{del}(AOP_{n_j}^j)}+2^{\text{del}(S_{n_j}^j)}\right)+2^{\text{del}(AOP_{n_0}^0)}+2^0\nonumber\\
   \klgleich[\eqref{eq36},\eqref{eq38}]&\sum_{j=0}^{l-1}\left(2^{\log_2W_{P^{(j)}}+\log_2\log_2(2n_j)+\log_2\log_2\log_2(2n_j) + 4.3} \right)+
   \sum_{j=1}^{l-1}\left( 2^{\log_2W_{P^{(j)}}+3\log_2\log_2n_j+ 0.007}\right) + 1\nonumber\\
   \klgleich[$ n_j\leq k $]&\sum_{j=0}^{l-1}\left(W_{P^{(j)}}\log_2(2k)\log_2\log_2(2k) * 2^{4.3}+ W_{P^{(j)}}\log_2^3k* 2^{0.007}\right)\nonumber\\
   \klgleich[Lem. \ref{numlem4} (i)]& \sum_{j=0}^{l-1}\big(W_{P^{(j)}}\log_2^3k*2.97 + W_{P^{(j)}}\log_2^3k* 2^{0.007}\big)\nonumber\\
   \klgleich& \sum_{j=0}^{l-1} 3.975*W_{P^{(j)}}\log_2^3k\nonumber\\
   \klgleich& 3.975W\log_2^3k.\label{eq::bound_W_tilde_1}
\end{align}}\noindent
For the delay of $ B_l $ on the inputs with weight $ \widetilde{W} $ we get 
\setlength{\mywidth}{\widthof{$\stackrel{l\leq n..}{\leq}$}} 
\begin{align}
   \widetilde{\text{del}}(B_l)\klgleich[\eqref{eq40}]&\log_2\widetilde{W}+3\log_2\log_2l+5.007\nonumber\\
   \klgleich[$ \substack{l\leq n,\\ (\ref{eq::bound_W_tilde_1})} $]&\log_2W + 3\log_2\log_2n + 3\log_2\log_2k + 6.998.\label{eq42}
\end{align}
 		Now, we note that 
 		\setlength{\mywidth}{\widthof{$\stackrel{W_{P^{(j)}}\leq W}{\leq}$}} 
 		\begin{align}
 			\text{del}(A_{n_j}^{j})\klgleich[\eqref{eq34}]&\log_2W_{P^{(j)}} + \log_2n_j + \log_2\log_2n_j + 2.65\nonumber\\
 			\klgleich[$ \substack{n_j\leq k,\\ W_{P^{(j)}}\leq W} $]&\log_2W + \log_2k + \log_2\log_2k + 2.65\nonumber\\
 			\klgleich&\log_2W + 3\log_2\log_2n + 3\log_2\log_2k + 6.998 \label{eq43}
 		\end{align}
 		for $ j\in\{0,\dots,l-1\} $ and 
 		\setlength{\mywidth}{\widthof{$\stackrel{W_{P^{(j)}}\leq W}{\leq}$}} 
 		\begin{align}
 			\text{del}(S_{n_j}^{j})\klgleich[\eqref{eq38}]&\log_2W_{P^{(j)}} + 3\log_2\log_2n_j + 0.007\nonumber\\
 			\klgleich[$ \substack{n_j\leq k,\\ W_{P^{(j)}}\leq W} $]&\log_2W + 3\log_2\log_2k + 0.007\nonumber\\
 			\klgleich&\log_2W + 3\log_2\log_2n + 3\log_2\log_2k + 6.998 \label{eq44}
 		\end{align}
 		for $ j\in\{1,\dots,l-1\} $. Finally, we use Theorem~\ref{thmalg2} to derive the total delay bound 
 		\setlength{\mywidth}{\widthof{$\stackrel{\eqref{eq42},\eqref{eq43},\eqref{eq44}......}{\leq}$}} 
\begin{align*}
 \text{del}(C_n)\klgleich&\max\Big\{\text{del}(A_{n_0}^{0}),\, \widetilde{\text{del}}(B_l)+2,\, \max_{j\in\{1,\dots,l-1\}}\{\text{del}(A_{n_j}^{j})+1, \text{del}(S_{n_j}^{j})+2\}\Big\}\\
 \klgleich[\eqref{eq42},\eqref{eq43},\eqref{eq44}]&\log_2W + 3\log_2\log_2n + 3\log_2\log_2k + 8.998\\
 \klgleich[Lem. \ref{numlem4} (iii)]&\log_2W + 3\log_2\log_2n + 3\log_2\log_2\log_2n + d_1
\end{align*}
 		for $ d_1\geq 9.067 $.
 		
 		Next, we compute the size of the circuit $ C_n $ by first adding the sizes of our constructed sub-circuits separately, starting with the adder circuits $ A_{n_j}^{(j)} $. We get 
 		\setlength{\mywidth}{\widthof{$\stackrel{\eqref{eq35}..}{\leq}$}} 
 		\begin{align}
 			\sum_{j=0}^{l-1}\text{s}(A_{n_j}^{j})\klgleich[\eqref{eq35}]& \sum_{j=0}^{l-1} 6.2n_j\log_2n_j\stackrel{n_j\leq k}{\leq}\sum_{j=0}^{l-1} 6.2n_j\log_2k
 			\gleich  6.2n\log_2k\stackrel{\text{Lem. \ref{numlem4} (ii)}}{\leq} 6.3n\log_2\log_2n. \label{eq48}
 		\end{align}
 		Summing up the sizes of the \textsc{And}-prefix circuits $ S_{n_j}^{(j)} $ results in 
 		\setlength{\mywidth}{\widthof{$\stackrel{\eqref{eq39}..}{\leq}$}} 
 		\begin{align}
 			\sum_{j=1}^{l-1}\text{s}(S_{n_j}^{j})\klgleich[\eqref{eq39}]& \sum_{j=1}^{l-1} 3.114n_j\log_2\log_2n_j\stackrel{n_j\leq k}{\leq}\sum_{j=1}^{l-1} 3.114n_j\log_2\log_2k\nonumber\\
 			\gleich& 3.114n\log_2\log_2k\stackrel{\text{Lem. \ref{numlem4} (iv)}}{\leq} 1.533n\log_2\log_2n. \label{eq49}
 		\end{align}
 		The combined sizes of the \textsc{And-Or}-path circuits $ AOP_{n_j}^{(j)} $ are bounded by 
 		\setlength{\mywidth}{\widthof{$\stackrel{Lem. \ref{numlem4} (v)...}{\leq}$}} 
 		{\allowdisplaybreaks
 		\begin{align}
 			\sum_{j=0}^{l-1}\text{s}(AOP_{n_j}^{j})
 			\klgleich[\eqref{eq37}]& \sum_{j=0}^{l-1} \Big(2n_j\log_2(2n_j)+2n_j\log_2\log_2(2n_j)\:+
 			2n_j\log_2\log_2\log_2(2n_j)+ 6.6n_j-1\Big)\nonumber\\
 			\klgleich[$ n_j\leq k $]& \sum_{j=0}^{l-1} 2n_j\Big(\log_2(2k)+\log_2\log_2(2k)+ \log_2\log_2\log_2(2k)+3.3\Big)\nonumber\\
 			\gleich& 2n\Big(\log_2(2k)+\log_2\log_2(2k)+ \log_2\log_2\log_2(2k)+3.3\Big)\nonumber\\
 			\klgleich[Lem. \ref{numlem4} (v)]& 5.702n\log_2\log_2n. \label{eq50}
 		\end{align}}\noindent
 		Finally, the size of the adder circuit $ B_l $ is bounded by 
 		\begin{equation}
 			\text{s}(B_l)\stackrel{\eqref{eq41}}{\leq}2.422l\log_2^2l\stackrel{\text{Lem. \ref{numlem5}}}{\leq}2.422n\log_2n-4.336n\log_2\log_2n.\label{eq51}
 		\end{equation}
 		Now, Theorem~\ref{thmalg2} implies the total size bound 
 		\setlength{\mywidth}{\widthof{$\stackrel{Lem. \ref{numlem1},.}{\leq}$}}
 		{\allowdisplaybreaks
 		\begin{align*}
 			\text{s}(C_n)\klgleich& \text{s}(A_{n_0}^{0})+\text{s}(AOP_{n_{0}}^{0})\,+\,
 			 \sum_{j=1}^{l-1}\big(\text{s}(A_{n_j}^{j})+\text{s}(AOP_{n_{j}}^{j})+\text{s}(S_{n_j}^{(j)})\big) + \text{s}(B_l)+2n\\
 			\klgleich[$ \substack{\eqref{eq48},\eqref{eq49},\\ \eqref{eq50},\eqref{eq51}} $]& 2.422n\log_2n + 9.199n\log_2\log_2n +2n\\
 			\klgleich[$ \substack{\text{Lem. \ref{numlem1},}\\n\geq 2^{21}} $]& n\log_2n\Big(2.422+\frac{9.199\log_221}{21}+\frac{2}{21}\Big)\\
 			\klgleich& d_2 n\log_2n
 		\end{align*}}\noindent
 		for $ d_2\geq 4.442 $.
 
 In summary this leads to the delay and size bounds for the circuit $ C_n $ 
stated in the Theorem.
 	
 		The only thing left to prove is the runtime. It suffices to consider the case where $ n>2^{21} $. Here, we use Algorithm \ref{alg2} complemented with Lemma~\ref{AOPpathlem} where we compute a polynomial amount of sub-circuits which all have a polynomial runtime (see Theorem~\ref{daddercirc}, Theorem~\ref{andorcirc}, Theorem~\ref{andprecirc}, Theorem~\ref{mainthm1}). The rest of the algorithm is also done in polynomial runtime which leads to a polynomial runtime overall.
	\end{proof}



\begin{remark}
Remember that the circuit which we used as $ B_l $ in this construction has a fanout in $ \mathcal{O}(\sqrt{l}) $ (see the remark at the end of Section~\ref{deloptadder}). Since we chose $ k= \lceil\log_2n\rceil $ in Algorithm \ref{alg2}, the newly constructed adder has a fanout in $ \mathcal{O}\Big(\sqrt{\frac{n}{\log_2n}}\Big) $.
\end{remark}

\subsection{Linear Size Adder Circuits}
	\label{linadder}
	
Now, we further reduce the size of our circuits and finally achieve a linear size adder.
Note that Theorem~\ref{mainthm2} holds analogously if we drop the factor of $ \log_2l $ in the size of the circuit $ B_l $. 
In this case, the size of the resulting circuit would be in $ \mathcal{O}(n\log_2\sigma(n)) $ instead of $ \mathcal{O}(n\log_2n) $ and its delay bound would only differ in the additive constant. Hence, we can expect to construct a circuit with a similar delay bound as the circuit in Theorem~\ref{mainthm3} but with a size in $ \mathcal{O}(n\log_2\log_2n) $. 
This is done by using the same construction as in Theorem~\ref{mainthm3} but with $ k\in\mathcal{O}(\log_2^2n) $ instead of $ k\in\mathcal{O}(\log_2n) $. 
We elaborate this idea in Section~\ref{genconstr2}. 
Then, we will construct a circuit on $ n $ input pairs with weight $ W $ with a delay of at most $ \log_2W + 3\log_2\log_2n + 3\log_2\log_2\log_2n + \text{const} $ 
and a size of $\mathcal{O}(n\log_2\log_2n)$ in Section~\ref{sizenloglogn}. 
In Section~\ref{linsizeadderconstr}, we apply the linearization framework to the new adder 
with size in $ \mathcal{O}(n\log_2\log_2n) $ to achieve a linear size adder with a delay of at most $ \log_2W + 3\log_2\log_2n + 4\log_2\log_2\log_2n + \text{const}$.
	
\subsubsection{General Construction}
	\label{genconstr2}

\linebox{
	\begin{theorem}
		\label{mainthm4}
		Let $ l\in\mathbb{N} $ with $ l\geq 2 $. Assume that for any given input pairs $ p_0, g_0, \dots, p_{l-1},g_{l-1} $ with arrival times $ a:\{p_0, g_0\dots p_{l-1}, g_{l-1}\}\to \mathbb{N} $ and weight $ W_l $, we can construct an adder circuit $ B_l $ with delay 
		\begin{align*}
			\mathrm{del}(B_l)&\in [\log_2W_l+\delta(l)-1,\log_2W_l+\delta(l)] \intertext{and size} 
			\mathrm{s}(B_l)&\leq l\sigma(l)
		\end{align*} for all $ l\in\mathbb{N} $, where $ \delta,\sigma:\mathbb{N}\to \mathbb{R}_{\geq 0} $ are monotonely increasing functions. Then, for any $ n\in\mathbb{N} $ and input pairs $ p_0, g_0, \dots, p_{n-1},g_{n-1} $ with arrival times $ a:\{p_0, g_0\dots p_{n-1}, g_{n-1}\}\to \mathbb{N} $ and weight $ W $, we can construct the following two circuits: 
\begin{enumerate}
   \item There is an adder circuit $ C_n^{(1)} $ with size in $ \mathcal{O}(n\log_2\sigma(n)) $ and delay
      \begin{align*}
				\mathrm{del}(C_n^{(1)})\leq\,&\max\{\log_2W+\log_2\sigma(n)+\log_2\log_2\sigma(n), \mathrm{del}(B_n)+3\log_2\log_2\sigma(n)\}+ const.
      \end{align*}
   \item There is an adder circuit $ C_n^{(2)} $ with size in $ \mathcal{O}(n) $ and delay
      \begin{align*}
         \mathrm{del}(C_n^{(2)})\leq\,&\max\{\log_2W + \log_2\sigma(n) + \log_2\log_2\sigma(n)\:+\log_2\log_2\log_2\sigma(n),\\&
         \phantom{\text{max}\{ }\mathrm{del}(B_n)+ \log_2\sigma(n)\}+const.
      \end{align*}
\end{enumerate}
	\end{theorem}
}

	\begin{proof}
		The proof is similar to the proof of Theorem~\ref{mainthm2}. In both cases, we apply Algorithm \ref{alg2} combined with Lemma~\ref{AOPpathlem} to our inputs with the given adder family $ (B_l)_{l\in\mathbb{N}} $ and we choose $ k=\lceil\sigma(n)\rceil $ in Algorithm \ref{alg2}. For $ C_n^{(1)} $ we even choose the same other sub-circuits as in the proof of Theorem~\ref{mainthm2}. The proof of the size and delay bounds then work analogously to Theorem~\ref{mainthm2}. The only difference is that the size of $ B_l $ is now bounded by \[\text{s}(B_l)\leq l\sigma(l)\overset{l=\lceil\frac{n}{k}\rceil}{\leq} \Big\lceil\frac{n}{\sigma(n)}\Big\rceil\sigma(n)\in\mathcal{O}(n).\] Consequently, the overall size of $ C_n^{(1)} $ is in $ \mathcal{O}(n\log_2\sigma(n)) $.
		
		We now focus on the second statement. In order to achieve a linear size for the circuit $ C_n^{(2)} $, we need to choose linear sub-circuits in our construction:
\begin{itemize}
   \item For the family of adder circuits $ (A_k)_{k\in\mathbb{N}} $, we use the linear size, 
depth-optimizing circuits from Theorem~\ref{daddercirclin}, with the delay and size bounds
			\setlength{\mywidth}{\widthof{$\stackrel{Cor. \ref{cordaddercirclin}~(c)..}{\leq}$}}
			\begin{align}
				\text{del}(A_k)\klgleich[Cor. \ref{cordaddercirclin}~(c)]&\log_2W +\log_2k + \log_2\log_2k + \log_2\log_2\log_2k+7.6,\label{eqq51}\\
				\text{s}(A_k)\klgleich&16.7k.\nonumber
			\end{align}
			\item For the \textsc{And-Or}-path circuits $ (AOP_k)_{k\in\mathbb{N}} $, we use the construction by Spirkl from Theorem~\ref{andorcircspir}. Since the circuit $ AOP_k $  has $ 2k-1 $ inputs, we get the delay and size bounds
			\begin{align}
				\text{del}(AOP_k)&\leq\log_2W+2\sqrt{2\log_2(2k)}+6,\label{eqq52}\\\text{s}(AOP_k)&\leq 6k.\nonumber
			\end{align}
			\item For the family of \textsc{And}-prefix circuits $ (S_k)_{k\in\mathbb{N}} $ we use the linear size, depth-optimizing circuits by Ladner and Fischer from Theorem~\ref{thmlf} with $ f=0 $. They guarantee the delay and size bounds 
			\setlength{\mywidth}{\widthof{$\stackrel{Thm. \ref{corthmall}~(a)..}{\leq}$}}
			\begin{align}
				\text{del}(S_k)\klgleich[Cor. \ref{corthmall}~(a)]& \log_2W+\lceil\log_2k\rceil,\label{eqq53}\\
				\text{s}(S_k)\klgleich &4k. \nonumber
			\end{align}
		\end{itemize}
		
		As in the first statement, the size of $ B_l $ is linear. Since the other sub-circuits are also linear, we get the linear overall size bound
		\setlength{\mywidth}{\widthof{$\stackrel{Thm. \ref{thmalg2}..}{\leq}$}} 
		\begin{align*}
			\text{s}(C_n^{(2)})\klgleich[Thm. \ref{thmalg2}]&\text{s}(A_{n_0}^{0})+\text{s}(AOP_{n_{0}}^{0})+
			\sum_{j=1}^{l-1}\big(\text{s}(A_{n_j}^{j})+\text{s}(AOP_{n_{j}}^{j})+\text{s}(S_{n_j}^{j})\big) + \text{s}(B_l)+2n
			\stackrel{n_j\leq k}{\in} \mathcal{O}(l*k+n)
			\,=\, \mathcal{O}(n).
		\end{align*}

		Next, we again have to compute the weight $ \widetilde{W} $ of the input pairs
		\[\text{out}(AOP_{n_0}^{0}),\text{out}_{n_1}(S_{n_1}^{1}), \text{out}(AOP_{n_1}^{1}),\dots,\text{out}_{n_{l-2}}(S_{n_{l-2}}^{l-2}), \text{out}(AOP_{n_{l-2}}^{l-2}) \] of the circuit $ B_l $ in order to compute the delay of $ B_l $. We have 
		\setlength{\mywidth}{\widthof{$\stackrel{k=\lceil \sigma(n)\rceil...}{\leq}$}} 
		{\allowdisplaybreaks
		\begin{align*}
			\widetilde{W}\klgleich& \sum_{j=1}^{l-1}\Big(2^{\text{del}(AOP_{n_j}^j)}+2^{\text{del}(S_{n_j}^j)}\Big)+2^{\text{del}(AOP_{n_0}^0)}+2^0\\
			\klgleich[\eqref{eqq52},\eqref{eqq53}]&\sum_{j=0}^{l-1}\left(2^{\log_2W_{P^{(j)}}+2\sqrt{2\log_2(2n_j)}+6}\right) + \sum_{j=1}^{l-1}\left(2^{\log_2W_{P^{(j)}}+\lceil\log_2n_j\rceil}\right)+ 1 \\
			\klgleich[$ n_j\leq k $]&\sum_{j=0}^{l-1}\Big(2^{\log_2W_{P^{(j)}}+2\sqrt{2\log_2(2k)}+6}+ 2^{\log_2W_{P^{(j)}}+\log_2k+1}\Big)\\
			\klgleich[$ (*) $]& \sum_{j=0}^{l-1}\big(W_{P^{(j)}}\big(2^{\log_2k+const}+2^{\log_2k+1}\big)\big)\\
			\klgleich& const*Wk\\
			\klgleich[$ k=\lceil \sigma(n)\rceil $]&const*W\sigma(n),
		\end{align*}}\noindent
		where we used at $ (*) $ that $ \log_2k $ grows faster in $ k $ than $ 2\sqrt{2\log_2(2k)} $. For the delay of $ B_l $ we get 
		\setlength{\mywidth}{\widthof{$\stackrel{l\leq n..}{\leq}$}} 
		\begin{align}
			\widetilde{\text{del}}(B_l)\klgleich &\log_2\widetilde{W}+\delta(l)\nonumber\\
			\klgleich &const+\log_2W+\log_2\sigma(n)+\delta(l)\nonumber\\
			\klgleich[$ l\leq n $]&\log_2W+\delta(n)+\log_2\sigma(n)+const\nonumber\\
			\klgleich& \text{del}(B_n)+\log_2\sigma(n)+const. \label{eqq54}
		\end{align}
		Again, we can use Theorem~\ref{thmalg2} to compute the overall delay bound
		\setlength{\mywidth}{\widthof{$\stackrel{\eqref{eqq51},\eqref{eqq53},\eqref{eqq54}.......\hspace{1.55pt}}{\leq}$}} 
{\allowdisplaybreaks
\begin{align*}
\text{del}(C_n^{(2)})\klgleich&\max\big\{\text{del}(A_{n_0}^{(0)}),\, \widetilde{\text{del}}(B_l)+2,\, \max_{j\in\{1,\dots,l-1\}}\{\text{del}(A_{n_j}^{(j)})+1, \text{del}(S_{n_j}^{(j)})+2\}\big\}\\
  \klgleich[$ \eqref{eqq51},\eqref{eqq53},\eqref{eqq54} $]&\max\big\{\log_2W_{P^{(0)}} + \log_2n_0 + \log_2\log_2n_0\: + \log_2\log_2\log_2n_0+ 7.6,\\
  &\phantom{\max\big\{} \text{del}(B_n)+\log_2\sigma(n)+const +2,\\
  &\phantom{\max\big\{}\max_{j\in\{1,\dots,l-1\}}\{\log_2W_{P^{(j)}} + \log_2n_j + \log_2\log_2n_j\\
  &\phantom{\max\big\{\max_{j\in\{1,\dots,l-1\}} }+\log_2\log_2\log_2n_j + 8.6,\log_2W_{P^{(j)}}+\lceil\log_2n_j\rceil \}\big\}\\
\klgleich[$ \substack{n_j\leq k,\\ W_{P^{(j)}}\leq W} $]&\max\{\log_2W + \log_2k + \log_2\log_2k+\log_2\log_2\log_2k, \\
  &\phantom{\max\big\{}\text{del}(B_n)+ \log_2\sigma(n), \log_2W+\log_2k\}+const\\
  \klgleich[$ k=\lceil\sigma(n)\rceil $]&\max\{\log_2W + \log_2\sigma(n) + \log_2\log_2\sigma(n)\:+\\
  &\phantom{\max\big\{}\log_2\log_2\log_2\sigma(n),
			\text{del}(B_n)+ \log_2\sigma(n)\}+const.\tag*{\qedhere}
\end{align*}}\noindent
	\end{proof}

Now, we could already use this linearization framework (Theorem~\ref{mainthm4}, part 2.)
on one of our previously constructed adders. Assume\vspace{-1pt} for example that we built an adder like $ C_n^{(2)} $ in the previous theorem using our adder with the delay bound $ \log_2W+3\log_2\log_2n+5.007 $ and size $\mathcal{O}(n\log_2^2n)$ (Theorem~\ref{mainthm1}) as $ B_l $. Then, we would get a linear size adder circuit with a delay bound of 
$\log_2W+5\log_2\log_2n+\text{const}$ since $ \sigma(n)\in\mathcal{O}(\log_2^2n) $. 
If we used the adder from Theorem~\ref{mainthm3} as $ B_l $, we would get a linear size adder with a delay bound of $ \log_2W+4\log_2\log_2n+3\log_2\log_2\log_2n+\text{const}$ 
because, in this case, $ \sigma(n)\in\mathcal{O}(\log_2n) $. Both of these circuits asymptotically improve on the previously best known delay bound for linear size adder circuits. 
However, we can even get a linear size 
and a delay of at most $ \log_2W + 3\log_2\log_2n + 4\log_2\log_2\log_2n+\text{const}$
(see Section~\ref{linsizeadderconstr}).

\subsubsection{Size $ \mathcal{O}(n\log_2\log_2n) $ Adder Construction}	
	\label{sizenloglogn}
	
	Now, we want to compute the adder $ C_n^{(1)} $ as described in Theorem~\ref{mainthm4} with the adder from Theorem~\ref{mainthm1} as $ B_l $. Because the construction is very similar to the one executed in Section~\ref{sizenlogn}, the delay and size computations are also similar. Hence, some of the upcoming numerical lemmas may look familiar.
	
	\begin{lemma}
		\label{numlem51}
		Assume that $ n\in \mathbb{N} $ with $ 3\leq n\leq 2^{16} $. Then
		\[\log_2n\leq 2\log_2\log_2n + 2\log_2\log_2\log_2n + 4.\]
	\end{lemma}
	
	\begin{proof}
		This lemma is proven similarly to Lemma~\ref{numlem3}. 
The statement can be easily verified for $n=3$. If $n \geq 4$ then both $\log_2\log_2n$
and $\log_2\log_2\log_2n$ are non-negative, so the statement is clear for $\log_2n \leq 4$.
If $4 < \log_2 n \leq 2^e \leq 6.6$ then the statement follows from
$2\log_2 \log_2(n) > 2 \log_2(4) = 4$. Hence, assume $\log_2 n \geq 2^e$. 
Then, by \Cref{numlem1}, both $n \mapsto \frac{\log_2 \log_2 n}{\log_2 n}$ and 
$n \mapsto \frac{\log_2 \log_2 \log_2 n}{\log_2 \log_2 n}$ are decreasing.
Thus, we can conclude:
\begin{align*}
   2\log_2\log_2n + 3\log_2\log_2\log_2n 
      & \geq 2  \frac{\log_2 \log_2 (2^{16})}{\log_2(2^{16})} \log_2 n
           + 2  \frac{\log_2 \log_2 \log_2 (2^{16})}{\log_2(2^{16})} \log_2 n
      \, = \, \left(\frac{1}{2} + \frac{1}{4}\right) \log_2 n\\
      & = \log_2 n - 0.25 \log_2 n
      \, \geq \, \log_2 n - 0.25\cdot 16
        = \log_2 n -4 \tag*{\qedhere}
\end{align*}
	\end{proof}

	\begin{lemma}
		\label{numlem6}
		Assume that $ n,k\in \mathbb{N} $ with $ n>2^{16} $ and $ k=\lceil\log_2^2n\rceil $. Then, the following inequalities hold:
		\begin{enumerate}[(i)]
			\item $ 2^{4.3}\log_2(2k)\log_2\log_2(2k)\leq 1.097\log_2^3k $,
			\item $ \log_2k\leq 2.002\log_2\log_2n $,
			\item $ \log_2\log_2k\leq \log_2\log_2\log_2n+1.002 $,
			\item $ \log_2\log_2k\leq 0.751\log_2\log_2n $,
			\item $ \log_2(2k)+\log_2\log_2(2k)+\log_2\log_2\log_2(2k)+3.3\leq 4.288 \log_2\log_2n$.
		\end{enumerate}
	\end{lemma}

	\begin{proof}
		These statements look similar to the ones from Lemma~\ref{numlem4} and are proven in a similar way. First, we note that $ k\geq 257 $. 
\begin{enumerate}[(i)]
\item
We can simply replace 22 by 257 in the proof of Lemma~\ref{numlem4}~(i) and get
\begin{align*}
2^{4.3}\log_2(2k)\log_2\log_2(2k) \leq 
   2^{4.3} \frac{\log_2 \log_2(2\cdot257)}{\log_2(2\cdot257)\log_2(257)}
            \left(1+\frac{1}{\log_2(257)}\right)^2\log^3_2(k) \leq  1.097 \log^3_2(k)
\end{align*}

\item		For the next statement, we calculate
{\allowdisplaybreaks
		\begin{align*}
			\log_2k&\leq \log_2(\log_2^2n+1)\stackrel{\log_2^2n\geq 256}{\leq} \log_2\left(\left(1+\frac{1}{256}\right)\log_2^2n\right)
			\,\leq\, 0.006 + 2\log_2\log_2n \\ &\leq\, \left(1+\frac{0.006}{2\log_216}\right)2\log_2\log_2n \,\leq\, 2.002\log_2\log_2n.
		\end{align*}
}	
\item		Via statement (ii) we can now easily prove the third statement. We have 
		\[ \log_2\log_2k\stackrel{(ii)}{\leq}\log_2(2.002\log_2\log_2n)\leq \log_2\log_2\log_2n+1.002.\]
		
\item		The next statement is again easily proven with the help of statement (ii) and Lemma~\ref{numlem1}. We have
		\begin{equation*}
			\log_2\log_2k\stackrel{\substack{\text{Lem. \ref{numlem1}},\\ k\geq 257}}{\leq}\frac{\log_2\log_2257}{\log_2257}\log_2k\stackrel{(ii)}{\leq}0.751\log_2\log_2n.
		\end{equation*}
	
\item		The last statement is proven by examining the expression term by term. We get
		\setlength{\mywidth}{\widthof{$\stackrel{Lem. \ref{numlem1}..}{\leq}$}} 
		{\allowdisplaybreaks
		\begin{align}
			\log_2(2k)\klgleich[(ii)]&2.002\log_2\log_2n+1,\label{eqqq1}\\
			\log_2\log_2(2k)\klgleich[$\substack{\text{Lem. \ref{numlem1},} \\ k\geq 257 }$]& \frac{\log_2\log_2514}{\log_2514} \log_2(2k)
			\stackrel{\eqref{eqqq1}}{\leq} 0.705\log_2\log_2n+0.353, \label{eqqq2}\\
			\log_2\log_2\log_2(2k)\klgleich[$ \substack{\text{Lem. \ref{numlem1}},\\k\geq 257} $]& \frac{\log_2\log_2\log_2514}{\log_2\log_2514}\log_2\log_2(2k)
			\stackrel{\eqref{eqqq2}}{\leq}  0.371\log_2\log_2n+0.186. \nonumber
		\end{align}}\noindent
		Finally, we can sum up the above inequalities to achieve the desired bound
		\setlength{\mywidth}{\widthof{$\stackrel{n\geq 2^{20}..}{\leq}$}} 
		\begin{align*}
			\log_2(2k)+\log_2\log_2(2k)+\log_2\log_2\log_2(2k)+3.3
			\klgleich& 3.078\log_2\log_2n + 4.839\\
			\klgleich[$ n\geq 2^{16} $]&\Big(3.078+\frac{4.839}{\log_216}\Big)\log_2\log_2n\\
			\klgleich& 4.288\log_2\log_2n.\tag*{\qedhere}
		\end{align*}
\end{enumerate}
	\end{proof}

	\begin{lemma}
		\label{numlem7}
		Assume that $ n,k,l\in\mathbb{N} $ with $ n>2^{16} $, $ k=\lceil\log_2^2n\rceil $ and $ l=\lceil\frac{n}{k}\rceil $. Then, we have
		\[2.422l\log_2^2l\leq 2.432n.\]
	\end{lemma}

	\begin{proof}
		We first bound $ l $ in terms of $ n $. We have 
		\[l\leq \frac{n}{k}+1\leq \frac{n}{\log_2^2n}+1\stackrel{n\geq 2^{16}}{\leq}\Big(1+\frac{256}{2^{16}}\Big)\frac{n}{\log_2^2n}\leq 1.004\frac{n}{\log_2^2n}.\]
		Now, we use this bound to calculate
		\[2.422l\log_2^2l\leq 2.432\frac{n}{\log_2^2n}\log_2^2l\stackrel{l\leq n}{\leq}2.432n.\tag*{\qedhere}\]
	\end{proof}

	Next we can move on to our third specific adder construction.

\linebox{
	\begin{theorem}
		\label{mainthm5}
		Let $ n\in \mathbb{N} $ with $ n\geq 3 $ and input pairs $ p_0, g_0, \dots, p_{n-1},g_{n-1} $ with arrival times $ a:\{p_0, g_0,\dots ,p_{n-1}, g_{n-1}\}\to \mathbb{N} $ be given and let $ W = \big(\sum_{i=0}^{n-1}2^{a(p_{i})} + 2^{a(g_{i})}\big) $ be the weight of the input pairs. We can construct an adder circuit $ C_{n} $ on these input pairs with delay
		\begin{align*}
			\mathrm{del}(C_{n})&\leq \log_2W + 3\log_2\log_2n + 3\log_2\log_2\log_2n + e_1 \intertext{and size} \mathrm{s}(C_n)&\leq e_2n\log_2\log_2n
		\end{align*} 
		in polynomial time for $ e_1 = 11.085 $ and $ e_2 = 24.436$.
	\end{theorem}
}

	\begin{proof}
		First, we assume that $ 3\leq n\leq 65536=2^{16} $. In this case, we use the linear size, depth-optimizing adder circuit from Theorem~\ref{daddercirclin0}, as $ C_n $. We get the desired bounds
		\setlength{\mywidth}{\widthof{$\stackrel{Cor. \ref{cordaddercirclin}~(b).}{\leq}$}}
		\begin{align*}
			\text{del}(C_n)\klgleich[Cor. \ref{cordaddercirclin}~(b)]&\log_2W + \log_2n + \log_2\log_2n + \log_2\log_2\log_2n + 6.6\\
			\klgleich[Lem. \ref{numlem51}]&\log_2W + 3\log_2\log_2n + 3\log_2\log_2\log_2n + e_1
			\intertext{and}
			\text{s}(C_n)\klgleich&e_2n\log_2\log_2n
		\end{align*}
		for $ e_1\geq 10.6 $ and $ e_2\geq 21.6 $.
		
		Now, assume that $ n>2^{16} $. Here, we apply Algorithm \ref{alg2} combined with Lemma~\ref{AOPpathlem} with the same families of sub-circuits used in the proof of Theorem~\ref{mainthm3}.
Hence, the bounds (\ref{eq34}) to (\ref{eq41}) from the proof of Theorem~\ref{mainthm3}
are again valid for the subcircuits.

		As hinted at before, we now choose $ k:=\lceil\log_2^2n\rceil $ which implies $ k\stackrel{n>2^{15}}{\geq} 257 $ and $ l=\lceil\frac{n}{k}\rceil\leq \lceil\frac{n}{\log_2^2n}\rceil $.
		
		We proceed similarly to the proof of Theorem~\ref{mainthm3} by first bounding the weight $ \widetilde{W} $ of the input pairs 
		\[\text{out}(AOP_{n_0}^{0}),\text{out}_{n_1}(S_{n_1}^{1}), \text{out}(AOP_{n_1}^{1}),\dots,\text{out}_{n_{l-2}}(S_{n_{l-2}}^{l-2}), \text{out}(AOP_{n_{l-2}}^{l-2}) \] of $ B_l $. We get 
		\setlength{\mywidth}{\widthof{$\stackrel{Lem. \ref{numlem6} (i)..}{\leq}$}}
		{\allowdisplaybreaks 
\begin{align}
			\widetilde{W}\klgleich& \sum_{j=1}^{l-1}\Big(2^{\text{del}(AOP_{n_j}^j)}+2^{\text{del}(S_{n_j}^j)}\Big)+2^{\text{del}(AOP_{n_0}^0)}+2^0\nonumber\\
			\klgleich[\eqref{eq36},\eqref{eq38}]&\sum_{j=0}^{l-1}\Big(2^{\log_2W_{P^{(j)}}+\log_2\log_2(2n_j)+\log_2\log_2\log_2(2n_j) + 4.3}\Big) +
			\sum_{j=1}^{l-1}\Big(2^{\log_2W_{P^{(j)}}+3\log_2\log_2n_j+ 0.007}\Big) + 1\nonumber\\
			\klgleich[$ n_j\leq k $]&\sum_{j=0}^{l-1}\Big(W_{P^{(j)}}\log_2(2k)\log_2\log_2(2k) * 2^{4.3}+ W_{P^{(j)}}\log_2^3k* 2^{0.007}\Big)\nonumber\\
			\klgleich[Lem. \ref{numlem6} (i)]& \sum_{j=0}^{l-1}\Big(W_{P^{(j)}}\log_2^3k*1.097 + W_{P^{(j)}}\log_2^3k* 2^{0.007}\Big)\nonumber\\
			\klgleich& 2.102W\log_2^3k.\label{eq::bound_W_tilde_2}
\end{align}}\noindent
The circuit $ B_l $ then has delay
		\setlength{\mywidth}{\widthof{$\stackrel{l\leq n..}{\leq}$}} 
		\begin{align}
			\widetilde{\text{del}}(B_l)\klgleich[\eqref{eq40}]&\log_2\widetilde{W}+3\log_2\log_2l+5.007\nonumber\\
			\klgleich[$ \substack{l\leq n,\\(\ref{eq::bound_W_tilde_2})} $]&\log_2W + 3\log_2\log_2n + 3\log_2\log_2k + 6.079.\label{eqq63}
		\end{align}
		This delay again dominates the delay of the other sub-circuits since 
		\setlength{\mywidth}{\widthof{$\stackrel{k\leq \log_2^2n+1..}{\leq}$}} 
{\allowdisplaybreaks
		\begin{align}
			\text{del}(A_{n_j}^{j})\klgleich[\eqref{eq34}]&\log_2W_{P^{(j)}} + \log_2n_j + \log_2\log_2n_j + 2.65\nonumber\\
			\klgleich[$ \substack{n_j\leq k,\\ W_{P^{(j)}}\leq W} $]&\log_2W + \log_2k + \log_2\log_2k + 2.65\nonumber\\
			\klgleich[$ k\leq \log_2^2n+1 $]&\log_2W + 2\log_2\log_2n + \log_2\log_2k + 3.65\nonumber\\
			\klgleich&\log_2W + 3\log_2\log_2n + 3\log_2\log_2k + 7.079 \label{eqq64}
		\end{align}
}
		for $ j\in\{0,\dots,l-1\} $ and 
		\setlength{\mywidth}{\widthof{$\stackrel{W_{P^{(j)}}\leq W}{\leq}$}} 
		\begin{align}
			\text{del}(S_{n_j}^{j})\klgleich[\eqref{eq38}]&\log_2W_{P^{(j)}} + 3\log_2\log_2n_j + 0.007\nonumber\\
			\klgleich[$ \substack{n_j\leq k,\\ W_{P^{(j)}}\leq W} $]&\log_2W + 3\log_2\log_2k + 0.007\nonumber\\
			\klgleich&\log_2W + 3\log_2\log_2n + 3\log_2\log_2k + 6.079 \label{eqq65}
		\end{align}
		for $ j\in\{1,\dots, l-1\} $. Hence, Theorem~\ref{thmalg2} again implies the total delay bound
		\setlength{\mywidth}{\widthof{$\stackrel{Lem. \ref{numlem6} (iii)......}{\leq}$}} 
		\begin{align*}
			\text{del}(C_n)\klgleich&\max\Big\{\text{del}(A_{n_0}^{0}),\, \widetilde{\text{del}}(B_l)+2,\, \max_{j\in\{1,\dots,l-1\}}\{\text{del}(A_{n_j}^{j})+1, \text{del}(S_{n_j}^{j})+2\}\Big\}\\
			\klgleich[\eqref{eqq63},\eqref{eqq64},\eqref{eqq65}]&\log_2W + 3\log_2\log_2n + 3\log_2\log_2k + 8.079\\
			\klgleich[Lem. \ref{numlem6} (iii)]&\log_2W + 3\log_2\log_2n + 3\log_2\log_2\log_2n + e_1
		\end{align*}
		for $ e_1\geq 11.085 $.
		
		For the size bound, we again add the sizes of our computed sub-circuits separately. For the adder circuits $ A_{n_j}^{j} $ we get
		\setlength{\mywidth}{\widthof{$\stackrel{\eqref{eq35}..}{\leq}$}} 
		\begin{align}
			\sum_{j=0}^{l-1}\text{s}(A_{n_j}^{j})\klgleich[\eqref{eq35}]& \sum_{j=0}^{l-1} 6.2n_j\log_2n_j\stackrel{n_j\leq k}{\leq}\sum_{j=0}^{l-1} 6.2n_j\log_2k
			\gleich 6.2n\log_2k\stackrel{\text{Lem. \ref{numlem6} (ii)}}{\leq} 12.413n\log_2\log_2n. \label{eqq66}
		\end{align}
		Summing up the sizes of the \textsc{And}-prefix circuits $ S_{n_j}^{j} $ results in 
		\setlength{\mywidth}{\widthof{$\stackrel{\eqref{eq39}..}{\leq}$}} 
		\begin{align}
			\sum_{j=1}^{l-1}\text{s}(S_{n_j}^{j})\klgleich[\eqref{eq39}]& \sum_{j=1}^{l-1} 3.114n_j\log_2\log_2n_j\stackrel{n_j\leq k}{\leq}\sum_{j=1}^{l-1} 3.114n_j\log_2\log_2k\nonumber\\
			\gleich& 3.114n\log_2\log_2k\stackrel{\text{Lem. \ref{numlem6} (iv)}}{\leq} 2.339n\log_2\log_2n. \label{eqq67}
		\end{align}
		The sizes of the \textsc{And-Or}-path circuits $ AOP_{n_j}^{j} $ are summed to at most
		\setlength{\mywidth}{\widthof{$\stackrel{Lem. \ref{numlem6} (v)...}{\leq}$}} 
		{\allowdisplaybreaks
		\begin{align}
			\sum_{j=0}^{l-1}\text{s}(AOP_{n_j}^{j})
			\klgleich[\eqref{eq37}]& \sum_{j=0}^{l-1} \Big(2n_j\log_2(2n_j)+2n_j\log_2\log_2(2n_j)+ 
			2n_j\log_2\log_2\log_2(2n_j)+ 6.6n_j-1\Big)\nonumber\\
			\klgleich[$ n_j\leq k $]& \sum_{j=0}^{l-1} 2n_j\Big(\log_2(2k)+\log_2\log_2(2k)+ \log_2\log_2\log_2(2k)+3.3\Big)\nonumber\\
			\gleich& 2n\Big(\log_2(2k)+\log_2\log_2(2k)+ \log_2\log_2\log_2(2k)+3.3\Big)\nonumber\\
			\klgleich[Lem. \ref{numlem6} (v)]& 8.576n\log_2\log_2n. \label{eqq68}
		\end{align}}\noindent
		The size of the adder circuit $ B_l $ is bounded by 
		\begin{equation}
			\text{s}(B_l)\stackrel{\eqref{eq41}}{\leq}2.422l\log_2^2l\stackrel{\text{Lem. \ref{numlem7}}}{\leq}2.432n.\label{eqq69}
		\end{equation}
		Combining all these inequalities, Theorem~\ref{thmalg2} implies the total size bound 
		\setlength{\mywidth}{\widthof{$\stackrel{n\geq 2^{13}.....}{\leq}$}}
		\begin{align*}
			\text{s}(C_n)\klgleich& \text{s}(A_{n_0}^{0})+\text{s}(AOP_{n_{0}}^{0})+
			\sum_{j=1}^{l-1}\Big(\text{s}(A_{n_j}^{j})+\text{s}(AOP_{n_{j}}^{j})+\text{s}(S_{n_j}^{j})\Big) + \text{s}(B_l)+2n\\
			\klgleich[$ \substack{\eqref{eqq66},\eqref{eqq67},\\ \eqref{eqq68},\eqref{eqq69}} $]& 23.328n\log_2\log_2n +4.432n
			\klgleich[$n\geq 2^{16} $] n\log_2\log_2n\Big(23.328+\frac{4.432}{\log_216}\Big)\\
			\klgleich& e_2 n\log_2\log_2n
		\end{align*}
		for $ e_2\geq 24.436 $.

Summarized, we can guarantee the delay and size bounds for the circuit $ C_n $ for constants
$e_1\geq 11.085$ and $e_2\geq 24.436$.
	
For the runtime, It suffices to consider the case $ n> 2^{16} $. 
Here, we have a polynomial runtime analogously to Theorem~\ref{mainthm3} because we use the same algorithm with the same sub-circuits.
The only difference is that here we choose $ k=\lceil\log_2^2n\rceil $ instead of $ k=\lceil\log_2n\rceil $ which does not prevent a polynomial runtime.
	\end{proof}



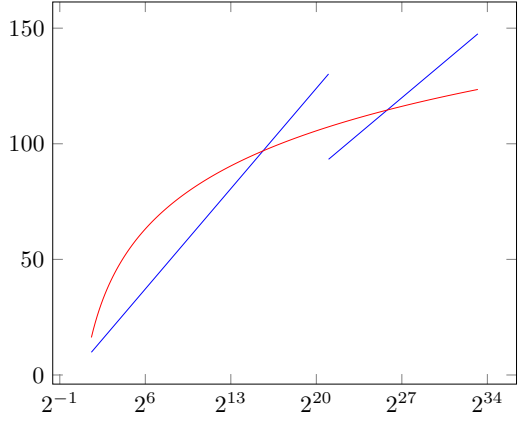
\begin{figure}[htb]
		\centering
		\begin{tikzpicture}[scale=0.89]
			\begin{axis}[xmode=log, log basis x={2}]
				\addplot[samples=200,blue,domain=3:2097152] {6.2*log2(x)};
				\addplot[samples=200,blue,domain=2097152:10000000000] {4.442*log2(x)};
				\addplot[samples=200,red,domain=3:10000000000] {24.436*log2(log2(x))};
			\end{axis}
		\end{tikzpicture}
		\caption[Comparison of the sizes of two adder circuits.]{Comparison of the sizes of the adders from Theorem~\ref{mainthm3} (blue) and Theorem~\ref{mainthm5} (red).}
		\label{figsize}
	\end{figure}

	\begin{remark}
		Figure~\ref{figsize} depicts the size bounds of the circuits from Theorem~\ref{mainthm3} (blue curves) and Theorem~\ref{mainthm5} (red curve) without the factor $ n $ in each bound. We use a logarithmic scale for the number of inputs (x-axis) such that the blue curve is linear. Note that our adder with size in $ \mathcal{O}(n\log_2\log_2n) $ only improves on our adder with size in $ \mathcal{O}(n\log_2n) $ for $ 50829\leq n\leq 2097152=2^{21} $ and for $ n\geq 58212864 $, so only for large instances.
	\end{remark}

	\begin{remark}
		The fanout of this circuit is in $ \mathcal{O}(\frac{\sqrt{n}}{\log_2n}) $ due to the same reason as for the adder from Theorem~\ref{mainthm3} (see the last remark in Section~\ref{smalleradder}) with the difference that, in this case, we have $ k=\lceil\log_2^2n\rceil $.
	\end{remark}

\subsubsection{Linear Size Adder Construction}
	\label{linsizeadderconstr}
	
Finally, we apply the linearization framework to our circuit from Theorem~\ref{mainthm5}.
This gives us a further delay increase of $ \log_2\log_2\log_2n $ plus some constant, according to Theorem~\ref{mainthm4}, because $ \sigma(n)\in\mathcal{O}(\log_2\log_2n) $. 
	
	But first, we examine what happens if we iteratively apply construction~1 of Theorem~\ref{mainthm4} to the respective next resulting circuit, starting with the adder from Theorem~\ref{mainthm5}. In the following, we write, for any $ k\in\mathbb{N} $,
	\[ \log_2^{(k)}x=\begin{cases}
		x & \text{for } k=0,\\
		\log_2(\log_2^{(k-1)}x) & \text{otherwise.}
	\end{cases} \]
	
	For the first iteration, we have $ \sigma(n)\in\mathcal{O}(\log_2\log_2n) $ and Theorem~\ref{mainthm4} implies an adder circuit with size in $ \mathcal{O}(n\log_2\log_2\log_2n) $ and a delay of at most $ \log_2W + 3\log_2\log_2n + 3\log_2\log_2\log_2n + 3\log_2\log_2\log_2\log_2n + const $. 
For the next iteration, we would have $ \sigma(n)\in\mathcal{O}(\log_2\log_2\log_2n) $, 
a size in $ \mathcal{O}(n\log_2\log_2\log_2\log_2n) $ and a delay increase of $ 3\log_2^{(5)}n $, 
disregarding the constants. Hence, we get the following result.
	\begin{corollary}
		\label{coralmostlinadders}
		For any $ j\in\mathbb{N}_{>2} $, we can construct a circuit with size in $ \mathcal{O}(n\log_2^{(j)}n) $ and a delay of at most \[ \log_2W + \sum_{i=2}^{j+1}3\log_2^{(i)}n + const \] on $ n $ input pairs with weight $ W $. \qed
	\end{corollary}

	If we apply the linearization framework (part 2 of Theorem~\ref{mainthm4}) to these circuits, we get the following result.	
	\begin{corollary}
		\label{corlinadders}
		For any $ j\in\mathbb{N}_{>2} $, we can construct a linear size adder circuit with a delay of at most \[ \log_2W + \sum_{i=2}^{j}3\log_2^{(i)}n + 4\log_2^{(j+1)}n + const \] on $ n $ input pairs with weight $ W $. \qed
	\end{corollary}
	
	The next corollary is another immediate consequence.	
	\begin{corollary}
		An asymptotic upper bound on the minimum delay of a linear size adder circuit is given by \[ \log_2W + \sum_{i=2}^{\infty}3\log_2^{(i)}n.\tag*{\qed}\]
	\end{corollary}
	
	A concrete computation of the corresponding constants of these circuits would quickly become unfeasible. Therefore, we only compute specific delay and size bounds for the direct linearization of our circuit from Theorem~\ref{mainthm5}. For this, we again need some numerical lemmas. 
	
	\begin{lemma}
		\label{numlem71}
		Assume that $ n\in \mathbb{N} $ with $ 3\leq n\leq 2^{28} $. Then
		\[\log_2n\leq 2\log_2\log_2n + 3\log_2\log_2\log_2n + 11.59.\]
	\end{lemma}

	\begin{proof}
		This lemma is proven analogously to Lemma~\ref{numlem3} by simply exchanging $ 2^{21} $ with $ 2^{28} $.
	\end{proof}

	\begin{lemma}
		\label{numlem8}
		Assume that $ n,k\in \mathbb{N} $ with $ n>2^{28} $ and $ k=\lceil\log_2\log_2n\rceil $. Then, the following inequalities hold:
		\begin{enumerate}[(i)]
			\item if $ k\leq 1336 $, then $ \log_2(2k)+\log_2\log_2(2k)\leq \log_2k+4.509 $,
			\item if $ k\geq 1337 $, then $ 2\sqrt{2\log_2(2k)}\leq \log_2k-0.841 $,
			\item $ \log_2k\leq \log_2\log_2\log_2n+ 0.279 $.
		\end{enumerate}
	\end{lemma}

	\begin{proof}
		For the first statement, it suffices to verify that $ \log_2\log_2(2k)\leq 3.509 $ for $ k= 1336 $ and, hence, for $ k\leq1336 $.
		
		We prove the second statement by verifying it for $ k=1337 $ and showing that the function \[f(x)=\log_2x-2\sqrt{2\log_2(2x)}-0.841\] is increasing for $ x\geq 1337 $. This is the case since the derivative of $ f $ is given by 
		\begin{equation*}
			f'(x)=\frac{1}{x\ln2}-\frac{1}{\sqrt{2\log_2(2x)}}*\frac{2}{x\ln2}=\frac{1}{x\ln2}\bigg(1-\frac{2}{\sqrt{2\log_2(2x)}}\bigg)
		\end{equation*}
		which is positive for $ x\geq 1337 $.
		
		For the last statement, we calculate
		\begin{align*}
			\log_2k&\leq\log_2(\log_2\log_2n+1)\stackrel{\log_2n\geq 28}{\leq} \log_2\left(\left(1+\frac{1}{\log_228}\right)\log_2\log_2n\right)
			\leq \log_2\log_2\log_2n+0.273.\tag*{\qedhere}
		\end{align*}
	\end{proof}

	\begin{lemma}
		\label{numlem9}
		Assume that $ n,k,l\in\mathbb{N} $ with $ n>2^{28} $, $ k=\lceil\log_2\log_2n\rceil $ and $ l=\lceil\frac{n}{k}\rceil $. Then, we have
		\[24.436l\log_2\log_2l\leq 24.437n.\]
	\end{lemma}

	\begin{proof}
		We first calculate
		\begin{align*}
			l&\leq \frac{n}{k}+1\leq \frac{n}{\log_2\log_2n}+1\stackrel{n\geq 2^{28}}{\leq}\Big(1+\frac{\log_228}{2^{28}}\Big)\frac{n}{\log_2\log_2n}
			\,\leq\, 1.00001\frac{n}{\log_2\log_2n}.
		\end{align*}
		Hence, we get
		\[24.436l\log_2\log_2l\leq 24.437\frac{n}{\log_2\log_2n}\log_2\log_2l\stackrel{l\leq n}{\leq}24.437n.\tag*{\qedhere}\]
	\end{proof}

	Now, we can explicitly construct our fourth and final adder circuit.

\linebox{
	\begin{theorem}
		\label{mainthm6}
		Let $ n\in \mathbb{N} $ with $ n\geq 3 $ and input pairs $ p_0, g_0, \dots, p_{n-1},g_{n-1} $ with arrival times $ a:\{p_0, g_0,\dots ,p_{n-1}, g_{n-1}\}\to \mathbb{N} $ be given and let $ W = \big(\sum_{i=0}^{n-1}2^{a(p_{i})} + 2^{a(g_{i})}\big) $ be the weight of the input pairs. We can construct an adder circuit $ C_{n} $ on these input pairs with delay
		\begin{align*}
			\mathrm{del}(C_{n})&\leq \log_2W + 3\log_2\log_2n + 4\log_2\log_2\log_2n + f_1 \intertext{and size} \mathrm{s}(C_n)&\leq f_2n
		\end{align*} 
		in polynomial time for $f_1=18.596$ and $f_2=53.877$.
	\end{theorem}
}

	\begin{proof}
		We first assume that $ 3\leq n\leq 268435456=2^{28} $. We again use the linear size, depth-optimizing adder from Theorem~\ref{daddercirclin} as $ C_n $. Due to Lemma~\ref{numlem71}, we get our desired bounds
		\setlength{\mywidth}{\widthof{$\stackrel{Thm. \ref{cordaddercirclin}~(b)..}{\leq}$}}
		\begin{align*}
			\text{del}(C_n)\klgleich[Thm. \ref{cordaddercirclin}~(b)]&\log_2W + \log_2n + \log_2\log_2n + \log_2\log_2\log_2n + 6.6\\
			\klgleich[Lem. \ref{numlem71}]&\log_2W + 3\log_2\log_2n + 4\log_2\log_2\log_2n + f_1
			\intertext{and}
			\text{s}(C_n)\klgleich&f_2n
		\end{align*}
		for $ f_1\geq 18.19 $ and $ f_2\geq 21.6 $.
		
		Hence, assume that $ n>2^{28} $. We again apply Algorithm \ref{alg2} together with Lemma~\ref{AOPpathlem}. In order to achieve a linear size, we use the same sub-circuits as in the proof of Theorem~\ref{mainthm4}:
		\begin{itemize}
			\item For the family of adder circuits $ (A_k)_{k\in\mathbb{N}} $, we use the circuits from Theorem~\ref{daddercirclin} yielding the bounds
			\setlength{\mywidth}{\widthof{$\stackrel{Cor. \ref{cordaddercirclin}~(c)..}{\leq}$}}
			\begin{align}
				\text{del}(A_k)\klgleich[Cor. \ref{cordaddercirclin}~(c)]&\log_2W +\log_2k + \log_2\log_2k + \log_2\log_2\log_2k+7.6,\label{eqq70}\\
				\text{s}(A_k)\klgleich&16.7k.\label{eqq71}
			\end{align}
			\item For the \textsc{And-Or}-path circuits $ (AOP_k)_{k\in\mathbb{N}} $, we make a case distinction. If $ k\leq 1336 $, we use the depth-optimizing circuits from Theorem~\ref{thmdaop}. Because the circuit $ AOP_k $  has $ 2k-1 $ inputs, we get the delay and size bounds 
			\setlength{\mywidth}{\widthof{$\stackrel{Cor. \ref{corthmall}~(b)..}{\leq}$}}
			\begin{align}
				\text{del}(AOP_k)\klgleich[Cor. \ref{corthmall}~(b)]&\log_2W+\log_2(2k)+\log_2\log_2(2k)+0.65,\nonumber\\ 
				\text{s}(AOP_k)\klgleich &7.34k-2.\nonumber
			\end{align}
			If $ k\geq 1337 $, we use the construction by Spirkl from Theorem~\ref{andorcircspir} guaranteeing the delay and size bounds
			\setlength{\mywidth}{\widthof{$\stackrel{}{\leq}$}}
			\begin{align}
				\text{del}(AOP_k)&\leq\log_2W+2\sqrt{2\log_2(2k)}+6,\nonumber\\ 
				\text{s}(AOP_k)&\leq6k.\nonumber
			\end{align}
			Due to Lemma~\ref{numlem8} (i) and (ii), we get in both cases that
			\setlength{\mywidth}{\widthof{$\stackrel{}{\leq}$}}
			\begin{align}
				\text{del}(AOP_k)&\leq\log_2W+\log_2k+5.159,\label{eqq72}\\ 
				\text{s}(AOP_k)&\leq7.34k-2.\label{eqq73}
			\end{align}
			\item For the family of \textsc{And}-prefix circuits $ (S_k)_{k\in\mathbb{N}} $, we use the construction by Ladner and Fischer from Theorem~\ref{thmlf} with $ f=0 $ which gives us the delay and size bounds 
			\setlength{\mywidth}{\widthof{$\stackrel{Thm. \ref{corthmall}~(a)..}{\leq}$}}
			\begin{align}
				\text{del}(S_k)\klgleich[Cor. \ref{corthmall}~(a)]& \log_2W+\lceil\log_2k\rceil,\label{eqq74}\\
				\text{s}(S_k)\klgleich &4k. \label{eqq75}
			\end{align}
			\item For the family of adder circuits $ (B_l)_{l\in\mathbb{N}} $, introduced in Lemma~\ref{AOPpathlem}, we use our latest constructed adder from Theorem~\ref{mainthm5} resulting in the delay and size bounds
			\begin{align}
				\text{del}(B_l)&\leq \log_2W + 3\log_2\log_2l + 3\log_2\log_2\log_2l + 11.085 \label{eqq751}\\
				\text{s}(B_l)&\leq 24.436l\log_2\log_2l.\label{eqq752}
			\end{align} 
		\end{itemize}
	
		In order to achieve a linear size for the sub-circuit $ B_l $, it suffices to choose $ k:=\lceil\log_2\log_2n\rceil $. 
		
		We again compute the delay by first bounding the weight $ \widetilde{W} $ of the input pairs 
		\[\text{out}(AOP_{n_0}^{0}),\text{out}_{n_1}(S_{n_1}^{1}), \text{out}(AOP_{n_1}^{1}),\dots,\text{out}_{n_{l-2}}(S_{n_{l-2}}^{l-2}), \text{out}(AOP_{n_{l-2}}^{l-2}) \] of $ B_l $. We have
		\setlength{\mywidth}{\widthof{$\stackrel{\eqref{eqq72},\eqref{eqq74}....}{\leq}$}} 
{\allowdisplaybreaks
\begin{align*}
			\widetilde{W}\klgleich& \sum_{j=1}^{l-1}\Big(2^{\text{del}(AOP_{n_j}^j)}+2^{\text{del}(S_{n_j}^j)}\Big)+2^{\text{del}(AOP_{n_0}^0)}+2^0\\
			\klgleich[\eqref{eqq72},\eqref{eqq74}]&\sum_{j=0}^{l-1}\Big(2^{\log_2W_{P^{(j)}}+\log_2n_j+5.159}\Big) + \sum_{j=1}^{l-1}\Big(2^{\log_2W_{P^{(j)}}+\lceil\log_2n_j\rceil}\Big) + 1\\
			\klgleich[$ n_j\leq k $]&\sum_{j=0}^{l-1}\Big(2^{\log_2W_{P^{(j)}}+\log_2k+5.159}+ 2^{\log_2W_{P^{(j)}}+\log_2k+1}\Big)\\
			\klgleich& (2^{5.159}+2)*Wk\\
			\klgleich& 37.729Wk.
\end{align*}
}
		Then, the delay of $ B_l $ is bounded by
		\setlength{\mywidth}{\widthof{$\stackrel{l\leq n..}{\leq}$}} 
		\begin{align}
			\widetilde{\text{del}}(B_l)\klgleich[\eqref{eqq751}]&\log_2\widetilde{W}+ 3\log_2\log_2l + 3\log_2\log_2\log_2l + 11.085\nonumber\\
			\klgleich[$ l\leq n $]&\log_2W + 3\log_2\log_2n + 3\log_2\log_2\log_2n + \log_2k+16.323.\label{eqq76}
		\end{align}
		This delay again dominates the delay of the other sub-circuits, since 
		\setlength{\mywidth}{\widthof{$\stackrel{k\leq \log_2\log_2n+1..}{\leq}$}} 
		\begin{align}
			\text{del}(A_{n_j}^{j})\klgleich[\eqref{eqq70}]&\log_2W_{P^{(j)}} +\log_2n_j + \log_2\log_2n_j
			+ \log_2\log_2\log_2n_j+7.6\nonumber\\
			\klgleich[$ \substack{n_j\leq k,\\ W_{P^{(j)}}\leq W} $]&\log_2W +\log_2k + \log_2\log_2k + \log_2\log_2\log_2k+7.6\nonumber\\
			\klgleich[$ k\leq \log_2\log_2n+1 $]&\log_2W + 3\log_2\log_2n + 3\log_2\log_2\log_2n
			 + \log_2k+17.323 \label{eqq77}
		\end{align}
		for $ j\in\{0,\dots, l-1\} $ and 
		\setlength{\mywidth}{\widthof{$\stackrel{W_{P^{(j)}}\leq W}{\leq}$}} 
		{\allowdisplaybreaks
		\begin{align}
			\text{del}(S_{n_j}^{j})\klgleich[\eqref{eqq74}]&\log_2W_{P^{(j)}} +\lceil\log_2n_j\rceil\nonumber\\
			\klgleich[$ \substack{n_j\leq k,\\ W_{P^{(j)}}\leq W} $]&\log_2W +\log_2k +1\nonumber\\
			\klgleich&\log_2W + 3\log_2\log_2n + 3\log_2\log_2\log_2n
			 + \log_2k+16.323 \label{eqq78}
		\end{align}}\noindent
		for $ j\in\{1,\dots, l-1\} $. Hence, we get the total delay bound
		\setlength{\mywidth}{\widthof{$\stackrel{Lem. \ref{numlem8} (iii)......}{\leq}$}} 
		\begin{align*}
			\text{del}(C_n)\klgleich[Thm. \ref{thmalg2}]&\max\Big\{\text{del}(A_{n_0}^{0}),\, \widetilde{\text{del}}(B_l)+2,\, \max_{j\in\{1,\dots,l-1\}}\{\text{del}(A_{n_j}^{j})+1,\text {del}(S_{n_j}^{j})+2\}\Big\}\\
			\klgleich[\eqref{eqq76},\eqref{eqq77},\eqref{eqq78}]&\log_2W + 3\log_2\log_2n + 3\log_2\log_2\log_2n + \log_2k +18.323\\
			\klgleich[Lem. \ref{numlem8} (iii)]&\log_2W + 3\log_2\log_2n + 4\log_2\log_2\log_2n + f_1
		\end{align*}
		for $ f_1\geq 18.596 $.
		
		For the size, we can directly calculate
		\setlength{\mywidth}{\widthof{$\stackrel{Lem. \ref{numlem9}..}{\leq}$}}
{\allowdisplaybreaks
		\begin{align*}
			\text{s}(C_n)\klgleich[Thm. \ref{thmalg2}]& \text{s}(A_{n_0}^{0})+\text{s}(AOP_{n_{0}}^{0})+
			\sum_{j=1}^{l-1}\Big(\text{s}(A_{n_j}^{j})+\text{s}(AOP_{n_{j}}^{j})+\text{s}(S_{n_j}^{j})\Big) + \text{s}(B_l)+2n\\
			\klgleich[$ \substack{\eqref{eqq71},\eqref{eqq73},\\ \eqref{eqq75},\eqref{eqq752}} $]&\sum_{j=0}^{l-1}(16.1n_j+7.34n_j-2+4n_j)+24.436l\log_2\log_2l+2n \\
			\klgleich& 29.44n+24.436l\log_2\log_2l\\
			\klgleich[Lem. \ref{numlem9}]& 29.44n+ 24.437n\\
			\klgleich& f_2n
		\end{align*}
}
		for $ f_2\geq 53.877 $.

Hence, we get the delay and size bounds for the circuit $ C_n $ for constants
$f_1 \geq 18.596$ and $f_2 \geq 53.877$.
	
		For the runtime, we again note that we use Algorithm \ref{alg2} combined with Lemma~\ref{AOPpathlem} with a polynomial amount of sub-circuits which all have polynomial runtime (see Theorem~\ref{daddercirclin}, Theorem~\ref{thmdaop}, Theorem~\ref{andorcircspir}, Theorem~\ref{thmlf}, Theorem~\ref{mainthm5}). Analogously to the proof of Theorem~\ref{mainthm3}, we get a polynomial runtime overall.
	\end{proof}



	\begin{remark}
		As for the fanout, we chose the circuit from Theorem~\ref{mainthm5} with a fanout in $ \mathcal{O}(\frac{\sqrt{l}}{\log_2l}) $ (see the last remark in Section~\ref{sizenloglogn}) as $ B_l $, and we chose $ k=\lceil\log_2\log_2n\rceil $ in Algorithm \ref{alg2}. 
Therefore, the circuit from \Cref{mainthm6} has a fanout in 
$ \mathcal{O}\Big(\frac{\sqrt{n}}{\sqrt{\log_2\log_2n}(\log_2n-\log_2\log_2\log_2n)}\Big) $.
	\end{remark}

\section{Conclusions}\label{sec::conclusions}

We have constructed several fast and small adder circuits.


	
Using the \textsc{And}-prefix circuits by Rautenbach, Szegedy and Werber \cite{RSW} 
and the \textsc{And-Or} path circuits by Brenner and Hermann \cite{BH19}, 
we constructed an adder circuit in polynomial runtime with a delay of at most \[ \log_2W+3\log_2\log_2n+const \]
and a size in $ \mathcal{O}(n\log_2^2n) $ in Section~\ref{deloptadder}. This is the best delay bound known for adders with a sub-quadratic size and exceeds the delay bound of the fastest known \textsc{And-Or} path circuit (and thus, adder circuit), provided in \cite{BH19} (Theorem~\ref{andorcirc}), only by $ 2\log_2\log_2n-\log_2\log_2\log_2n+const $.
	
We developed two size reduction frameworks for delay-optimizing adder circuits, based on a linearization framework from \cite{BS24} 
for depth-optimizing adders. Applying them to our previously constructed adder gave us several smaller adders with a slightly larger delay guarantee. In particular, we achieved two adder circuits with a delay bound of
\[ \log_2W+3\log_2\log_2n+3\log_2\log_2\log_2n+ const \]
and sizes in $ \mathcal{O}(n\log_2n) $ and $ \mathcal{O}(n\log_2\log_2n) $, respectively. Finally, we constructed a linear size adder with a delay bound of
\[ \log_2W+3\log_2\log_2n+4\log_2\log_2\log_2n+ const. \]
Hence, we were able to linearize our first constructed adder by the cost of a delay 
increase of order $ \mathcal{O}(\log_2\log_2\log_2n) $ and improve on the previously 
best known delay bound of $ \log_2W+\mathcal{O}(\sqrt{\log_2n}) $ for linear size adders,
provided by Spirkl \cite{Sp14}. In particular, the distance to the lower bound of 
$\log_2 W$ could be reduced significantly from $\Theta(\sqrt{\log_2n})$ to
$ \Theta(\log_2\log_2\log_2n) $
	
	Lastly, we also presented a method to construct circuits with a delay of at most $ \log_2W + \sum_{i=2}^{j+1}3\log_2^{(i)}n + const $ and a size in $ \mathcal{O}(n\log_2^{(j)}n) $, as well as linear size circuits with a delay of at most \[ \log_2W + \sum_{i=2}^{j}3\log_2^{(i)}n + 4\log_2^{(j+1)}n + const \] for all $ j\in\mathbb{N}_{>2} $. This implies an asymptotic upper bound on the minimum delay of linear size adder circuits of $ \log_2W + \sum_{i=2}^{\infty}3\log_2^{(i)}n $.

	
	\addcontentsline{toc}{section}{Index}
	\printindex

	\bibliographystyle{alpha}

\small

	\addcontentsline{toc}{section}{References}
	\bibliography{LiteraturMasterarbeit}

\end{document}